\documentclass[11pt,a4paper]{article}
\pdfoutput=1

\usepackage{jheppub}
\usepackage{latexsym}
\usepackage{multirow}
\usepackage{color}
\usepackage[usenames,dvipsnames,table]{xcolor}
\usepackage{graphicx}% Include figure files
\usepackage{epsfig}  % Include figure files
\usepackage{epsf}    % Include figure files
\usepackage{grffile}
\usepackage{dcolumn}% Align table columns on decimal point
\usepackage{bm}% bold math
\usepackage{dcolumn}% Align table columns on decimal point
\usepackage{textcomp}% Align table columns on decimal point
\usepackage{float}
\usepackage{subfig}
\usepackage{hypcap}
\usepackage[]{hyperref}
\usepackage{makecell}
\usepackage{color}
\usepackage{pifont}
\usepackage{appendix}
\usepackage{amsmath}
\usepackage{multirow,bigdelim}  
\usepackage{lineno}
\usepackage[normalem]{ulem}

\hypersetup{
  bookmarks=true,         % show bookmarks bar?
  unicode=false,          % non-Latin characters in Acrobat?s bookmarks
  pdftoolbar=true,        % show Acrobat?s toolbar?
 pdfmenubar=true,        % show Acrobat?s menu?
 pdffitwindow=true,     % window fit to page when opened
 pdfstartview={FitH},    % fits the width of the page to the window
 pdfsubject={Neutrino Oscillations Phenomenology},   % subject of the document
 pdfnewwindow=true,      % links in new window
 pdfcreator={RevTeX},
 colorlinks=true,       % false: boxed links; true: colored links
 linkcolor=red,          % color of internal links
 citecolor=blue,        % color of links to bibliography
 filecolor=black,      % color of file links
 urlcolor=blue,           % color of external links
  }

\newcommand{\be}{\begin{equation}}
\newcommand{\ee}{\end{equation}}
\newcommand{\ba}{\begin{eqnarray}}
\newcommand{\ea}{\end{eqnarray}}
\def\epsmutau{\varepsilon_{\mu\tau}}

\newcommand{\capdef}{}
\newcommand{\mycaption}[2][\capdef]{\renewcommand{\capdef}{#2}
       \caption[#1]{{\footnotesize #2}}}
\makeatletter
\renewcommand{\fnum@table}{\textbf{\tablename~\thetable}}
\renewcommand{\fnum@figure}{\textbf{\figurename~\thefigure}}
\makeatother
\preprint{IP/BBSR/2020-8, TIFR/TH/20-49}

\title{A New Approach to Probe Non-Standard Interactions in Atmospheric Neutrino Experiments }

\author[a,b,c]{Anil Kumar,}
\author[d]{Amina Khatun,} 
\author[a,c,e]{Sanjib Kumar Agarwalla,}
\author[f]{Amol Dighe} 

\affiliation[a]{Institute of Physics, Sachivalaya Marg, Sainik School Post,
  Bhubaneswar 751005, India}
\affiliation[b]{Applied Nuclear Physics Division, Saha Institute of
  Nuclear Physics, Block AF, Sector 1, Bidhannagar, Kolkata 700064, India}
\affiliation[c]{Homi Bhabha National Institute, Anushakti Nagar,
  Mumbai 400094, India}
\affiliation[d]{Faculty of Mathematics, Physics and Informatics, Comenius University, Mlynsk\'{a} dolina F1, SK842 48 Bratislava, Slovakia }
\affiliation[e]{International Centre for Theoretical Physics,
  Strada Costiera 11, 34151 Trieste, Italy}
\affiliation[f]{Tata Institute of Fundamental Research, Homi Bhabha Road,
  Colaba, Mumbai 400005, India}

\emailAdd{anil.k@iopb.res.in (ORCID: 0000-0002-8367-8401)}
\emailAdd{amina.khatun@fmph.uniba.sk (ORCID: 0000-0003-3493-607X)}
\emailAdd{sanjib@iopb.res.in (ORCID: 0000-0002-9714-8866)}
\emailAdd{amol@theory.tifr.res.in (ORCID: 0000-0001-6639-0951)}

\abstract
{We propose a new approach to explore the  neutral-current non-standard
  neutrino interactions (NSI) in atmospheric neutrino experiments using
  oscillation dips and valleys in reconstructed muon observables,
  at a detector like ICAL that can identify the muon charge.
  We focus on the flavor-changing NSI parameter $\varepsilon_{\mu\tau}$,
  which has the maximum impact on the muon survival probability
  in these experiments.
  We show that non-zero $\varepsilon_{\mu\tau}$ shifts the oscillation
  dip locations in $L/E$ distributions of the up/down event ratios of
  reconstructed $\mu^-$ and $\mu^+$ in opposite directions.
  We introduce a new variable $\Delta d$ representing the difference
  of dip locations in $\mu^-$ and $\mu^+$, which is sensitive to the
  magnitude as well as the sign of $\varepsilon_{\mu\tau}$,
  and is independent of the value of $\Delta m^2_{32}$.
  We further note that the oscillation valley in the ($E$, $\cos \theta$)
  plane of the reconstructed muon observables bends in the presence of NSI,
  its curvature having opposite signs for $\mu^-$ and $\mu^+$.
  We demonstrate the identification of NSI with this curvature, 
  which is feasible for detectors like ICAL having excellent muon energy
  and direction resolutions.
  We illustrate how the measurement of contrast in the curvatures
  of valleys in $\mu^-$ and $\mu^+$ can be used to estimate $\epsmutau$.  
  Using these proposed oscillation dip and valley measurements,
  the achievable precision on $|\varepsilon_{\mu\tau}|$ at 90\% C.L.
  is about 2\% with 500 kt$\cdot$yr exposure. 
  The effects of statistical fluctuations, systematic errors,
  and uncertainties in oscillation parameters have been incorporated using
  multiple sets of simulated data.
  Our method would provide a direct and robust measurement of
  $\varepsilon_{\mu\tau}$ in the multi-GeV energy range.}

\arxivnumber{2101.02607}
\begin{document}
\maketitle
\flushbottom

%==========================
\section{Introduction and Motivation}
\label{sec:intro}
%==========================

The phenomenon of neutrino oscillations has now been well-established,
from measurements at the solar, atmospheric, reactor as well as accelerator
experiments with short and long baselines \cite{Zyla:2020zbs}. 
Neutrino oscillations are the consequences of mixing among different neutrino
flavors and non-degenerate values of neutrino masses, with at least
two neutrino masses nonzero.
However, neutrinos are massless in the Standard Model (SM) of particle
physics, and therefore, physics beyond the SM (BSM) is needed to accommodate
nonzero neutrino masses and mixing. Many models of BSM physics suggest new
non-standard interactions (NSI) of neutrinos~\cite{Wolfenstein:1977ue},
which may affect neutrino production, propagation, and detection in a
given experiment. 
The possible impact of these NSI at neutrino oscillation experiments
have been studied extensively, for example 
see Refs.~\cite{Valle:1987gv,Guzzo:1991hi,Guzzo:2000kx,Huber:2001zw,Gago:2001si,Escrihuela:2011cf,GonzalezGarcia:2011my,Ohlsson:2012kf,Gonzalez-Garcia:2013usa,Farzan:2017xzy,Miranda:2015dra,Dev:2019anc}.
In this paper, we propose a new method for identifying NSI at atmospheric
neutrino experiments which can reconstruct the energy, direction, as
well as charge of the muons produced in the detector due to charged-current
interactions of $\nu_\mu$ and $\bar\nu_\mu$.

We shall focus on the neutral-current NSI, which may be described at low
energies via effective four-fermion dimension-six operators
as~\cite{Wolfenstein:1977ue}   
\begin{equation}
 {\mathcal L}_{\rm{NC-NSI}} = -2\sqrt 2 \, G_F \,\varepsilon^{f}_{\alpha\beta, C} 
 \, (\bar\nu_{\alpha}\gamma^\rho P_L\nu_{\beta}) \, 
 (\bar f \gamma_\rho P_C f) \, ,
\end{equation}
where $G_{F}$ is the Fermi constant. The dimensionless parameters
$\varepsilon^{f}_{\alpha \beta, C}$ describe the strength of NSI,
where the superscript $f \in \{e,u,d\}$ denotes the matter fermions
($e$: electron, $u$: up-quark, $d$: down-quark), and the indices
$\alpha, \beta \in \{e,\mu,\tau \}$ refer to the neutrino flavors.
The subscript $C \in \{L,R \}$ represents the chiral projection operator
$P_L = (1 - \gamma_5)/2$ or $P_R = (1 + \gamma_5)/2$.
The hermiticity of the interactions demands
$\varepsilon^{f}_{\beta\alpha, C} = (\varepsilon^{f}_{\alpha \beta, C})^*$.

The effective NSI parameter relevant for the neutrino propagation through
matter is 
\begin{equation}
  \varepsilon_{\alpha\beta} \equiv \sum_{f=e,u,d} \left(\varepsilon_{\alpha\beta}^{fL}
  + \varepsilon_{\alpha\beta}^{fR}\right) \,\,\frac{N_f}{N_e}
  \equiv \sum_{f=e,u,d} \varepsilon_{\alpha\beta}^{f}\,\, \frac{N_f}{N_e}\,,
  \label{eq:eps-definition}
\end{equation}
where $N_f$ is the number density of fermion $f$. In the approximation of
a neutral and isoscalar Earth, the number densities of electrons, protons,
and neutrons are identical, which implies $N_u \approx N_d \approx 3 N_e$.
Thus,
\begin{equation}
  \varepsilon_{\alpha\beta} \approx \varepsilon^e_{\alpha\beta}
  + 3 \, \varepsilon^u_{\alpha\beta} + 3 \, \varepsilon^d_{\alpha\beta}\,. 
\end{equation}
In the presence of NSI, the modified effective Hamiltonian for neutrino
propagation through matter is 
\begin{equation}
 H_{\rm eff} = \frac{1}{2E} \, U \left(\begin{array}{ccc}
   0 & 0 & 0 \\
   0 & \Delta m^2_{21} & 0 \\
   0 & 0 & \Delta m^2_{31}\end{array}
 \right) U^\dag + V_{\rm CC}\left(\begin{array}{ccc}
   1 + \varepsilon_{ee} & \varepsilon_{e\mu} & \varepsilon_{e\tau} \\
   \varepsilon_{e\mu}^* & \varepsilon_{\mu\mu} & \varepsilon_{\mu\tau} \\
   \varepsilon_{e\tau}^* & \varepsilon_{\mu\tau}^* & \varepsilon_{\tau\tau}
 \end{array}\right)\,,
\end{equation}
where $U$ is the Pontecorvo-Maki-Nakagawa-Sakata (PMNS) matrix
that parametrizes neutrino mixing, and
$\Delta m^2_{ij} \equiv m_i^2 - m_j^2$ are the mass-squared differences.
The quantity $V_{\rm CC} \equiv \sqrt{2} G_F N_e$ is the effective matter
potential due to the coherent elastic forward scattering of neutrinos
with electrons inside the medium via the SM gauge boson $W$. Thus, the effective potential due to NSI would be ${(V_{\rm NSI})}_{\alpha\beta} = \sqrt{2} G_F N_e \, \varepsilon_{\alpha\beta}$. 
For antineutrinos, $V_{\rm CC} \rightarrow -V_{\rm CC}$, $U \rightarrow U^*$,
and $\varepsilon_{\alpha\beta} \to \varepsilon_{\alpha\beta}^*$.

In the present study, we suggest a novel approach to unravel the presence of
flavor-changing neutral-current NSI parameter $\varepsilon_{\mu\tau}$,
via its effect on the propagation of multi-GeV atmospheric neutrinos
and antineutrinos through Earth matter.
We choose $\varepsilon_{\mu\tau}$ as the NSI parameter to focus on,
since it can significantly affect the evolution of the mixing angle
$\theta_{23}$ and the mass-squared difference $\Delta m^2_{32}$
in matter~\cite{KumarAgarwalla:2021twp}, which in turn would alter the survival
probabilities of atmospheric muon neutrinos and antineutrinos 
substantially~\cite{GonzalezGarcia:2004wg,Mocioiu:2014gua}.  
In general, $\varepsilon_{\mu\tau}$ can be complex, {\it i.e.}
$\epsmutau \equiv |\epsmutau| e^{i\phi_{\mu\tau}}$.
However, in the disappearance channels $\nu_\mu \to \nu_\mu$ and
$\bar\nu_\mu \to  \bar\nu_\mu$ that dominate in our
analysis\footnote{The appearance channels $\nu_e \to \nu_\mu$ and
  $\bar\nu_e \to \bar\nu_\mu$ also contribute to the muon events in our
  analysis, however these channels are not affected by $\epsmutau$ to
  leading order~\cite{Kopp:2007ne}.}, 
$\epsmutau$ appears only as $|\epsmutau| \cos\phi_{\mu\tau}$ at the
leading order~\cite{Kopp:2007ne}.  
Thus, a complex phase only changes the effective value of $\epsmutau$
to a real number between $-|\epsmutau|$ and $+|\epsmutau|$ at the
leading order. 
We take advantage of this observation, and restrict ourselves to real
values of $\epsmutau$ in the range $-0.1 \leq \epsmutau \leq 0.1$.
From the arguments given above, this covers the whole range of complex
values of $\epsmutau$ with $|\epsmutau| \leq 0.1$.

Based on the global neutrino data analysis in~\cite{Esteban:2018ppq},
where the possible contributions to NSI from only up and down quark have
been included, the bound on $|\epsmutau|$ turns out to be 
$|\varepsilon_{\mu\tau}|<$ 0.07 at 2$\sigma$ confidence level. A phenomenological study to constrain $\epsmutau$ using preliminary IceCube and DeepCore data  has been performed in~\cite{Esmaili:2013fva}. 
The existing bounds on $\varepsilon_{\mu\tau}$ from various neutrino
oscillation experiments are listed in Table~\ref{tab:existing-limit}.
An important point to note is the energies of neutrinos involved in these
measurements: the IceCube results are obtained using high energy events
($>300$ GeV)~\cite{Salvado:2016uqu}, the energy threshold of DeepCore
is around 10 GeV~\cite{Aartsen:2017xtt}, while the Super-K experiment
is more efficient in the sub-GeV energy range~\cite{Mitsuka:2011ty}.

%%%%%%%%%%%%%%%%%%%%%%%%%%%%%%%%%%%%%%%%%%%%%%%%%%%%%%%%%%%%%%%%%%%% 
\begin{table}[h!]
  \begin{center}
    \begin{tabular}{|c|c|c|}
      \hline \hline 
      \multirow{2}{*}{Experiment}    & \multicolumn{2}{c|}{$90\%$ C.L. bounds} \\
      \cline{2-3}
      & Convention in~\cite{Salvado:2016uqu,Aartsen:2017xtt,Mitsuka:2011ty} & Our convention~\cite{Ohlsson:2012kf,Gonzalez-Garcia:2013usa,Choubey:2015xha,Farzan:2017xzy,Khatun:2019tad}\\ 
      \hline
      IceCube \cite{Salvado:2016uqu} & 
      $-0.006 < \tilde{\varepsilon}_{\mu\tau} < 0.0054 $ &
      $-0.018<\varepsilon_{\mu\tau}<0.0162 $ \\
      DeepCore \cite{Aartsen:2017xtt} &
      $-0.0067<\tilde{\varepsilon}_{\mu\tau}< 0.0081$  & 
      $-0.0201<\varepsilon_{\mu\tau} <0.0243 $ \\ 
      Super-K \cite{Mitsuka:2011ty}& 
      $|\tilde{\varepsilon}_{\mu\tau}| < 0.011$ & 
      $|\varepsilon_{\mu\tau}| < 0.033$ \\
      \hline \hline
    \end{tabular}
  \end{center}
  \mycaption{Existing bounds on $\varepsilon_{\mu\tau}$ at $90\%$ confidence level. Note that the bounds presented in~\cite{Salvado:2016uqu,Aartsen:2017xtt,Mitsuka:2011ty} are on $\tilde{\varepsilon}_{\mu\tau}$ that is defined according to the convention $V_{\rm NSI} = \sqrt{2} G_F N_d\, \tilde{\varepsilon}_{\mu\tau}$, while we use the convention  $V_{\rm NSI} = \sqrt{2} G_F N_e \,\epsmutau$ ($\epsmutau$ is defined in Eq.~\ref{eq:eps-definition}). Since $N_d \approx 3 N_e$ in Earth, the bounds in \cite{Salvado:2016uqu,Aartsen:2017xtt,Mitsuka:2011ty} on $\tilde{\varepsilon}_{\mu\tau}$ have been converted to the bounds on $\epsmutau$, using $\epsmutau = 3\, \tilde{\varepsilon}_{\mu\tau}$, as shown in the third column.}
    \label{tab:existing-limit}
\end{table}
%%%%%%%%%%%%%%%%%%%%%%%%%%%%%%%%%%%%%%%%%%%%%%%%%%%%%%%%%%%%%%%%%%%

The proposed 50 kt magnetized Iron Calorimeter (ICAL) detector at the
India-based Neutrino Observatory (INO)~\cite{Kumar:2017sdq,INO}
would be sensitive to multi-GeV neutrinos, since it can efficiently
detect muons in the energy range 1--25 GeV. 
Note that the MSW
resonance~\cite{Mikheev:1986wj,Mikheev:1986gs,Wolfenstein:1977ue}
due to Earth matter takes place for neutrino energies around 4--10 GeV,
so ICAL would also be in a unique position to detect any interplay between
the matter effects and NSI. 
Another important feature is that ICAL can explore physics in
neutrinos and antineutrinos separately, unlike Super-K and IceCube/DeepCore.
The studies of physics potential of ICAL for detecting NSI have shown that,
using the reconstructed muon momentum, it would be possible to obtain a bound
of $|\varepsilon_{\mu\tau}| < 0.015$ at 90$\%$ C.L.~\cite{Choubey:2015xha}
with 500 kt$\cdot$yr exposure.
When information on the reconstructed hadron energy in each event is also
included, the expected 90$\%$ C.L. bound improves to
$|\varepsilon_{\mu\tau}|<0.010$~\cite{Khatun:2019tad}.
The results in~\cite{Choubey:2015xha,Khatun:2019tad} are obtained using
a $\chi^2$ analysis with the
pull method~\cite{Huber:2002mx,Fogli:2002au,GonzalezGarcia:2004wg}. 

The wide range of neutrino energies and baselines available in
atmospheric neutrino experiments offer an opportunity to study the
features of ``oscillation dip'' and ``oscillation valley'' in the
reconstructed $\mu^-$ and $\mu^+$ observables, as demonstrated
in~\cite{Kumar:2020wgz}.
These features can be clearly identified in the ratios of upward-going
and downward-going muon events at ICAL. If the muon
neutrino disappearance is solely due to non-degenerate masses and
non-zero mixing of neutrinos, then the valley in both $\mu^-$
and $\mu^+$ is approximately a straight line.
The location of the dip, and the alignment of the valley, can be
used to determine $\Delta m^2_{32}$ \cite{Kumar:2020wgz}.
These features may undergo major changes in the presence of NSI,
and can act as smoking gun signals for NSI.

The novel approach, which we propose in this paper, is to probe the  NSI
parameter $\varepsilon_{\mu\tau}$ based on the elegant features associated
with the oscillation dip and valley, both of which arise from the same physics phenomenon, viz. the first oscillation minimum in the muon neutrino survival probability.
For the oscillation dip feature, we note that non-zero $\epsmutau$ shifts
the oscillation dip location in opposite directions for $\mu^-$ and $\mu^+$.
We demonstrate that this opposite shift in dip location due to NSI
can be clearly seen in the $\mu^-$ and $\mu^+$ data by reconstructing
$L_\mu^{\rm rec}/E_\mu^{\rm rec}$ distributions, thanks to the excellent
energy and direction resolutions for muons at ICAL.
We develop a whole new analysis methodology to extract
the information on $\varepsilon_{\mu\tau}$ using the dip locations.
For this, we define a new variable exploiting the contrast between the shifts
in reconstructed dip locations, which eliminates the dependence of our results
on the actual value of $\Delta m^2_{32}$.
For the oscillation valley feature, we notice that the valley becomes curved
in the presence of non-zero $\varepsilon_{\mu\tau}$, and the direction of this
bending is opposite for neutrino and antineutrino.
We then demonstrate that this opposite bending can indeed be observed in
expected $\mu^-$ and $\mu^+$ events.
We propose a methodology to extract the information on the bending
of the valley in terms of reconstructed muon variables, and use it for
identifying NSI.

In Sec.~\ref{sec:prob}, we discuss the oscillation probabilities of neutrino
and antineutrino in the presence of non-zero $\varepsilon_{\mu\tau}$, and
discuss the shifts in the dip locations as well as the bending of the
oscillation valleys in the survival probabilities of $\nu_\mu$ and $\bar\nu_\mu$.
In Sec.\,\ref{sec:evt-ical}, we investigate the survival of these two
striking features in the  reconstructed $L_\mu^{\rm rec}/E_\mu^{\rm rec}$
distributions and in the ($E_\mu^{\rm rec},\,\cos\theta_\mu^{\rm rec}$)
distributions of $\mu^-$ and $\mu^+$ events separately at ICAL. 
In Sec.~\ref{sec:reco-dip}, we propose a novel variable for identifying
the NSI, which is based on the contrast in the shifts of dip locations in
$\mu^-$ and $\mu^+$.
This variable leads to the calibration of $\varepsilon_{\mu\tau}$,
and is used to find the expected bound on $\varepsilon_{\mu\tau}$ from
a 500 kt$\cdot$yr exposure of ICAL.
In Sec.~\ref{subsec:dmsq-si}, we come up with a new procedure for
determining the alignment of the oscillation valley and estimating the value
of $\Delta m^2_{32}$ in the absence of NSI, which we extend
to the NSI analysis in Sec.~\ref{sec:reco-valley}. Here, we measure the
contrast in the curvatures of the oscillation valleys in $\mu^-$ and $\mu^+$
in the presence of NSI, and use it for determining the expected bound
on $\varepsilon_{\mu\tau}$ from the valley analysis at ICAL.
Finally, in Sec.~\ref{sec:conclusion}, we summarize our
findings and offer concluding remarks.

%%%%%%%%%%%%%%%%%%%%%%%%%%%%%%%%%%%%%%%%%%%%%%%%%%%%%%%%%%%%%%%
\section{Oscillation Dip and Valley in the Presence of NSI}
\label{sec:prob}
%%%%%%%%%%%%%%%%%%%%%%%%%%%%%%%%%%%%%%%%%%%%%%%%%%%%%%%%%%%%%%

In the limit of $\theta_{13} \to 0$, and the approximation of one mass scale
dominance scenario [$\Delta m^2_{21} L/(4E) \ll \Delta m^2_{32} L/(4E)$]
and constant matter density,
the survival probability of $\nu_\mu$ when traveling a distance $L_\nu$
is given by\,\cite{GonzalezGarcia:2004wg} 
\begin{equation}
  P_{\nu_\mu\rightarrow\nu_\mu} = 1 - \sin^2 2\theta_{\rm eff} \,
  \sin^2\left[ \xi \,\frac{\Delta m^2_{32} L_\nu}{4 E_\nu} \right]\,,
  \label{eq:pmumu-nsi-mutau-omsd}
\end{equation}
where
\begin{equation}
  \sin^2 2\theta_{\rm eff} =
  \frac{|\sin 2\theta_{23} +  2 \beta \, \eta_{\mu\tau}|^2}{\xi^2}\,,
  \label{eq:pmumu-thetaeff-omsd}
\end{equation}
\begin{equation}
  \xi = \sqrt{|\sin 2\theta_{23} + 2 \beta \, \eta_{\mu\tau}|^2
    + \cos^2 2\theta_{23}}\,,
  \label{eq:pmumu-xi-omsd}
\end{equation}
and
\begin{equation}
  \eta_{\mu\tau} = \frac{ 2 E_\nu \,V_{\rm CC} \,\varepsilon_{\mu\tau}}{|\Delta m^2_{32}|}
  \,,
\end{equation}
  where $\beta \equiv {\rm sgn}(\Delta m^2_{32})$. That is, $\beta = +1$
  for normal mass ordering (NO), and $\beta=-1$ for inverted mass ordering
  (IO).

In the limit of maximal mixing ($\theta_{23} = 45^{\circ}$),
Eq.\,\ref{eq:pmumu-nsi-mutau-omsd} reduces to the following simple
expression\,\cite{Mocioiu:2014gua}:
\begin{equation}
  P_{\nu_\mu\rightarrow\nu_\mu} = \cos^2 \left[L_\nu \left(
    \frac{\Delta m^2_{32}}{4E_\nu} + \varepsilon_{\mu\tau} V_{\rm CC} \right)\right]\,.
 \label{eq:pmumu-nu-final-omsd}
\end{equation}
We further find that the correction due to the deviation of $\theta_{23}$ from
maximality, $\chi \equiv \theta_{23}-\pi/4$, is second order in
the small parameter $\chi$, and hence can be neglected. 
Here, one notes that the NSI parameter $\epsmutau$ primarily modifies the
wavelength of neutrino oscillations. It also comes multiplied with $V_{\rm CC}$,
which increases at higher baselines inside the  Earth's matter.  Thus,
the modification in $\nu_\mu$ survival probabilities due to NSI varies
with the baseline $L_\nu$ (or neutrino zenith angle $\cos\theta_\nu$).

%%%%%%%%%%%%%%%%%%%%%%%%%%%%%%%%%%%%%%%%%%%%%%%%%%%%%%%%%%%%%%%%%
\subsection{Effect of $\epsmutau$ on the oscillation dip in the $L_\nu/E_\nu$}
\label{subsec:osc-dip}
%%%%%%%%%%%%%%%%%%%%%%%%%%%%%%%%%%%%%%%%%%%%%%%%%%%%%%%%%%%%%%%%

\begin{figure}[t]
  \centering
  \includegraphics[width=0.45\linewidth]{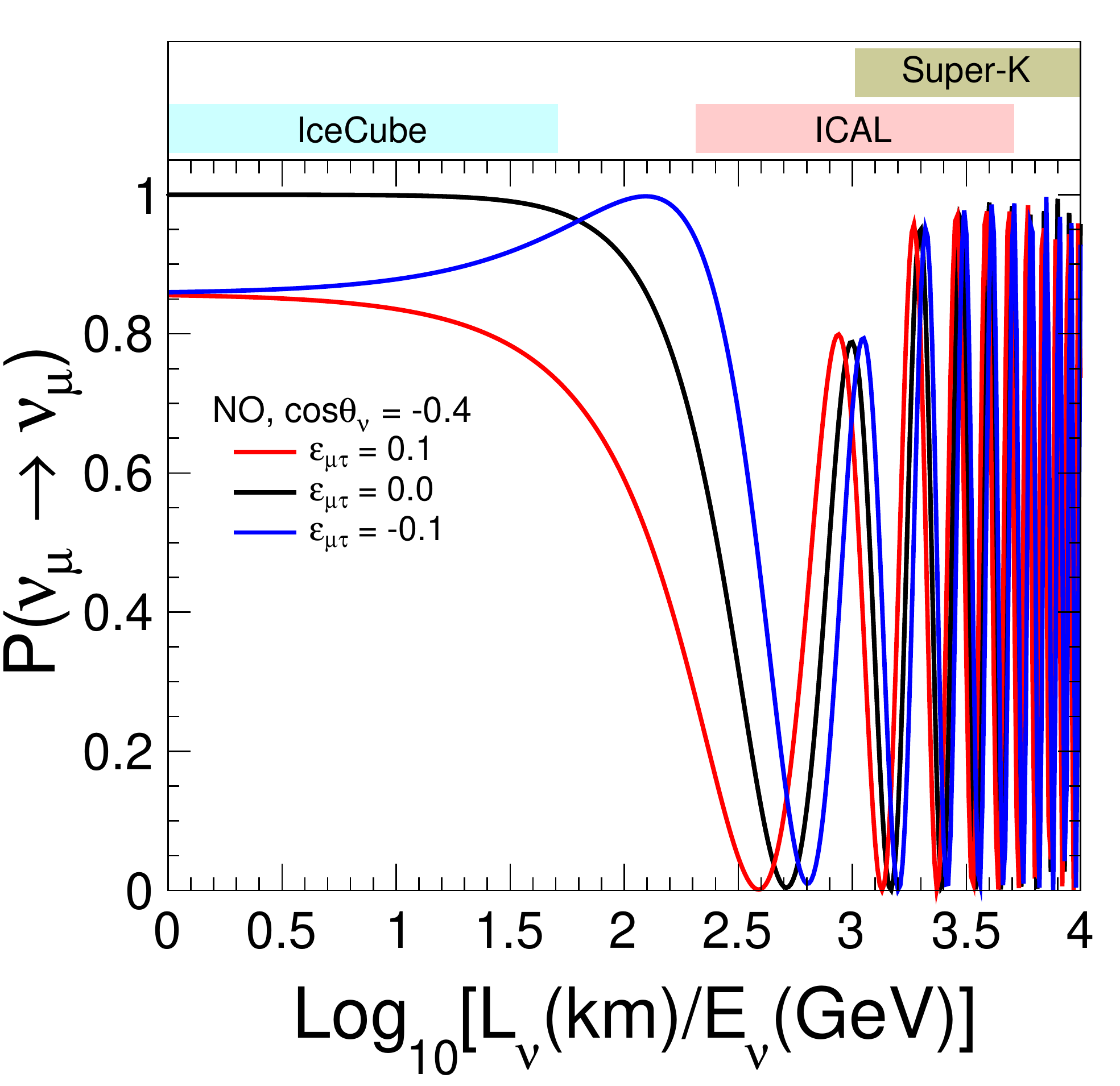}
  \includegraphics[width=0.45\linewidth]{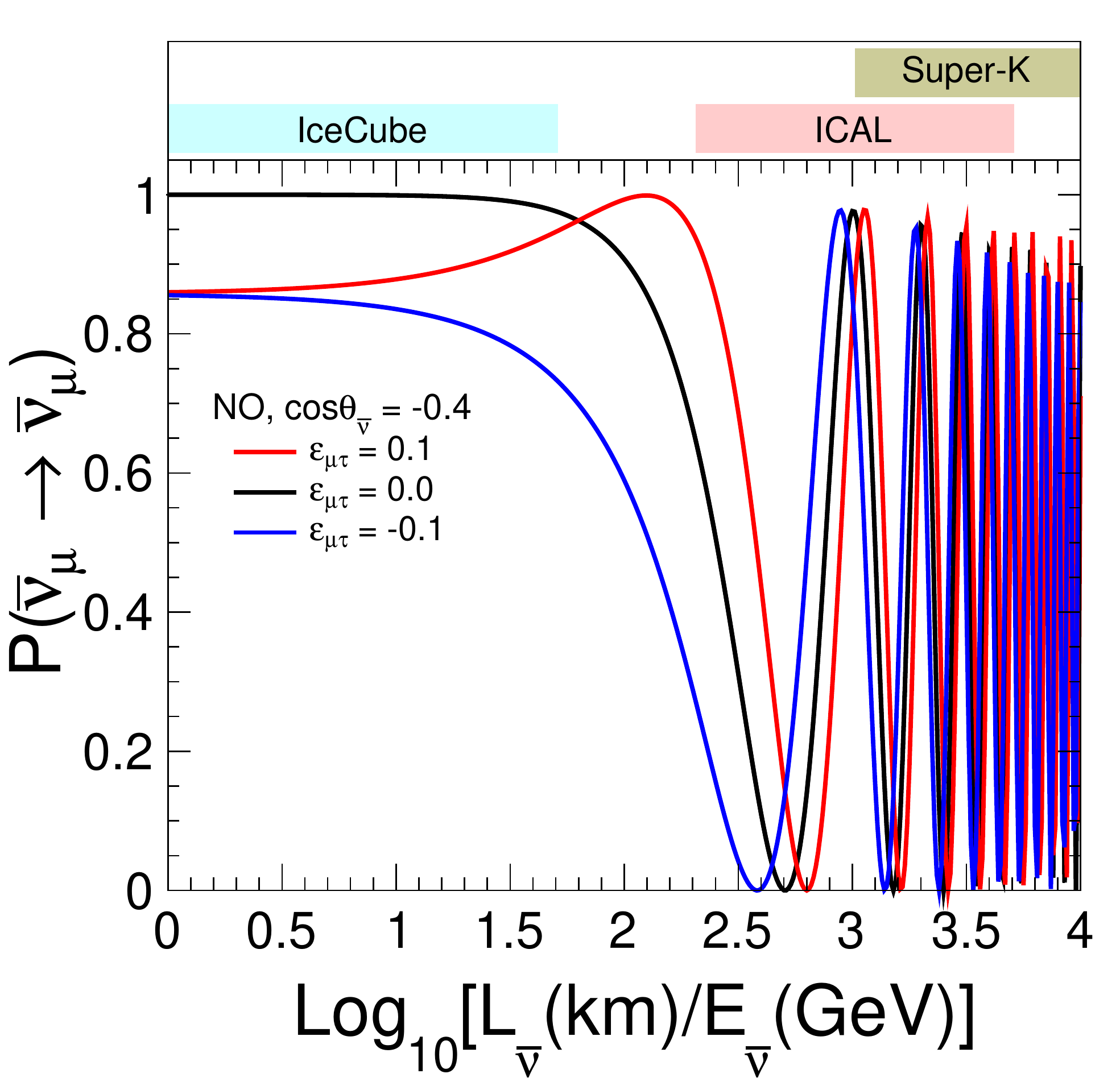}
  \includegraphics[width=0.45\linewidth]{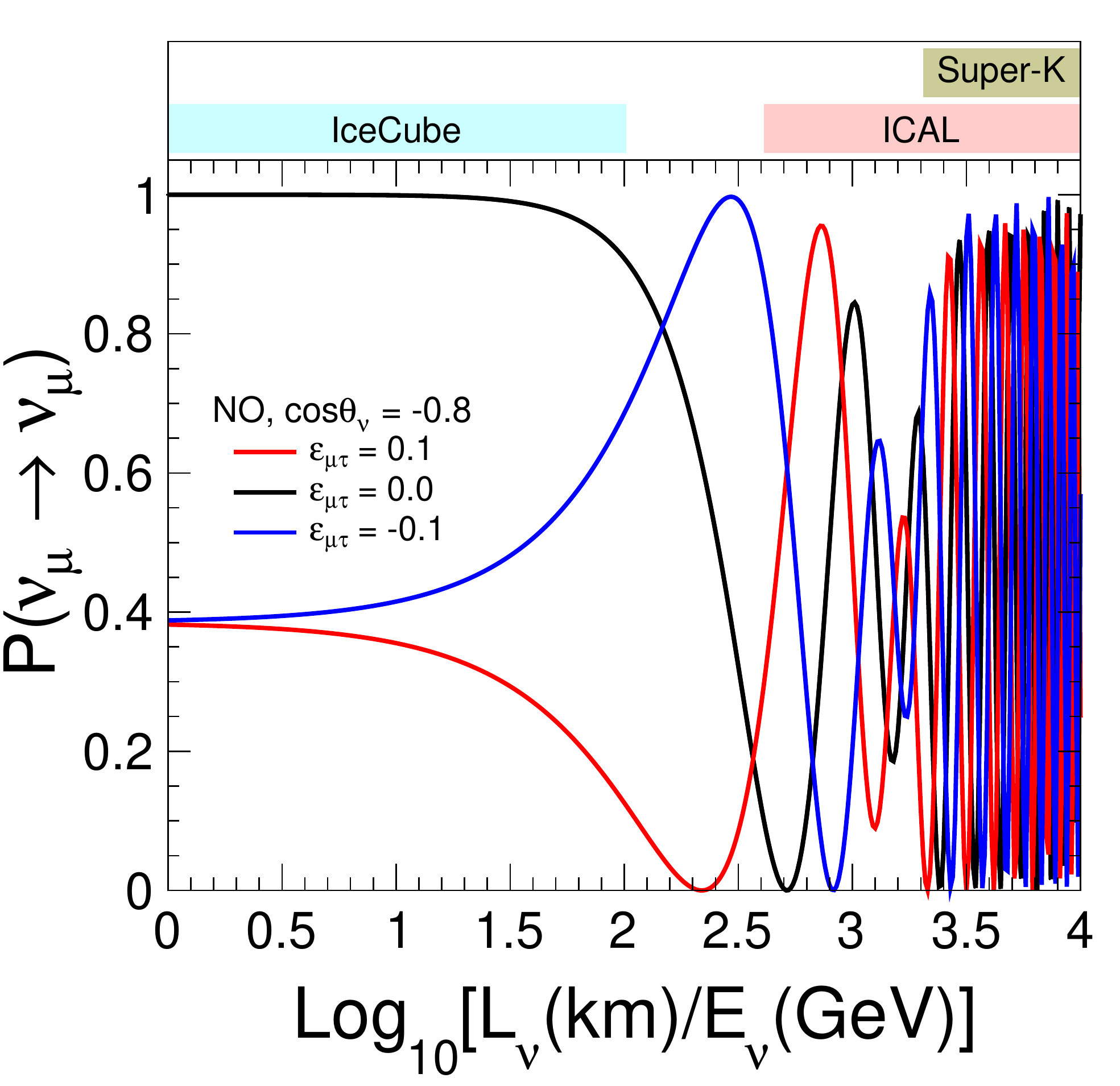}
  \includegraphics[width=0.45\linewidth]{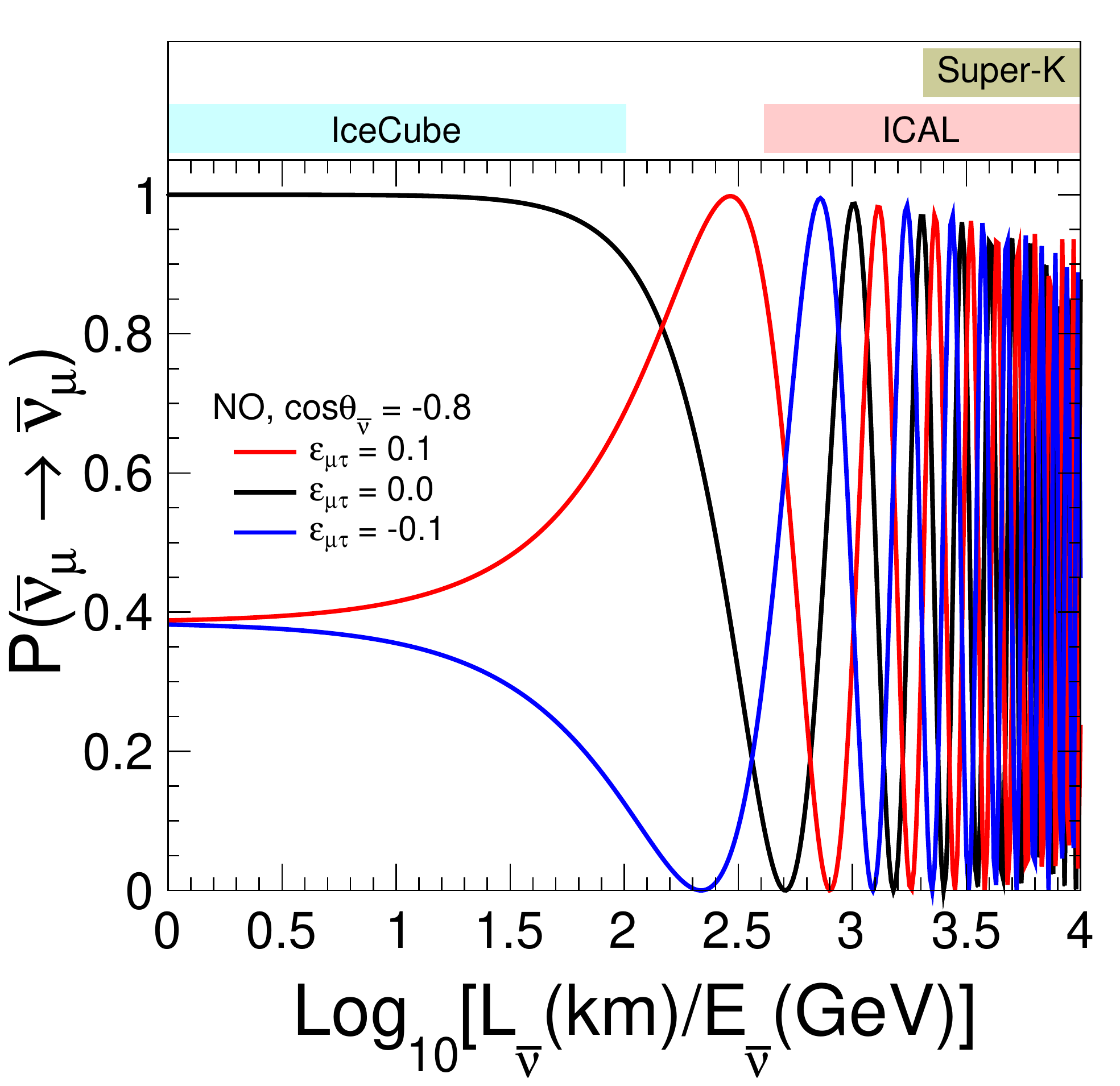}
  \mycaption{The survival probabilities of $\nu_\mu$ and $\bar\nu_\mu$
    as functions of $\log_{10}(L_\nu/E_\nu)$ in left and right panels,
    respectively. The black lines correspond to
    $\varepsilon_{\mu\tau} = 0$ (standard interactions: SI), whereas red and blue lines are with
    $\varepsilon_{\mu\tau} = 0.1$, and $-0.1$, respectively.
    The neutrino direction taken in upper (lower) panels is
    $\cos\theta_\nu = -0.4$ ($-0.8$). The horizontal bands shown here are
    the indicative $L_\nu/E_\nu$ ranges that these detectors are
    well-suited for.
    We consider normal mass ordering, and the benchmark oscillation
    parameters given in Table~\ref{tab:osc-param-value}.
    }
  \label{fig:Puu_NSI_NO}
\end{figure}

In Fig.\,\ref{fig:Puu_NSI_NO}, we present the survival probabilities of
$\nu_\mu$ and $\bar\nu_\mu$ as functions of $L_\nu/E_\nu$, for two fixed values
of $\cos\theta_\nu$ ({\it i.e.} $\cos \theta_\nu = -0.4, -0.8$), and for
three values of the NSI parameter $\varepsilon_{\mu\tau}$ ({\it i.e.}
$\epsmutau= +0.1, 0.0, -0.1$).
The other benchmark values of the oscillation parameters are
given in Table~\ref{tab:osc-param-value}.

%%%%%%%%%%%%%%%%%%%%%%%%%%%%%%%%%%%%%%%%%%%%%%%%%%%%%%%%%%%%%%%%
%%%%%%%%%%%%%%%%%%%%%%%%%%%%%%%%%%%%%%%%%%%%%%%%%%%%%%%%%%%%%%%%%

%%%%%%%%%%%%%%%%%%%%%%%%%%%%%%%%%%%%%%%%%%%%%%%%%%%%%%%%%%%%%%%%
\begin{table}[h]
  \centering
  \begin{tabular}{|c|c|c|c|c|c|c|}
    \hline
    $\sin^2 2\theta_{12}$ & $\sin^2\theta_{23}$ & $\sin^2 2\theta_{13}$ &
    $|\Delta m^2_{32}|$ (eV$^2$) & $\Delta m^2_{21}$ (eV$^2$) & $\delta_{\rm CP}$
    & Mass Ordering\\
    \hline
    0.855 & 0.5 & 0.0875 & $2.46\times 10^{-3}$ & $7.4\times10^{-5}$
    & 0 & Normal (NO)\\
    \hline 
  \end{tabular}
  \mycaption{The benchmark values of oscillation parameters that we use in
    the analysis.  Normal mass ordering corresponds to $m_1 < m_2 < m_3$.}
  \label{tab:osc-param-value}
\end{table}
%%%%%%%%%%%%%%%%%%%%%%%%%%%%%%%%%%%%%%%%%%%%%%%%%%%%%%%%%%%%%%%%

Figure\,\ref{fig:Puu_NSI_NO} also indicates the sensitivity ranges
for the atmospheric neutrino experiments Super-K, ICAL, and IceCube.
Note that these ranges are different for $\cos\theta_\nu=-0.4$ and $-0.8$,
which correspond to $L_\nu$ around 5100 km and 10200 km, respectively. 
While calculating these $L_\nu/E_\nu$ ranges, the energy ranges chosen are
those for which the detectors perform very well: we use the $E_\nu$ range of
100 MeV -- 5 GeV for Super-K~\cite{Ashie:2005ik}, 1--25 GeV for
ICAL~\cite{Kumar:2017sdq}, and 100 GeV -- 10 PeV for
IceCube~\cite{Aartsen:2013vja}. Note that these ranges are only indicative.
The following observations may be made from the figure.

\begin{itemize}

\item For $\log_{10}[L_\nu/E_\nu]$ in the range of [0 -- 1.5], the survival
  probabilities of both $\nu_\mu$ and $\bar\nu_\mu$ are observed to be
  suppressed in the presence of non-zero $\varepsilon_{\mu\tau}$.
  This is because, although the oscillations due to neutrino mass-squared
  difference do not develop for such small value of $L_\nu/E_\nu$,
  i.e. $\Delta m^2_{32}L_\nu / E_\nu \ll 1$, the disappearance of $\nu_\mu$
  is possible due to the $L_\nu \varepsilon_{\mu\tau} V_{\rm CC}$ term, which can be
  high for large baselines (for core passing neutrinos, $V_{\rm CC}$ is large).
  For example, log$_{10} [L_\nu/E_\nu] = 1$ may correspond to $L_\nu=5000$ km
  and $E_\nu = 500$ GeV, so that $\Delta m^2_{32}L_\nu/E_\nu \approx 0.025$.
  However, for this baseline, the average density of the Earth is
  $\rho \approx 3.9$ g/cc, and hence for $\epsmutau=0.1$, we have
  $L_\nu \epsmutau V_{\rm CC} \approx 0.93$, which takes the oscillation
  probability away from unity.
  The effect of NSI at such small $L_\nu/E_\nu$ is energy-independent,
  and can be seen at detectors like IceCube~\cite{Salvado:2016uqu} 
  due to its better performance at high energy. Note, however, that in the high energy limit, the $\nu_\mu$ survival probability depends only on the magnitude of $\epsmutau$ (Eq.~\ref{eq:pmumu-nu-final-omsd}). As a result, it may be difficult for IceCube to determine sgn($\epsmutau$), if indeed $|\epsmutau|$ turns out to be nonzero. 

\item As we go to higher value of log$_{10} [L_\nu/E_\nu]$ ($>2$), the term
  containing neutrino mass splitting becomes comparable to the
  $L \varepsilon_{\mu\tau} V_{\rm CC}$ term in Eq.\,\ref{eq:pmumu-nu-final-omsd},
  and the competition between these two terms leads to an energy dependence.
  When the oscillations due to the mass splitting are the dominating
  contribution, the change in oscillation wavelength due to non-zero
  $\varepsilon_{\mu\tau}$ results in a shift of the oscillation minima towards
  left or right side (lower or higher values of $L_\nu/E_\nu$, respectively),
  depending on the amplitude of $\varepsilon_{\mu\tau}$ and its sign.
  The direction of shift in the dip location depends on whether it is
  neutrino or antineutrino (since the sign of $V_{\rm CC}$ for them are opposite),
  and on the neutrino mass ordering.
  This effect on the shift in the dip location is discussed in next paragraph
  in detail. This region is relevant for Super-K, INO, and most of the
  long-baseline experiments.

\end{itemize}

We now discuss the modification of the oscillation dip due to non-zero
$\varepsilon_{\mu\tau}$. First, using the approximate expression of
$\nu_\mu$ survival probability from Eq.~\ref{eq:pmumu-nu-final-omsd},
we obtain the value of $L_\nu/E_\nu$ at the first oscillation minimum
(or the dip):  
\begin{equation}
  \left. \frac{L_\nu}{E_\nu} \right|_{\rm dip} = \frac{2\pi}{|\Delta m^2_{32}
    + 4\varepsilon_{\mu\tau} V_{\rm CC} E_\nu|}\,.
    \label{eq:osc-min-lbye}
\end{equation}
The above expression may be written by expressing $V_{\rm CC}$ in terms of the
line-averaged matter density\footnote{We can write $V_{\rm CC}$
approximately as a function of matter density $\rho$
\begin{equation}
  V_{\rm CC} \approx \pm 7.6\times Y_{e} \times \frac{\rho}{10^{14}\,
    {\rm g/cm^{3}}}\,\, {\rm eV}\,.
\label{eq:vcc}
\end{equation}
Here, $Y_e = \frac{N_e}{N_p + N_n}$ is the relative electron number density. 
In an electrically neutral and  isoscalar medium, $Y_e=0.5$.
The positive (negative) sign is for neutrino (antineutrino).}
$\rho$ and taking care of units, 
\begin{equation}
  \left. \frac{L_\nu [{\rm km]}}{E_\nu [\rm GeV]} \right|_{\rm dip}
  = \frac{\pi}{ \left| 2.54 \cdot \Delta m^2_{32} [{\rm eV^2}]
    \pm 7.7\times 10^{-4} \cdot \rho [{\rm g/cm^3}] \cdot Y_{e}
    \cdot\varepsilon_{\mu\tau} \cdot   E_\nu [{\rm GeV}] \right|  }\,.
  \label{eq:osc-min-lbye-units}
\end{equation}
Here, a positive sign in the denominator corresponds to neutrinos, whereas
a negative sign is for antineutrinos. This approximation is useful for
understanding the shift of dip position with non-zero $\varepsilon_{\mu\tau}$.
Let us take the case of neutrino with normal ordering. With positive
$\varepsilon_{\mu\tau}$, the denominator of Eq.\,\ref{eq:osc-min-lbye}
increases, thus the oscillation minimum appears at a lower value of
$L_\nu/E_\nu$ than that for the SI. On the other hand, with negative
$\varepsilon_{\mu\tau}$, the oscillation dip would occur at a higher value
of $L_\nu/E_\nu$. These two features can be seen clearly in the left panels of Fig.\,\ref{fig:Puu_NSI_NO}. 
For antineutrinos and the same mass ordering, since the matter potential has
the opposite sign, the shift of oscillation dip is in the opposite direction
as compared to that for neutrinos, given the same $\varepsilon_{\mu\tau}$.

Expanding Eq.\,\ref{eq:osc-min-lbye-units} to first order in $\epsmutau$,
one can write
\begin{equation}
  \left. \frac{L_\nu [{\rm km}]}{E_\nu [{\rm GeV}]} \right|_{\rm dip}
  = \frac{\pi}{\left| 2.54\times \Delta m^2_{32} [{\rm eV}^2] \right|}
  \mp \frac{ \pi \times 1.19 \times 10^{-4}\cdot \rho [{\rm g/cm^3}]
    \cdot Y_{e} \cdot   E_\nu [{\rm GeV}]}{\beta \,
    (\Delta m^2_{32})^2 [{\rm eV}^4]} \cdot \varepsilon_{\mu\tau}  \,,
  \label{eq:lbye-dip-calibration}
\end{equation}
where $\beta \equiv {\rm sgn}(\Delta m^2_{32})$, as defined earlier.
Here, the negative sign corresponds to neutrinos, whereas the positive sign
is for antineutrino.
This indicates that, for small values of $\epsmutau$, the shift in the dip
location will be linear in $\epsmutau$, and will have opposite sign for
neutrinos and antineutrinos, as well as for the two mass orderings.
We have checked that for $E_\nu<25$ GeV and typical
line-averaged density of the Earth, indeed $\varepsilon_{\mu\tau}<0.1$ is
small enough for the above approximation to hold.

%%%%%%%%%%%%%%%%%%%%%%%%%%%%%%%%%%%%%%%%%%%%%%%%%%%%%%%%%%%%%%%%%%%%%%%
\subsection{Effect of $\epsmutau$ on the oscillation valley in the
  ($E_\nu$, $\cos\theta_\nu$) plane}
\label{subsec:osc-valley}
%%%%%%%%%%%%%%%%%%%%%%%%%%%%%%%%%%%%%%%%%%%%%%%%%%%%%%%%%%%%%%%%%%%%%%

%%%%%%%%%%%%%%%%%%%%%%%%%%%%%%%%%%%%%%%%%%%%%%%%%%%%%%%%%%%%%%%%%%%%
\begin{figure}[t]
    \centering
    \includegraphics[width=0.45\linewidth]{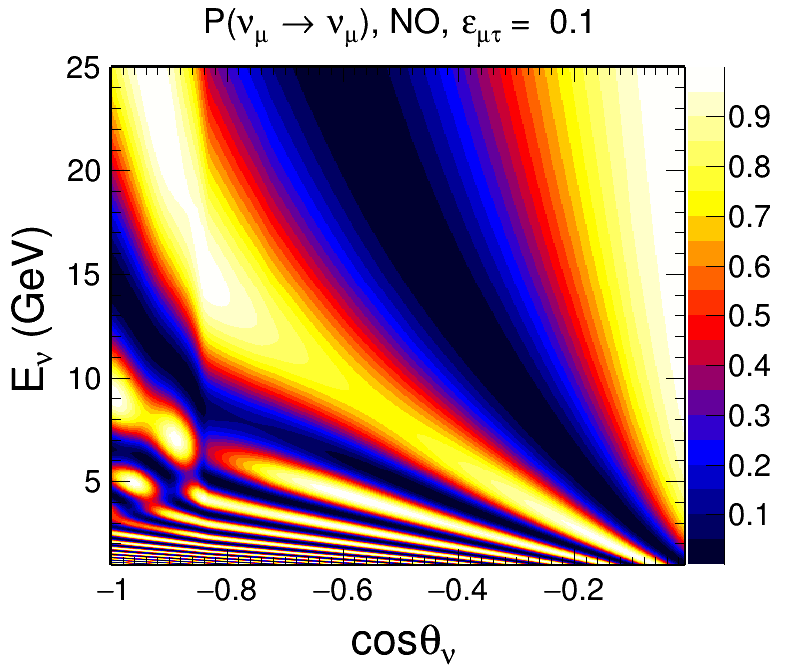}
    \includegraphics[width=0.45\linewidth]{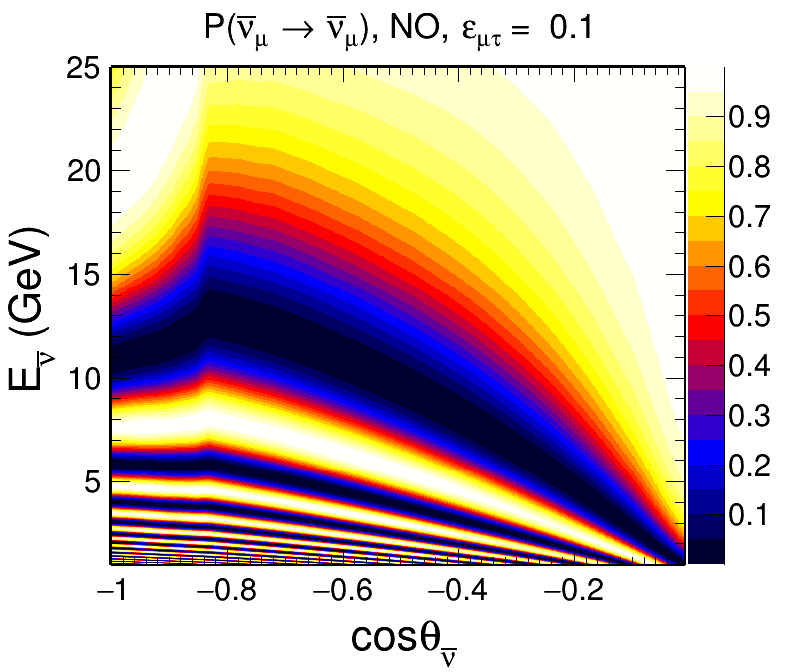}
    \includegraphics[width=0.45\linewidth]{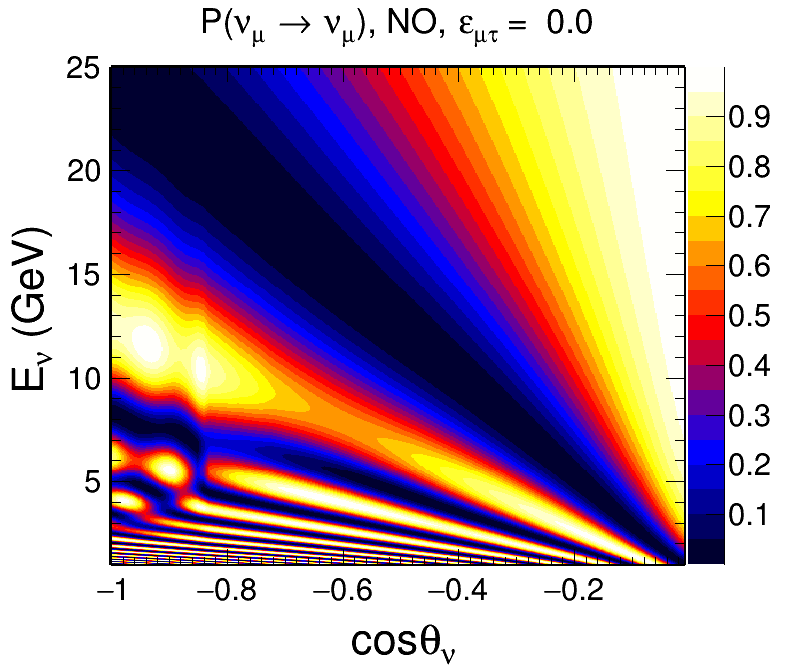}
    \includegraphics[width=0.45\linewidth]{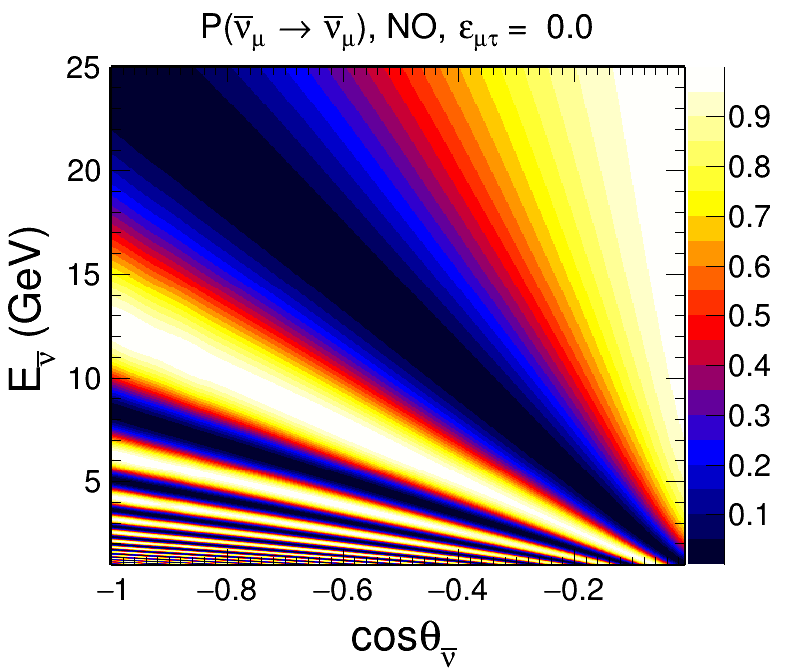}
    \includegraphics[width=0.45\linewidth]{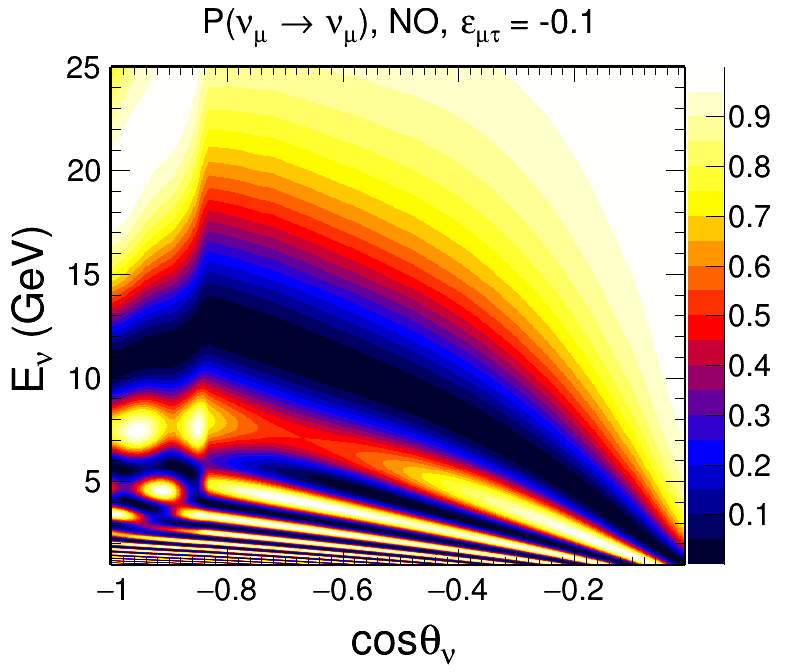} 
    \includegraphics[width=0.45\linewidth]{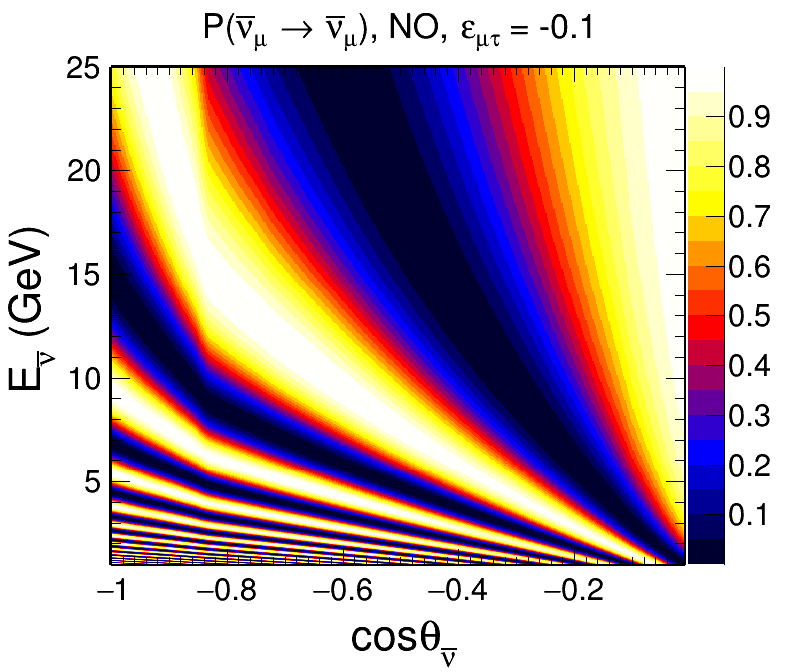}
    \mycaption{The oscillogram for the survival probabilities in
      ($E_\nu,\cos\theta_\nu$) plane for neutrino and antineutrino in
      left and right panels, respectively, with normal mass ordering and
      the benchmark oscillation parameters 
      from Table~\ref{tab:osc-param-value}.
      The value of $\varepsilon_{\mu\tau}$ is taken as +0.1, 0 (SI), and -0.1
      in top, middle and bottom panels, respectively.}
    \label{fig:oscillogram}
\end{figure}
%%%%%%%%%%%%%%%%%%%%%%%%%%%%%%%%%%%%%%%%%%%%%%%%%%%%%%%%%%%%%%%%%%%

In Fig.\,\ref{fig:oscillogram}, we show the oscillograms of survival
probabilities for $\nu_\mu$ and $\bar\nu_\mu$, in the plane of neutrino
energy and cosine of neutrino zenith angle ($\cos\theta_\nu$),
with NO as the neutrino mass ordering.
We show only upward-going neutrinos ($\cos \theta_\nu <0$), since for
downward-going neutrinos ($\cos\theta_\nu>0$), the baseline $L_\nu$ is 
very small\footnote{The zenith angle $\theta_\nu$ is related to the
  baseline $L_\nu$ by 
\be
L_\nu = \sqrt{(R+h)^2 - (R-d)^2\sin^2\theta_\nu} \,-\, (R-d)\cos\theta_\nu \, ,
\label{eq:zen-bl-rel}
\ee
where $R$, $h$, and $d$ are the radius of Earth, the average height from
the Earth surface at which neutrinos are created, and the depth of the
detector underground, respectively. In this study, we use  $R = 6371$ km,
$h = 15$ km, and $d=0$ km. A small change in $h$ and $d$ does not affect
the $\nu_\mu$ oscillation probabilities because $R \gg h \gg d$.}
and neutrino oscillations do not develop, making $P_{\mu\mu} \approx 1.0$.
The differences observed among the upper, middle, and lower panels are due
to different values of $\varepsilon_{\mu\tau}$. A major impact of NSI
is observed on the feature corresponding to the first oscillation minima,
seen in the figure as a broad blue/black diagonal band.
We refer to this broad band as the oscillation valley~\cite{Kumar:2020wgz}.
In the SI case, the oscillation valley appears like a triangle with straight
edges, while in the presence of NSI, its edges seem to acquire a curvature.
The sign of this curvature is opposite for neutrinos and antineutrinos,
as can be seen by comparing the left and right plots of upper and lower
panels of Fig.\,\ref{fig:oscillogram}. 

The modification in the oscillation valley due to NSI may be explained from
the following relation between $E_\nu$ and $\cos\theta_\nu$ at the first
oscillation minima. We rewrite Eq.\,\ref{eq:osc-min-lbye} with
$L_\nu \approx 2R |\cos\theta_\nu|$:
\begin{equation}
  E_\nu |_{\rm valley} \approx \frac{| \Delta m^2_{32}| }{
      (\pi / |R \cos\theta_\nu|) \;  
    - 4 \beta \, \varepsilon_{\mu\tau} V_{\rm CC}}\, .
     \label{eq:osc-min-Ecostheta-wounit}
\end{equation}
Taking care of units and expressing $V_{\rm CC}$ in terms of line-averaged constant
matter density $\rho$, the above expression may be written as 
\begin{equation}
  E_\nu[{\rm GeV}]|_{\rm valley} = \frac{|\Delta m^2_{32} [{\rm eV}^2]|}{
      (\pi /| 5.08 \cdot R [{\rm km}] \cdot \cos\theta_\nu|) \; 
    \mp 3.02 \times 10^{-4} \cdot \beta \cdot \varepsilon_{\mu\tau} \cdot Y_e
    \cdot \rho [{\rm g/cm^3}]  }  \,.
  \label{eq:osc-min-Ecostheta}
\end{equation}
Here, the negative sign in the denominator corresponds to neutrinos, whereas
the positive sign is for antineutrinos. 
Putting $\varepsilon_{\mu\tau}=0$ in Eq.\,\ref{eq:osc-min-Ecostheta} gives
the condition of the first oscillation minima to be
$E_\nu= | (5.08/\pi)\cdot \Delta m^2_{32}\cdot R [\rm km] \cdot \cos\theta_\nu|$,
and this relation  clearly shows that in the SI case, the minimum of the
oscillation valley is a straight line in the ($E_\nu\,,\cos\theta_\nu$) plane.

Now in the normal mass ordering scenario,
if $\varepsilon_{\mu\tau} >0$ in Eq.\,\ref{eq:osc-min-Ecostheta}, then
for neutrinos, $E_\nu$ increases for a given $\cos\theta_\nu$ as compared
to the SI case. As a result, the oscillation valley
(the broad blue/black band) bends
towards higher values of energies in top left panel.
On the other hand, for antineutrino, the oscillation valley tilts towards
lower energies for $\varepsilon_{\mu\tau}>0$, as can be seen in the top right
panel. For negative values of  $\varepsilon_{\mu\tau}$, the oscillation valley
bends in the opposite direction, for both neutrinos and antineutrinos,
as shown in the bottom panels of  Fig.\,\ref{fig:oscillogram}.   
In the inverse mass ordering scenario, the bending of the oscillation
valley will be in the direction opposite to that in the normal mass
ordering scenario.
  
From Eqs.~\ref{eq:lbye-dip-calibration} and~\ref{eq:osc-min-Ecostheta},
the shift in the oscillation minima and the bending in the oscillation valley
are in opposite directions for normal and inverted mass ordering.
The effect of $\epsmutau$, therefore, depends crucially on the mass ordering.
In this paper, we present our analysis in the scenario where the mass
ordering is known to be NO.
The analysis for IO may be performed in an exactly analogous manner.
Our results are presented with the exposure corresponding to 10 years
of ICAL data-taking. It is expected that the mass ordering will be
determined from the neutrino experiments (including ICAL) by this time.

%%%%%%%%%%%%%%%%%%%%%%%%%%%%%%%%%%%%%%%%%%%%%%%%%%%%%%%%%%%%%%%
%%%%%%%%%%%%%%%%%%%%%%%%%%%%%%%%%%%%%%%%%%%%%%%%%%%%%%%%%%%%%%
\section{Impact of NSI on the Event Distribution at ICAL}
\label{sec:evt-ical}
%%%%%%%%%%%%%%%%%%%%%%%%%%%%%%%%%%%%%%%%%%%%%%%%%%%%%%%%%%%%%
%%%%%%%%%%%%%%%%%%%%%%%%%%%%%%%%%%%%%%%%%%%%%%%%%%%%%%%%%%%%%%

While the effect of NSI on the neutrino and antineutrino survival
probabilities was discussed in the last section, it is important to
confirm whether the features present at the probability level can
survive in the observables at a detector, and if they can be reconstructed.
Here is where the response of the detector plays a crucial role. Neutrinos
cannot be directly detected in an experiment, however the charged leptons
produced from their charged-current interactions in the detector contains
information about their energy, direction, and flavor, which can be
recovered depending on the nature of the detector.

The upcoming ICAL detector at the India-based Neutrino Observatory (INO)
will be composed of 151 layers of magnetized iron plates; the 50 kt
iron acting as the target, and around 30,000 glass resistive plate chambers
as  detection elements. The magnetic field of strength 1.5 Tesla and the time
resolution of less than 1 ns~\cite{Dash:2014ifa, Bhatt:2016rek, Gaur:2017uaf}
will enable ICAL to distinguish $\mu^-$ from $\mu^+$ events in the multi-GeV
energy range. In this study, we generate the charged-current interactions of
$\nu_\mu$ and $\bar\nu_\mu$, similar to Ref.\,\cite{Kumar:2020wgz}, using
neutrino and antineutrino fluxes calculated at the Theni site by Honda
et.al.~\cite{Athar:2012it,PhysRevD.92.023004} and the neutrino event generator
NUANCE~\cite{Casper:2002sd}.  The reconstructed energy ($E_\mu^{\rm rec}$) and
zenith angle ($\theta_\mu^{\rm rec}$) of muons produced in neutrino and
antineutrino interactions will be used in the analysis. The detector properties of ICAL for muons as obtained in Ref.\cite{Chatterjee:2014vta} using the ICAL-GEANT4 simulation package, are incorporated.  With around 1 km rock coverage (3800 meter water equivalent), the downward-going cosmic muon background would get reduced by  $\sim 10^{6}$~\cite{Dash:2015blu}, most of which will be further vetoed by employing the fiducial volume cut. We do not consider the muon events induced by tau decay in the detector since the number of such events is small ($\sim 2\%$ of total upward-going muons from $\nu_\mu$ interactions in the energy range of interest~\cite{Pal:2014tre}), and these are mostly at lower energies and below the energy threshold of ICAL which is 1 GeV. 

The quantity $L_\mu^{\rm rec}$ associated with the reconstructed muon
  direction $\cos\theta_\mu^{\rm rec}$ is defined 
  (similar to Eq.\,\ref{eq:zen-bl-rel}) as
\be
L_\mu^{\rm rec} \equiv \sqrt{(R+h)^2 - (R-d)^2\sin^2\theta_\mu^{\rm rec}} \,
-\, (R-d)\cos\theta_\mu^{\rm rec} \, .
\label{eq:zen-bl-rel-mu}
\ee 
Note that $L_\mu^{\rm rec}$ is simply a proxy observable, and there is no
need to associate it directly with the distance traveled by the incoming
neutrino.

%%%%%%%%%%%%%%%%%%%%%%%%%%%%%%%%%%%%%%%%%%%%%%%%%%%%%%%%%%%%%%%%%%
\subsection{The $L_\mu^\text{rec}/E_\mu^\text{rec}$ distributions}
\label{sec:events-L/E}
%%%%%%%%%%%%%%%%%%%%%%%%%%%%%%%%%%%%%%%%%%%%%%%%%%%%%%%%%%%%%%%%%

%%%%%%%%%%%%%%%%%%%%%%%%%%%%%%%%%%%%%%%%%%%%%%%%%%%%%%%%%%%%%%%%%%%
\begin{figure}[t]
\includegraphics[width=0.49\linewidth]{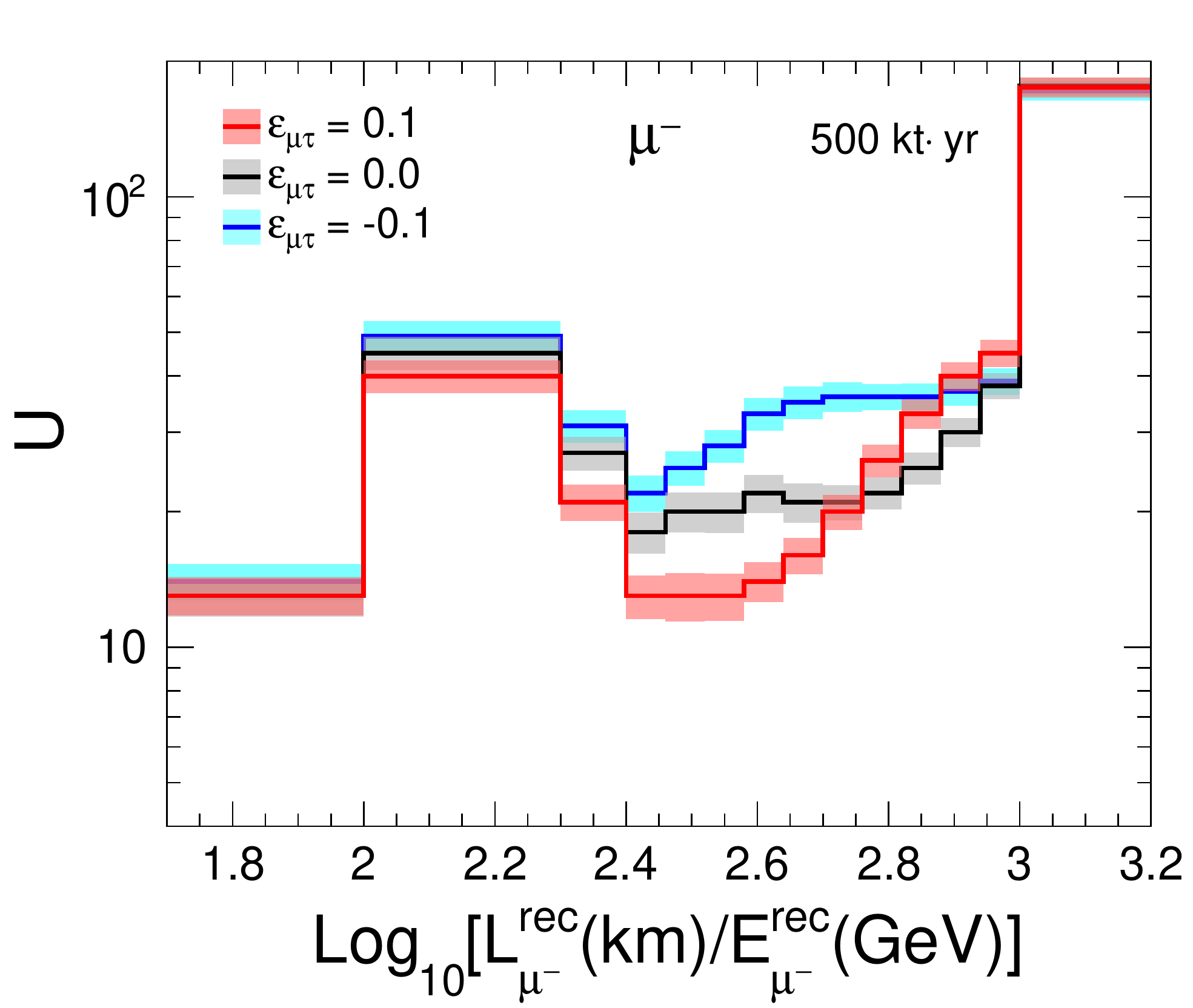}
 \includegraphics[width=0.49\linewidth]{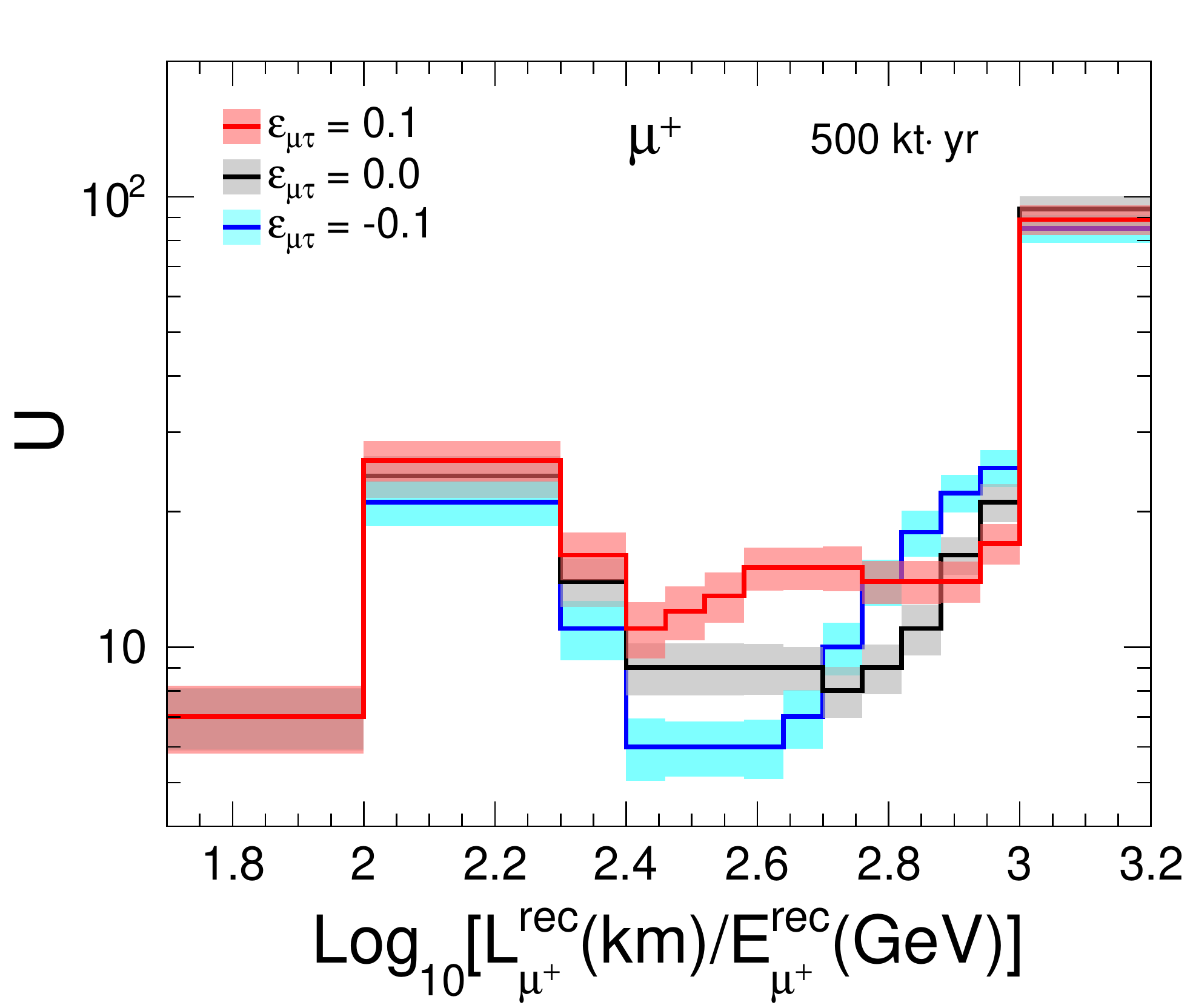}
 \mycaption{The $\log_{10}[L_\mu^\text{rec}/E_\mu^\text{rec}]$ distributions
   of $\mu^-$ and $\mu^+$ events in left and right panels, respectively,
   expected in 10 years at ICAL. The black  lines correspond to SI
   ($\varepsilon_{\mu\tau}=0$), whereas red and blue
    lines are for $\varepsilon_{\mu\tau} = 0.1$ and -0.1, respectively. The statistical
   uncertainties shown by shaded boxes are the root-mean square (rms)
   fluctuations of 100 independent 10-year data sets. 
   The solid lines show the mean of these 100 distributions.
   We consider normal mass ordering, and the benchmark oscillation
   parameters given in Table~\ref{tab:osc-param-value}.
 }
 \label{fig:event-1d}
\end{figure}
%%%%%%%%%%%%%%%%%%%%%%%%%%%%%%%%%%%%%%%%%%%%%%%%%%%%%%%%%%%%%%%%%%

%%%%%%%%%%%%%%%%%%%%%%%%%%%%%%%%%%%%%%%%%%%%%%%%%%%%%%%%%%%%%%%%%
\begin{table}[htb!]
\centering
 \begin{tabular}{|c|c|c|c c|}
  \hline
  Observable & Range & Bin width & \multicolumn{2}{c|}{Number of bins} \\
  \hline 
  \multirow{7}{*}{ $\log_{10}\left[\frac{L_\mu^{\rm rec}(\text{km})}{E_\mu^{\rm rec}
        (\text{GeV})}\right]$} & [0, 1] & 0.2 & 5 & \rdelim\}{7}{7mm}[34] \cr 
  & [1, 1.6] &  0.06 & 10  & \cr
  & [1.6, 1.7]& 0.1& 1  & \cr
  & [1.7, 2.3]&0.3 & 2  & \cr
  & [2.3, 2.4]& 0.1& 1  & \cr
  & [2.4, 3.0] &0.06 & 10  &\cr
  & [3, 4] & 0.2 & 5  & \cr
  \hline
 \end{tabular}
 \mycaption{The binning scheme of $\log_{10}[L_\mu^{\rm rec}/E_\mu^{\rm rec}]$
   for $\mu^-$ and $\mu^+$ events.}
 \label{tab:binning-1D-10years}
\end{table}
%%%%%%%%%%%%%%%%%%%%%%%%%%%%%%%%%%%%%%%%%%%%%%%%%%%%%%%%%%%%%%%

Fig.\,\ref{fig:event-1d} presents the
$\log_{10}[L_\mu^\text{rec}/E_\mu^\text{rec}]$ distribution\footnote{The values we mention
  for $\log_{10}[L_\mu^\text{rec}/E_\mu^\text{rec}]$ throughout the paper
  are calculated with $L_\mu^\text{rec}$ in the units of km
  and $E_\mu^\text{rec}$ in GeV.}  of the upward-going
$\mu^-$ and $\mu^+$ events (U, $\cos\theta_\mu^{\rm rec} <0$)
with SI ($\varepsilon_{\mu\tau} = 0$) and with SI + NSI
($\varepsilon_{\mu\tau} = \pm 0.1$) expected with 10-year exposure
at ICAL. We include the statistical fluctuation in the number of events
by simulating 100 independent data sets, and calculating the mean and
root-mean-square deviation for each bin.
For $\log_{10}[L_\mu^\text{rec}/E_\mu^\text{rec}]$, we use the same binning scheme
as used in Ref.~\cite{Kumar:2020wgz}, which is shown in
Table~\ref{tab:binning-1D-10years}. We have a total of 34 non-uniform bins of
$\log_{10}[L_\mu^\text{rec}/E_\mu^\text{rec}]$ in the range 0--4.
The downward-going events (D, $\cos\theta_\mu^\text{rec} >0$) are not affected
significantly by oscillations or by additional NSI interactions. The
$\log_{10}[L_\mu^\text{rec}/E_\mu^\text{rec}]$ distribution of downward-going $\mu^-$
and $\mu^+$ events is identical to the one shown in the upper panel of
Fig.~[4] in Ref.~\cite{Kumar:2020wgz}, and we do not repeat it here.

Fig.\,\ref{fig:event-1d} clearly shows that the
$\log_{10}[L_\mu^\text{rec}/E_\mu^\text{rec}]$ range of 2.4--3.0 would be important
for NSI studies, since non-zero $\varepsilon_{\mu\tau}$ has the largest effect
in this range. With positive $\varepsilon_{\mu\tau}$, the number of $\mu^-$
events is lower as compared to that of SI case, whereas the number
of $\mu^+$ events is higher. If $\varepsilon_{\mu\tau}$ is negative, then the
modifications of the number of $\mu^-$ and $\mu^+$ events are the other way
around. As a consequence, if a detector is not able to distinguish between
the $\mu^-$ and $\mu^+$ events, the difference between SI and NSI would
get diluted substantially. The charge identification capability of the
magnetized ICAL detector would be crucial in exploiting this observation.

%%%%%%%%%%%%%%%%%%%%%%%%%%%%%%%%%%%%%%%%%%%%%%%%%%%%%%%%%%%%%%%%%%%%%%%%%
\subsection{Distributions in the
  ($E_\mu^{\rm rec}$, $\cos\theta_\mu^{\rm rec}$) plane}
\label{sec:events-2d}
%%%%%%%%%%%%%%%%%%%%%%%%%%%%%%%%%%%%%%%%%%%%%%%%%%%%%%%%%%%%%%%%%%%%%%%%%

\begin{figure}[t]
 \includegraphics[width=0.49\linewidth]{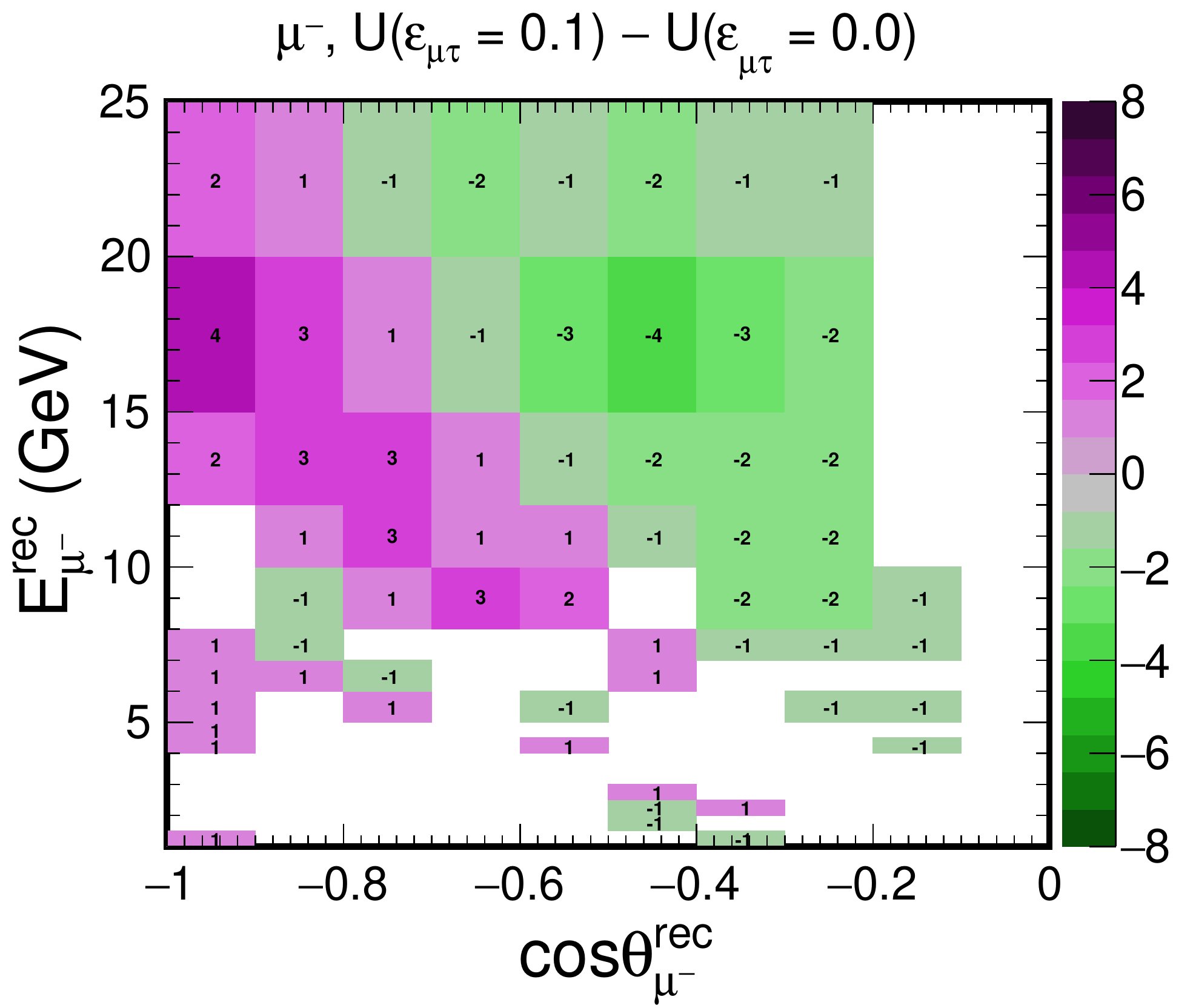}
 \includegraphics[width=0.49\linewidth]{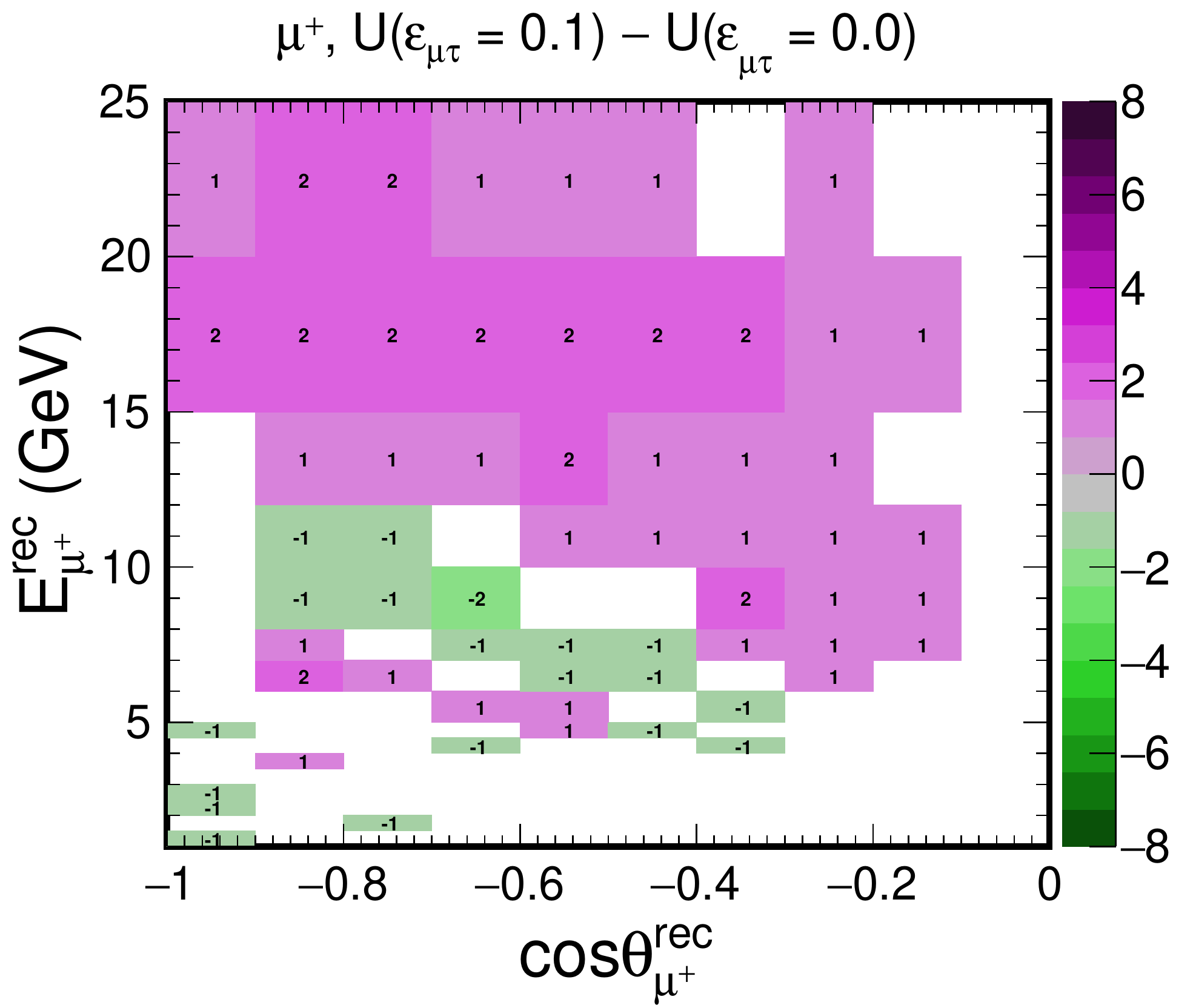}
  \includegraphics[width=0.49\linewidth]{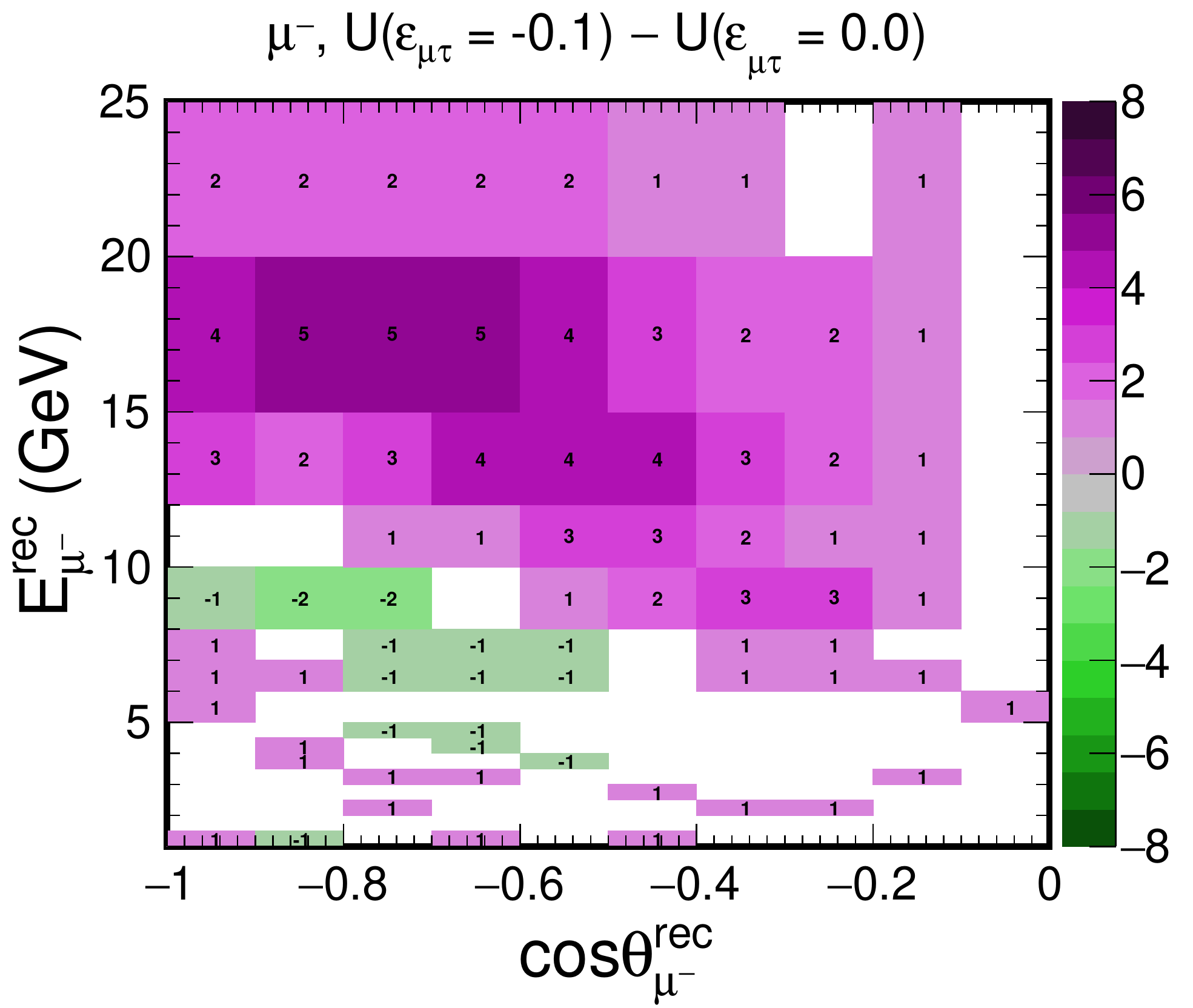}
 \includegraphics[width=0.49\linewidth]{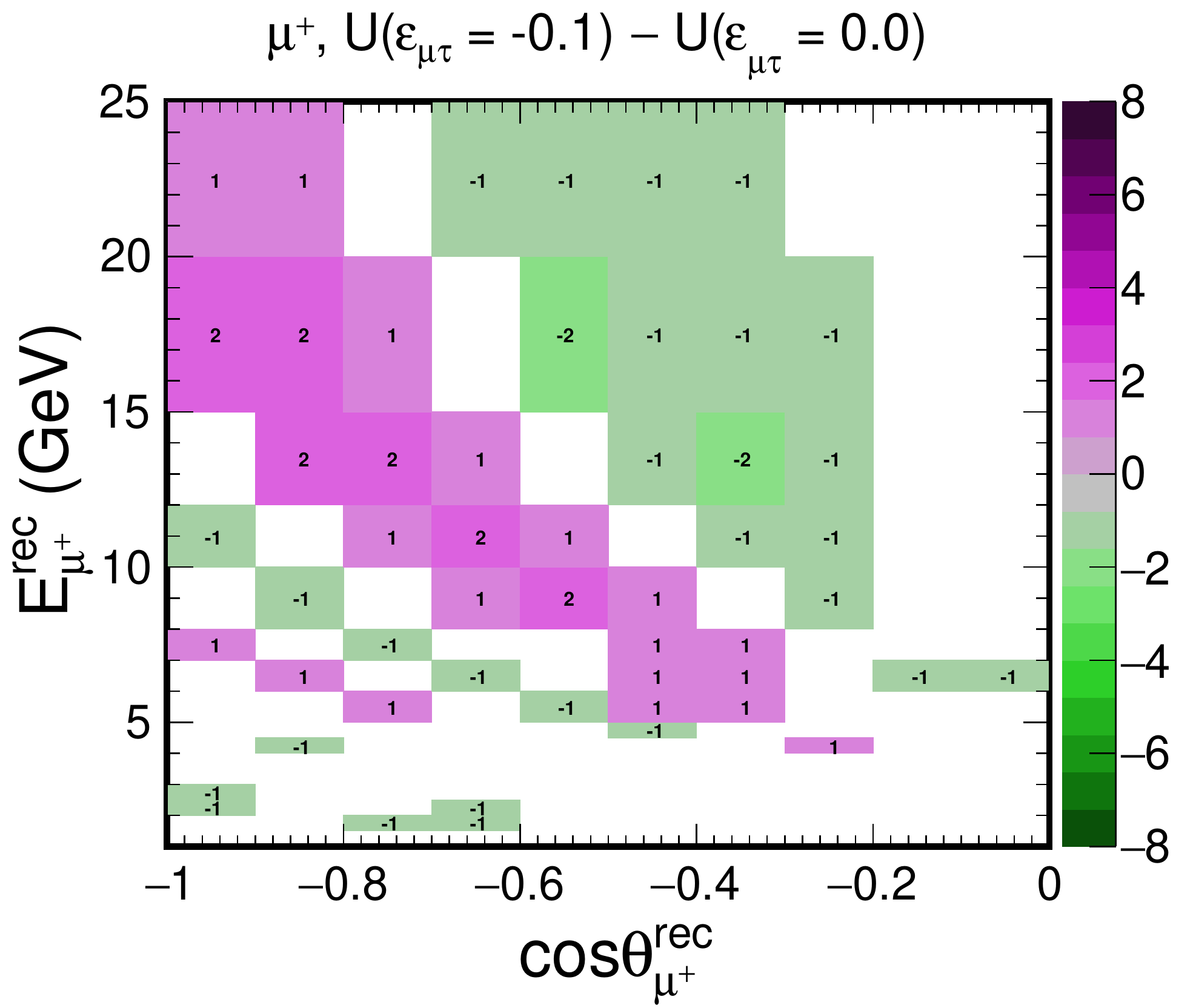}
 \mycaption{The $(E_\mu^\text{rec},\,\cos\theta_\mu^\text{rec})$ distributions of
   difference of events between including non-zero NSI and only SI
   ($\varepsilon_{\mu\tau} = 0$) expected with 500 kt$\cdot$yr of ICAL.
   Left and right panels are for $\mu^-$ and $\mu^+$ events, respectively,
   whereas upper and lower panels correspond to $\varepsilon_{\mu\tau}= 0.1$
   and $-0.1$, respectively.
       We consider normal mass ordering, and the benchmark oscillation
    parameters given in Table~\ref{tab:osc-param-value}.
}
 \label{fig:event-2d}
\end{figure}

In order to study the effect of NSI on the distribution of events in the
($E_\mu^{\rm rec}$, $\cos\theta_\mu^{\rm rec}$) plane, we bin the data in
reconstructed observables, $E_\mu^{\rm rec}$ and $\cos\theta_\mu^{\rm rec}$.
We have a total of 16 non-uniform $E_\mu^{\rm rec}$ bins in the range 1 -- 25 GeV,
and the whole range of $-1 \leq \cos\theta_\mu^{\rm rec}\leq 1$  is divided
into 20 uniform bins. The reconstructed $E_\mu^{\rm rec}$ and
$\cos\theta_\mu^{\rm rec}$ are binned with the same binning scheme
as used in~\cite{Kumar:2020wgz}, and is shown in
Table~\ref{tab:binning-2D-10years}. 
For demonstrating event distributions, we scale the 1000-year MC sample to an exposure of 10 years, and the difference in the number of events with SI + NSI ($|\varepsilon_{\mu\tau}| = 0.1$) and the number of events with SI, for $\mu^-$ and $\mu^+$ events are shown in Fig.\,\ref{fig:event-2d}.

%%%%%%%%%%%%%%%%%%%%%%%%%%%%%%%%%%%%%%%%%%%%%%%%%%%%%%%%%%%%%%%%%%%%%
%%%%%%%%%%%%%%%%%%%%%%%%%%%%%%%%%%%%%%%%%%%%%%%%%%%%%%%%%%%%%%%%%%%

%%%%%%%%%%%%%%%%%%%%%%%%%%%%%%%%%%%%%%%%%%%%%%%%%%%%%%%%%%%%%%%%%%%%
\begin{table}[htb!]
\centering
 \begin{tabular}{|c|c|c|c c|}
  \hline
  Observable & Range & Bin width & \multicolumn{2}{c|}{Number of bins} \\
  \hline 
  \multirow{4}{*}{ $E_\mu^{\rm rec}$ } & [1, 5]
  &  0.5 & 8 & \rdelim\}{5}{7mm}[16] \cr 
  & [5, 8] & 1 & 3  & \cr
  \multirow{2}{*}{ (GeV) }  & [8, 12] & 2 & 2  & \cr
  & [12, 15] & 3 & 1  & \cr
  & [15, 25] & 5 & 2  &\cr
  \hline 
  $\cos\theta_\mu^{\rm rec}$ & [-1.0, 1.0] & 0.1 & 20  & \cr
  \hline
 \end{tabular}
 \mycaption{The binning scheme adopted for $E_\mu^{\rm rec}$ and
   $\cos\theta_\mu^{\rm rec}$ of $\mu^-$ and $\mu^+$ events.  }
 \label{tab:binning-2D-10years}
\end{table}
%%%%%%%%%%%%%%%%%%%%%%%%%%%%%%%%%%%%%%%%%%%%%%%%%%%%%%%%%%%%%%%%%%%%%

From this figure, we can make the following observations.
\begin{itemize}
  
\item The events in the bins with $E_\mu^\text{rec} > 5$ GeV and
  $\cos\theta_\mu^\text{rec}<-0.2$ are particularly useful for NSI searches
  at ICAL. This is true for both $\mu^-$ and $\mu^+$ events.

\item There is almost no difference in the number of events between
  SI+NSI and SI at $E_\mu^\text{rec} < 5$ GeV. The events with reconstructed
  muon energy less than 5 GeV do not seem to be sensitive to NSI due to
  small NSI-induced matter effects: $L_\nu \epsmutau V_{\rm CC} \lesssim
  \Delta m^2/(2 E_\nu)$ at small energies.

\item As we go to higher energies, the mass-squared difference-induced
  neutrino oscillations die down, while the NSI-induced matter effect term
  $L_\nu \epsmutau V_{\rm CC}$ increases. Therefore, large effects of NSI
  are observed at high energies. 

\item The effect of NSI is also larger at higher baselines, since in this case,
  neutrinos pass through inner and outer cores that have very high matter
  densities (around 13 g/cm$^3$ and 11 g/cm$^3$, respectively), and hence
  higher values of $L_\nu \epsmutau V_{\rm CC}$.

\item In many of the $(E_\mu^\text{rec},\,\cos\theta_\mu^\text{rec})$ bins,
  NSI gives rise to an excess in $\mu^-$ events and a deficit in $\mu^+$ events, or vice versa. 
  If $\mu^-$ and $\mu^+$ events are not separated, this would lead to a dilution of information.  Therefore, muon charge information is a crucial
  ingredient in the search for NSI. We have also seen this feature in the
  $\log_{10}[L_\mu^\text{rec}/E_\mu^\text{rec}]$ distributions of
  $\mu^-$ and $\mu^+$ events in Sec.\,\ref{sec:events-L/E}.
  
\end{itemize}

%%%%%%%%%%%%%%%%%%%%%%%%%%%%%%%%%%%%%%%%%%%%%%%%%%%%%%%%%%%%%%%%%%
%%%%%%%%%%%%%%%%%%%%%%%%%%%%%%%%%%%%%%%%%%%%%%%%%%%%%%%%%%%%%%%%%
\section{Identifying NSI through the Shift in Oscillation Dip Location}
\label{sec:reco-dip}
%%%%%%%%%%%%%%%%%%%%%%%%%%%%%%%%%%%%%%%%%%%%%%%%%%%%%%%%%%%%%%%%%
%%%%%%%%%%%%%%%%%%%%%%%%%%%%%%%%%%%%%%%%%%%%%%%%%%%%%%%%%%%%%%%%%

To study the effect of NSI in oscillation dip, we focus on the ratio of
upward-going (U) and downward-going (D) muon events as the observable
to be studied. The up/down event ratio U/D is defined for
$\cos\theta_\mu^{\rm rec}<0$, as~\cite{Kumar:2020wgz}
\begin{equation}
  {\text{U/D}} (E_\mu^{\rm rec},\,\cos\theta_\mu^{\rm rec}) \equiv
  \frac{N(E_\mu^{\rm rec},\,-|\cos\theta_\mu^{\rm rec}|)}{N(E_\mu^{\rm rec},\,
    |\cos\theta_\mu^{\rm rec}|)}\,,
\end{equation}
where the numerator corresponds to the number of upward-going events and
the denominator is the number of downward-going events, with energy
$E_\mu^{\rm rec}$ and direction $|\cos\theta_\mu^{\rm rec}|$.
We associate the U/D ratio  with $\cos\theta_\mu^{\rm rec} <0$ as well as
the corresponding $L_\mu^{\rm rec}$ of upward-going events.  
This ratio is expected to serve as the proxy for the survival probabilities
of $\nu_\mu$ and $\bar\nu_\mu$ at ICAL, since around 98$\%$ muon events
at ICAL arise from the $\nu_\mu\rightarrow\nu_\mu$ oscillation
channel~\cite{Kumar:2020wgz}.

One advantage of the use of the U/D ratio would be to nullify the effects
of systematic uncertainties like flux normalization, cross sections,
overall detection efficiency, and energy dependent tilt error,
as can be seen later.
Note that the up-down symmetry of the detector geometry and detector response
play an important role in this.

%%%%%%%%%%%%%%%%%%%%%%%%%%%%%%%%%%%%%%%%%%%%%%%%%%%%%%%%%%%%%%%%%
\begin{figure}[t]
  \centering
  \includegraphics[width=0.48\linewidth]{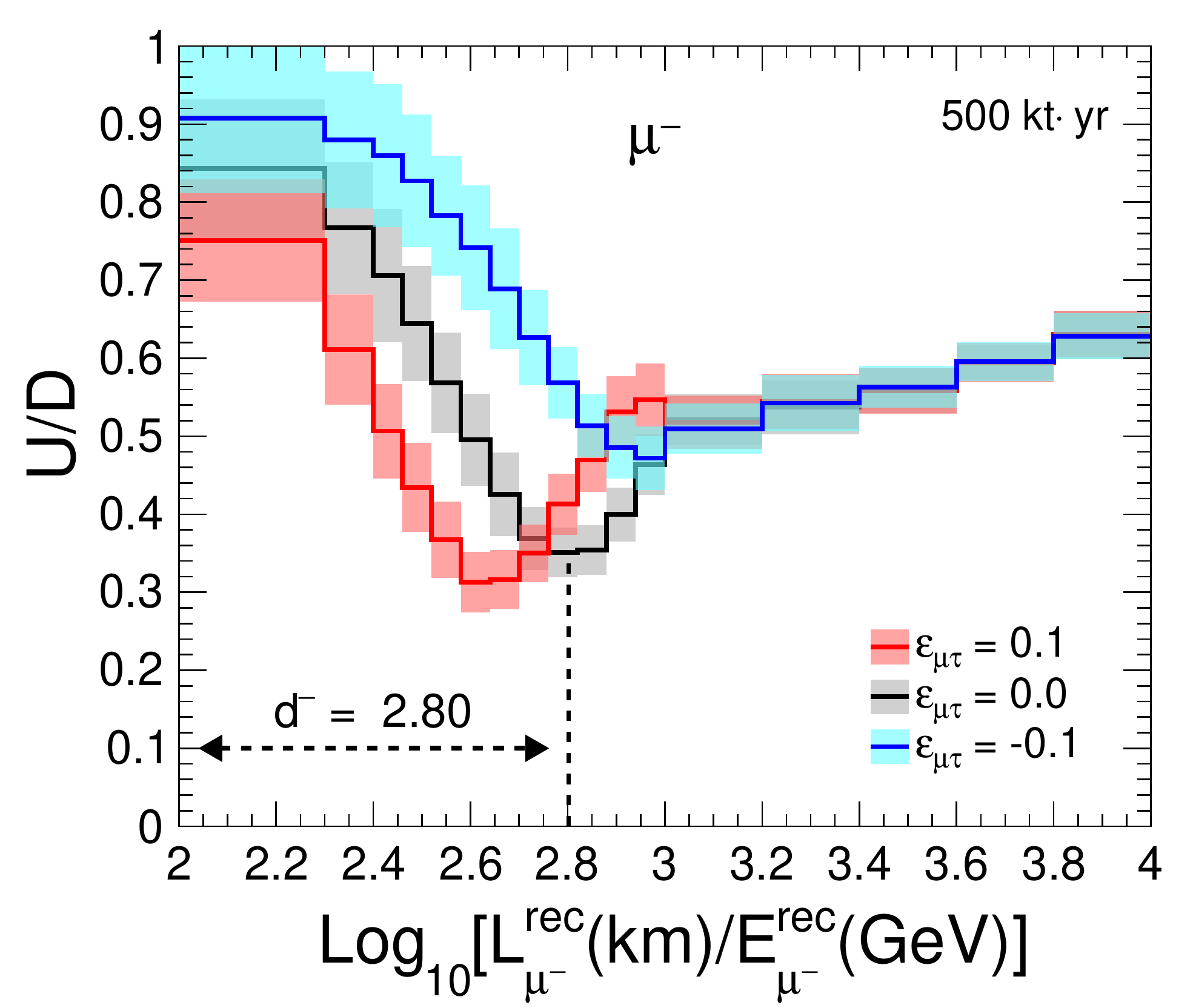}
  \includegraphics[width=0.48\linewidth]{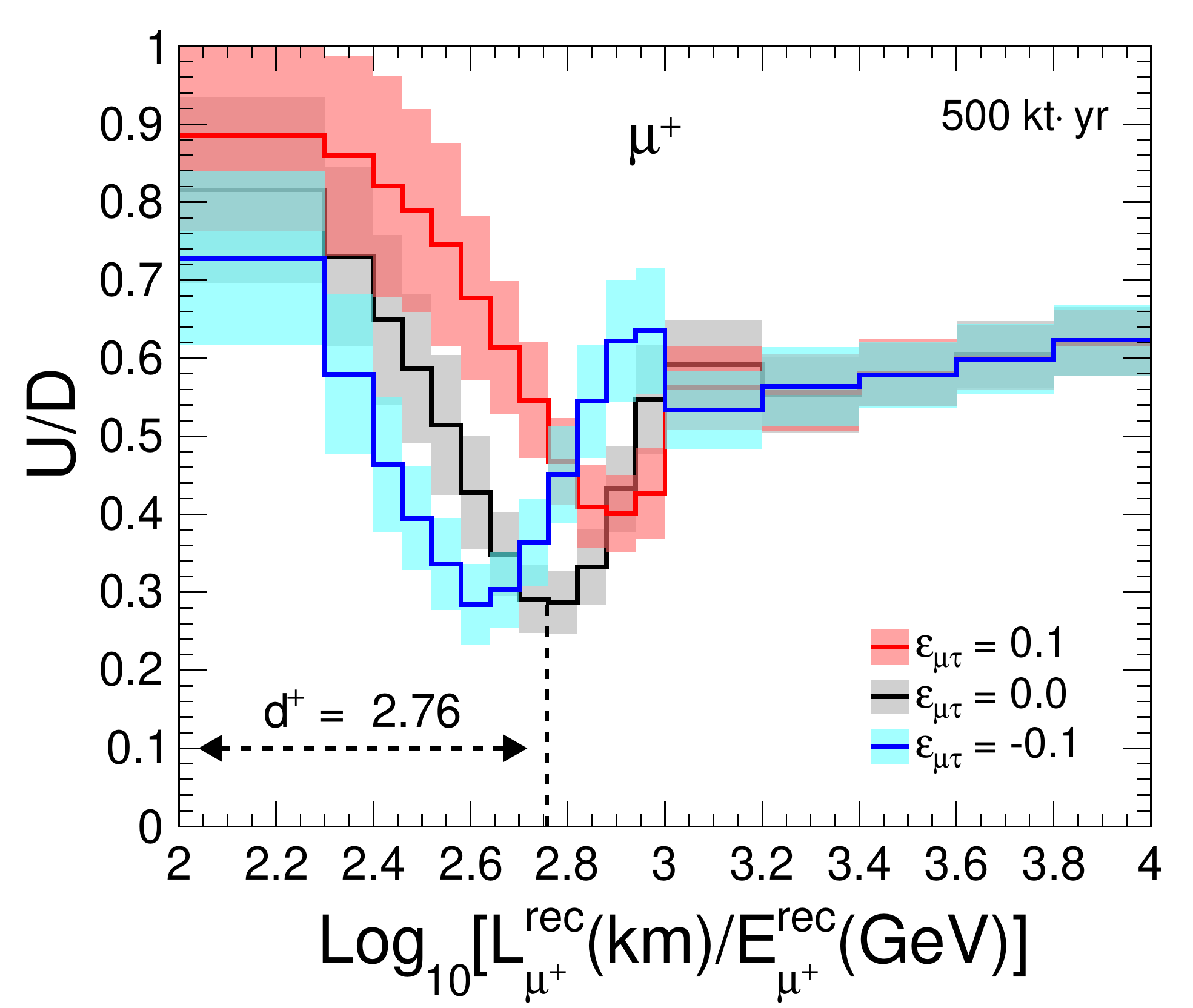}
  \mycaption{The $\log_{10}(L_\mu^\text{rec}/E_\mu^\text{rec})$ distributions of
    ratio of upward-going and downward-going $\mu^-$ (left panel) and $\mu^+$
    (right panel) events using 10-year data of ICAL. The black lines
    correspond to SI ($\varepsilon_{\mu\tau} = 0$), whereas red and blue
    lines are for $\varepsilon_{\mu\tau} = 0.1$ and -0.1, respectively. 
    The statistical fluctuations shown by shaded boxes are the
    root-mean-square deviation of 100 independent distributions of U/D
    ratio, each for 10 years, whereas the mean of these distributions are
    shown by solid lines.
    We consider normal mass ordering, and the benchmark oscillation
    parameters given in Table~\ref{tab:osc-param-value}.
  }
  \label{fig:updown-lbye-10}
\end{figure}
%%%%%%%%%%%%%%%%%%%%%%%%%%%%%%%%%%%%%%%%%%%%%%%%%%%%%%%%%%%%%%%%%%%%%

Fig.~\ref{fig:updown-lbye-10} presents the $L_\mu^{\text{rec}}/E_\mu^{\text{rec}}$
distributions of U/D with $\varepsilon_{\mu\tau} = 0$ (SI) and $\pm 0.1$,
for an exposure of 10 years at ICAL. The mass ordering is taken as NO.
The statistical fluctuations shown in the figure are the root-mean-square
deviation obtained from the simulations of 100 independent sets of 10 years
data. Around $\log_{10}[L_\mu^\text{rec}/E_\mu^\text{rec}] \sim 2.4 - 2.6$,
we see a significant modification in the U/D ratio due to the presence of
non-zero $\varepsilon_{\mu\tau}$ ($\pm 0.1$), particularly in the location of
the oscillation dip. The dip shifts towards left or right ({\it i.e.} to lower
or higher values of $L_\mu^{\rm rec}/E_\mu^{\rm rec}$, respectively) from that
of the SI case with non-zero $\varepsilon_{\mu\tau}$. This shift for
$\mu^-$ and $\mu^+$ events is observed to be in opposite directions, as
is expected from the discussions in Sec.\,\ref{subsec:osc-dip}. 
For example, for $\varepsilon_{\mu\tau} >0$, the location of oscillation dip in
$\mu^-$ events gets shifted towards smaller $L_\mu^\text{rec}/E_\mu^\text{rec}$
values. On other hand, in $\mu^+$ events, the oscillation dip shifts to
higher values of $L_\mu^\text{rec}/E_\mu^\text{rec}$.
For $\varepsilon_{\mu\tau} <0$, this shift is in the opposite directions.
Moreover, as the oscillation dip shifts towards the higher value of
$L_\mu^\text{rec}/E_\mu^\text{rec}$, the dip becomes shallower due to the
effect of rapid oscillation at high $L_\nu/E_\nu$.
This effect is visible in both, $\mu^-$ and $\mu^+$.  
It can be easily argued that the modification in oscillation dips of
$\mu^-$ and $\mu^+$ due to non-zero $\varepsilon_{\mu\tau}$ is the reflection
of how the survival probabilities of neutrino and antineutrino change in
the presence of non-zero $\varepsilon_{\mu\tau}$, as expected from
Eq.\,\ref{eq:osc-min-lbye} in Sec.~\ref{subsec:osc-dip}.

Note that for a neutrino detector that is blind to the charge of muons,
the dip itself would get diluted since different average inelasticities of
neutrino and antineutrino events at these energies would lead to
slightly different dip locations in $\mu^-$ and $\mu^+$ distributions.
Moreover, the shift of oscillation dip towards opposite directions in
$\mu^-$ and $\mu^+$ events due to non-zero $\varepsilon_{\mu\tau}$
would further contribute to the dilution of NSI effects on the dip location.
On the other hand, a detector like ICAL that has the charge identification
capability, not only provides independent undiluted measurements of dip
locations in $\mu^-$ and $\mu^+$ events, but also provides a unique novel
observable that can cleanly calibrate against the value of $\epsmutau$,
as we shall see in the next section.

%%%%%%%%%%%%%%%%%%%%%%%%%%%%%%%%%%%%%%%%%%%%%%%%%%%%%%%%%%%%%%%%%%%%%%
\subsection{A novel variable $\Delta d$ for determining $\epsmutau$}
\label{subsec:delta-d}
%%%%%%%%%%%%%%%%%%%%%%%%%%%%%%%%%%%%%%%%%%%%%%%%%%%%%%%%%%%%%%%%%%%%%%

%%%%%%%%%%%%%%%%%%%%%%%%%%%%%%%%%%%%%%%%%%%%%%%%%%%%%%%%%%%%%%%%%%%%
 \begin{figure}[t]
 \centering
  \includegraphics[width=0.48\linewidth]{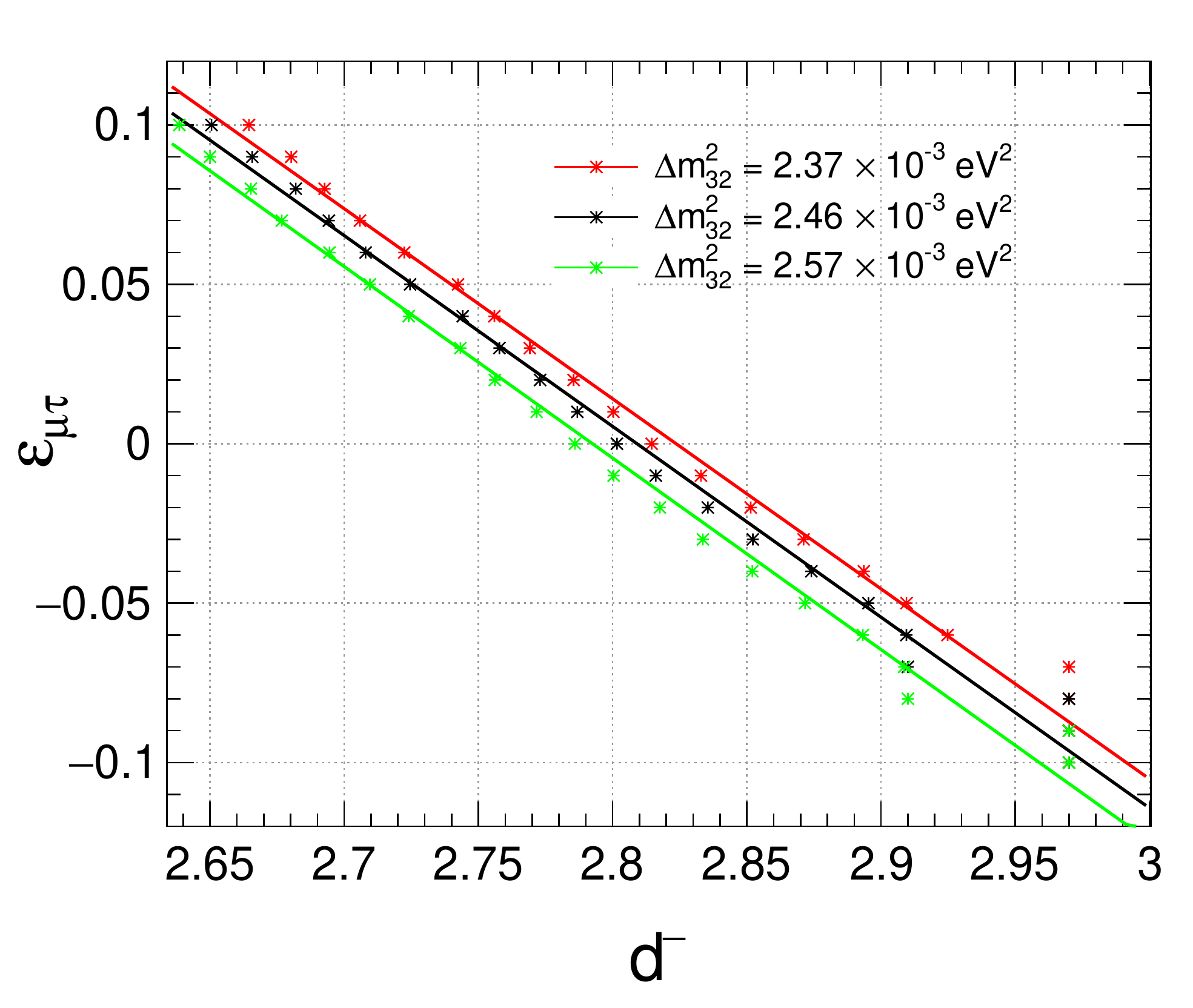}
  \includegraphics[width=0.48\linewidth]{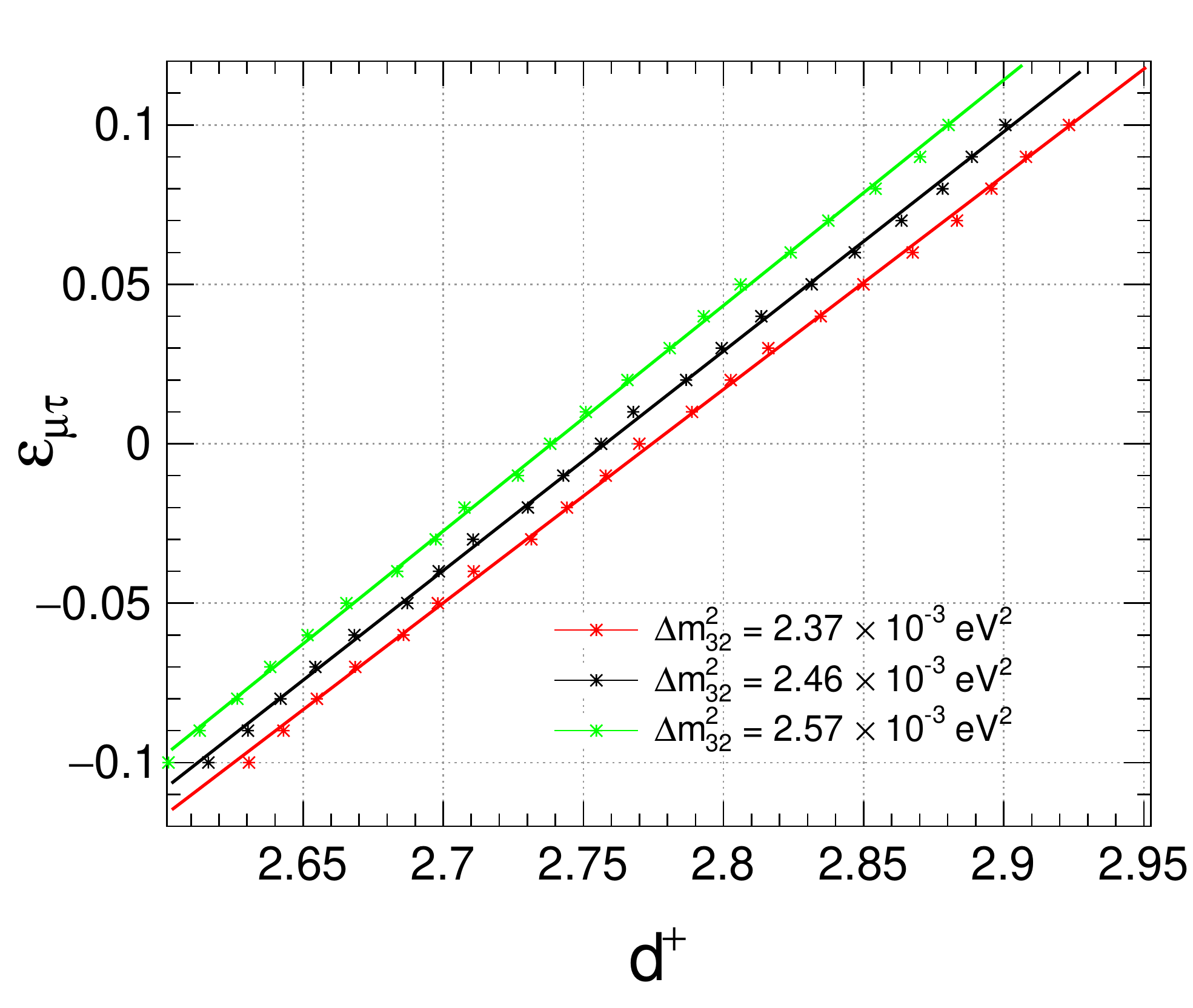}
  \includegraphics[width=0.48\linewidth]{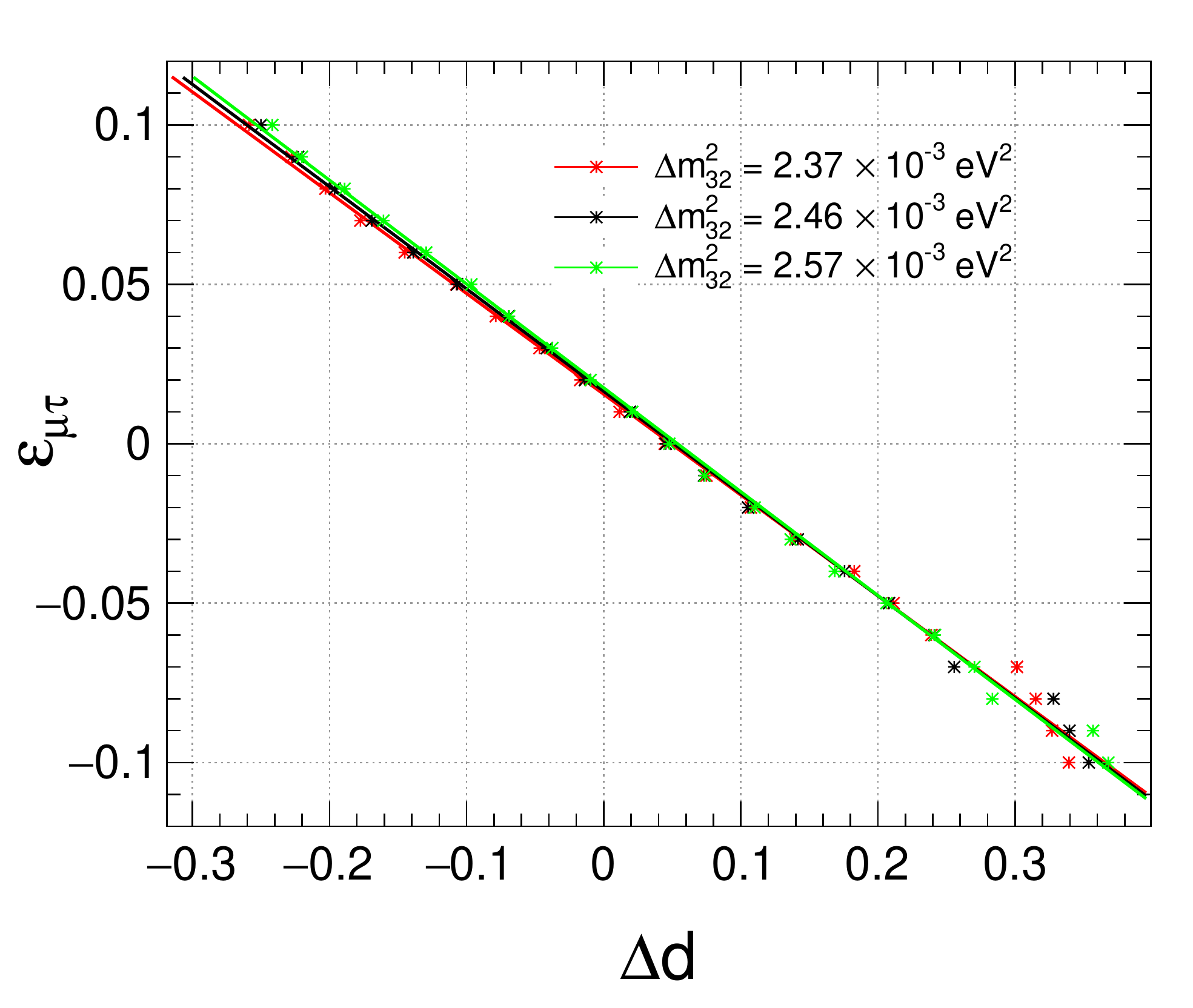}
  \mycaption{Upper panels: The calibration of $\varepsilon_{\mu\tau}$ with
    the dip location in U/D ratio of $\mu^-$ (left panel) and $\mu^+$
    (right panel) events separately. Lower panel: The calibration of
    $\varepsilon_{\mu\tau}$ with difference in dip location of $\mu^-$
    and $\mu^+$.   
    Note that for $-0.1<\varepsilon_{\mu\tau}<-0.07$, the dip location is
    simply the center of the $\log_{10}[L_\mu^\text{rec}/E_\mu^\text{rec}]$ bin with minimum U/D ratio since the fitting is not stable for
    these values.
    We consider normal mass ordering, and the benchmark oscillation
    parameters given in Table~\ref{tab:osc-param-value}.
  }
  \label{fig:novel-variable}
 \end{figure}
%%%%%%%%%%%%%%%%%%%%%%%%%%%%%%%%%%%%%%%%%%%%%%%%%%%%%%%%%%%%%%%%%%%%%%

Since the oscillation dip locations in the U/D distributions of both,
$\mu^-$ and $\mu^+$ events, shift to lower / higher values of
$L_\mu^{\text{rec}}/E_\mu^{\text{rec}}$ depending on the value of $\epsmutau$,
the dip location in either of these distributions may be used to
estimate $\varepsilon_{\mu\tau}$, independently.
We have already developed a dip-identification algorithm~\cite{Kumar:2020wgz},
where the dip location is not simply the lowest value of U/D ratio, but
takes into account information from the surrounding bins
that have a U/D ratio below a  threshold value. The algorithm essentially
identifies the cluster of contiguous bins that have a U/D ratio
lower than any surrounding bins, and fits these bins with a quadratic
function whose minimum corresponds to the dip.
The value of $\log_{10}[L_\mu^\text{rec}/E_\mu^\text{rec}]$ corresponding to this
minimum is termed as the dip location. We denote the dip location for
$\mu^-$ and $\mu^+$ events as $d^-$ and $d^+$, respectively.

We obtain the calibration of $\varepsilon_{\mu\tau}$ with the dip locations
for $\mu^-$ and $\mu^+$ events separately, using 1000-year MC data.
In the top panels of Fig.\,\ref{fig:novel-variable}, we present the calibration
curves of $\epsmutau$ with $d^-$ (left panel) and $d^+$ (right panel),
for three different values of $\Delta m^2_{32}$. The straight line nature of
the calibration curves can be understood qualitatively using
Eq.~\ref{eq:osc-min-lbye}. 
For a majority of events ($E_\mu^\text{rec} <25$ GeV), the
$\epsmutau V_{\rm CC} E_\nu$ term is much smaller than $\Delta m^2_{32}$ term,
and therefore the linear expansion as shown in
Eq.~\ref{eq:lbye-dip-calibration} is valid.

In Eq.~\ref{eq:lbye-dip-calibration}, the dominating $\Delta m^2_{32}$
dependence of the dip location is through the first term.
We can get rid of this dependence by introducing a new variable
\begin{equation}
  \Delta d = d^- - d^+\,,
  \label{eq:def-deltad}
\end{equation}
which is expected to be proportional to $\epsmutau$, and can be used to
calibrate for $\epsmutau$. As may be seen in the bottom panel of
Fig.\,\ref{fig:novel-variable}, this indeed removes the dominant 
$\Delta m^2_{32}$-dependence in the calibration of $\epsmutau$.
Note that some $\Delta m^2_{32}$-dependence still survives, which may
be seen in the slightly different slopes of the calibration curves
for different $\Delta m^2_{32}$ values. However, this dependence is clearly
negligible, as it appears in Eq.~\ref{eq:lbye-dip-calibration} as a
multiplicative correction to $\epsmutau$. We have thus obtained a
calibration for $\epsmutau$ that is almost independent of $\Delta m^2_{32}$.

Note that $\varepsilon_{\mu\tau} = 0$ does not imply $\Delta d =0$,
since different matter effects and different inelasticities in neutrino and
antineutrino channels cause the dips to arise at slightly different $\log_{10}[L_\mu^\text{rec}/E_\mu^\text{rec}]$ values in these two channels even in the absence of NSI.

%%%%%%%%%%%%%%%%%%%%%%%%%%%%%%%%%%%%%%%%%%%%%%%%%%%%%%%%%%%%%%%%%%%%%%%
\subsection{Constraints on $\varepsilon_{\mu\tau}$ from the measurement
  of $\Delta d$}
\label{subsec:eps-limit-delta-d}
%%%%%%%%%%%%%%%%%%%%%%%%%%%%%%%%%%%%%%%%%%%%%%%%%%%%%%%%%%%%%%%%%%%%%%

%%%%%%%%%%%%%%%%%%%%%%%%%%%%%%%%%%%%%%%%%%%%%%%%%%%%%%%%%%%%%%%%%%%%%
\begin{figure}[t]
  \centering
  \includegraphics[width=0.75\linewidth]{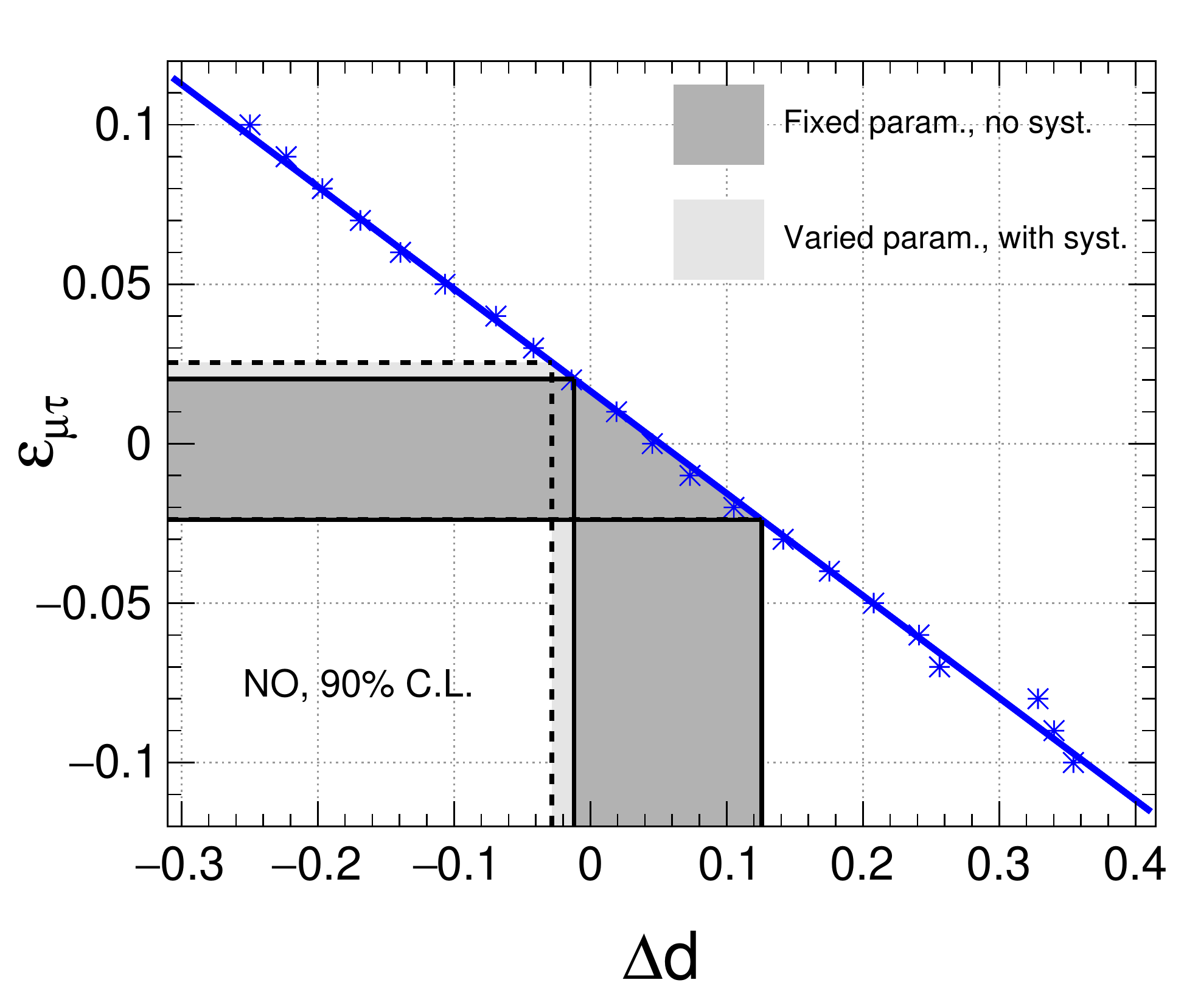}
  \mycaption{The blue stars and the blue solid line indicate the calibration curve
      of $\epsmutau$ against $\Delta d$, obtained using 1000-year MC data with normal mass ordering. 
      The gray bands show the expected 90\% C.L. limits on $\Delta d$
      (vertical bands), and hence on $\epsmutau$ (horizontal bands),
      with 500 kt$\cdot$yr of ICAL exposure, when the actual value of $\varepsilon_{\mu\tau}=0$. The light (dark) gray  bands
      show the limits when the errors on oscillation parameters,
      and the impact of systematic uncertainties 
      are included (excluded), as discussed in Sec.~\ref{subsec:eps-limit-delta-d}. For fixed-parameter case, we use the benchmark oscillation
      parameters given in Table~\ref{tab:osc-param-value}.
  }
  \label{fig:NSI_calib_dip_diff_band}
\end{figure}
%%%%%%%%%%%%%%%%%%%%%%%%%%%%%%%%%%%%%%%%%%%%%%%%%%%%%%%%%%%%%%%%%%%%%%%%

We saw in the last section that one can infer the value of
$\varepsilon_{\mu\tau}$ in the observed data from the calibration curve
between $\Delta d$ and $\varepsilon_{\mu\tau}$, independent of the actual value
of $\Delta m^2_{32}$. We obtain the calibration curve using the 1000-year
MC-simulated sample, and  show it with the solid blue line in Fig.~\ref{fig:NSI_calib_dip_diff_band}.

For estimating the expected constraints on $\varepsilon_{\mu\tau}$, we simulate
100 independent sets of data, each for 10-year exposure of ICAL with
the benchmark oscillation parameters as given in
Table~\ref{tab:osc-param-value}, and $\varepsilon_{\mu\tau} = 0$.
In Fig.~\ref{fig:NSI_calib_dip_diff_band}, the range indicated along the x-axis
by dark gray band is the expected 90\% C.L. range of the
measured value of $\Delta d$, while the range indicated along the 
y-axis by the dark gray band is the expected 90\% C.L. limit on
$\varepsilon_{\mu\tau}$. From the figure, we find
that one can expect to constrain $\epsmutau$ to
$-0.024 < \varepsilon_{\mu\tau} < 0.020$
at 90\% C.L. with 10 years of data.

Since the calibration was found to be almost independent of the actual
value of $\Delta m^2_{32}$, we expect that even if the actual value of
$\Delta m^2_{32}$ were not exactly known, the results will not change.
We have confirmed this by generating 100 independent sets of 10-year exposure,
where the $\Delta m^2_{32}$ values used as inputs follow a Gaussian
distribution $\Delta m^2_{32} = (2.46 \pm 0.03) \times 10^{-3}$ eV$^2$,
in accordance to the range allowed by the global
fit~\cite{deSalas:2020pgw,Esteban:2020cvm,Capozzi:2017ipn}
of available neutrino data.

We also checked the impact of uncertainties in
the values of the other oscillation parameters.
We first simulated 100 statistically independent unoscillated
data sets. Then for each of these data sets, we take 20
random choices of oscillation parameters, according to the Gaussian
distributions
\begin{equation}
  \Delta m^2_{21} = (7.4 \pm 0.2) \times 10^{-5} \mbox{ eV}^2 ~,
  \Delta m^2_{32} = (2.46 \pm 0.03) \times 10^{-5} \mbox{ eV}^2 ~,
\end{equation}
\begin{equation}
  \sin^2 2\theta_{12} = 0.855 \pm 0.020 ~, ~
  \sin^2 2\theta_{13} = 0.0875 \pm 0.0026 ~,~
  \sin^2 \theta_{23} = 0.50 \pm 0.03 ~,
\end{equation}
guided by the present global fit \cite{NUFIT}.
We keep $\delta_{\rm CP}=0$, since its effect on $\nu_\mu$ survival
probability is known to be highly suppressed in the multi-GeV
energy range~\cite{Kopp:2007ne}.
This procedure effectively enables us to consider the variation of our
results over 2000 different combinations of oscillation parameters,
to take into account the effect of their uncertainties.

We do not observe any significant dilution of results due to the
variation over oscillation parameters. This is expected, since
(i) solar oscillation parameters $\Delta m^2_{21}$ and $\theta_{12}$
do not contribute significantly to the $\nu_\mu$ survival probability
in the multi-GeV range,
(ii) the reactor mixing angle $\theta_{13}$ is already measured
to a high precision,
(iii) the mixing angle $\theta_{23}$ does not affect the location
of the oscillation minimum at the probability level, and
(iv) our procedure of using $\Delta d$ (see Eq.~\ref{eq:def-deltad})
to determine $\epsmutau$ minimizes the impact of $\Delta m^2_{32}$
uncertainty, as shown in Sec.~\ref{subsec:delta-d}.

In addition, we take into account the five major systematic
uncertainties in the neutrino fluxes and cross sections that are used
in the standard ICAL analyses~\cite{Kumar:2017sdq}.
These five uncertainties are
(i) 20\% in overall flux normalization, (ii) 10\% in cross sections,
(iii) 5\% in the energy dependence,
(iv) 5\% in the zenith angle dependence, and
(v) 5\% in overall systematics.
For each of the 2000 simulated data sets, we modify the number
of events in each $(E_\mu^{\rm rec}, \cos\theta_\mu^{\rm rec})$  bin as
\be
N = N^{(0)} (1 + \delta_1) (1 + \delta_2) 
(E_\mu^{\rm rec}/E_0)^{\delta_3}
(1 + \delta_4 \cos \theta_\mu^{\rm rec}) (1 + \delta_5)~,
\ee
where $N^{(0)}$ is the theoretically predicted number of events,
and $E_0 = 2$ GeV.
Here, $(\delta_1, \delta_2, \delta_3, \delta_4, \delta_5)$ is an
ordered set of random numbers, generated separately for each simulated
data set, with the Gaussian distributions centered around zero and
the $1\sigma$ widths given by $(20\%, 10\%, 5\%, 5\%, 5\%)$. 
The normalization uncertainties and energy tilt uncertainty are expected
to be canceled in the U/D ratio, while the zenith angle distribution
uncertainty affects the upward-going and downward-going events differently,
and hence would be expected to affect the U/D ratio. 
We explicitly check this and indeed find this to be true.
  
When the oscillation parameter uncertainties and all the five
systematic uncertainties mentioned above are included, it is found that the
method based on the shift in the dip locations may constrain $\epsmutau$ to
$- 0.025 < \epsmutau < 0.024$ at 90\% C.L..
The results denoting the effects of these uncertainties have been
shown in Fig.~\ref{fig:NSI_calib_dip_diff_band}.

%%%%%%%%%%%%%%%%%%%%%%%%%%%%%%%%%%%%%%%%%%%%%%%%%%%%%%%%%%%%%%%%%%%%%
%%%%%%%%%%%%%%%%%%%%%%%%%%%%%%%%%%%%%%%%%%%%%%%%%%%%%%%%%%%%%%%%%%%%%
\section{A New Method for Determining $\Delta m^2_{32}$
  in the Absence of NSI}
\label{subsec:dmsq-si}
%%%%%%%%%%%%%%%%%%%%%%%%%%%%%%%%%%%%%%%%%%%%%%%%%%%%%%%%%%%%%%%%%%%%
%%%%%%%%%%%%%%%%%%%%%%%%%%%%%%%%%%%%%%%%%%%%%%%%%%%%%%%%%%%%%%%%%%

The reconstruction of the oscillation valley using the distributions of the
upward-going and downward-going muons in
$(E^{\rm rec}_\mu,\, \cos\theta_\mu^{\rm rec})$ plane is discussed in detail
in Ref.~\cite{Kumar:2020wgz}. It has been shown that the alignment of
the valley can provide a measurement of $\Delta m^2_{32}$.
In this paper, we use an alternative procedure for getting the alignment
of the oscillation valley, which is more robust and more well-motivated
as compared to what was used in Ref.~\cite{Kumar:2020wgz}, and determine the
expected  precision that the ICAL detector can achieve on the atmospheric
oscillation parameter $\Delta m^2_{32}$, using this method.

%%%%%%%%%%%%%%%%%%%%%%%%%%%%%%%%%%%%%%%%%%%%%%%%%%%%%%%%%%%%%%%%%%%%%%%
\begin{figure}[t]
  \centering
  \includegraphics[width=0.48\linewidth]{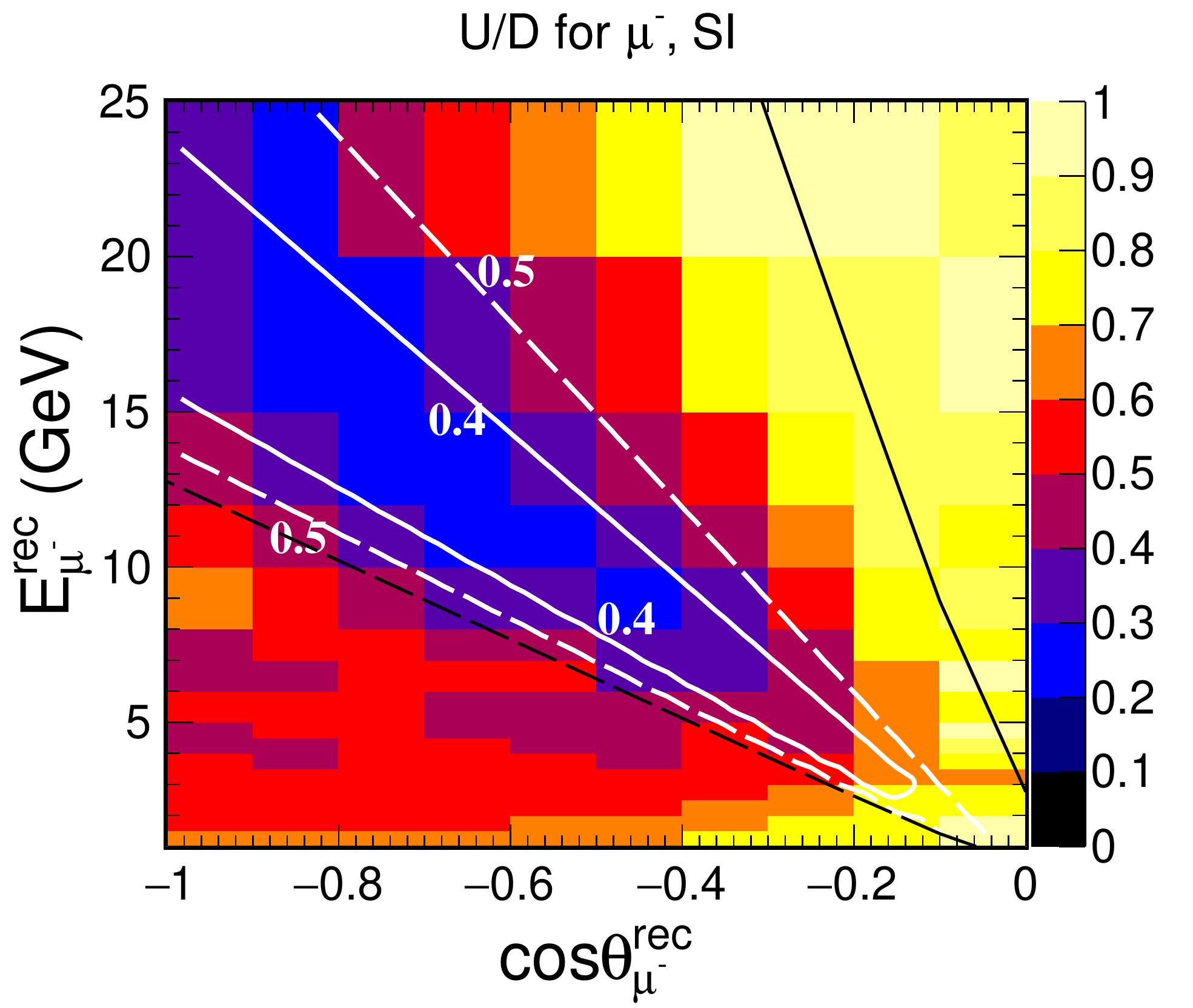}
  \includegraphics[width=0.48\linewidth]{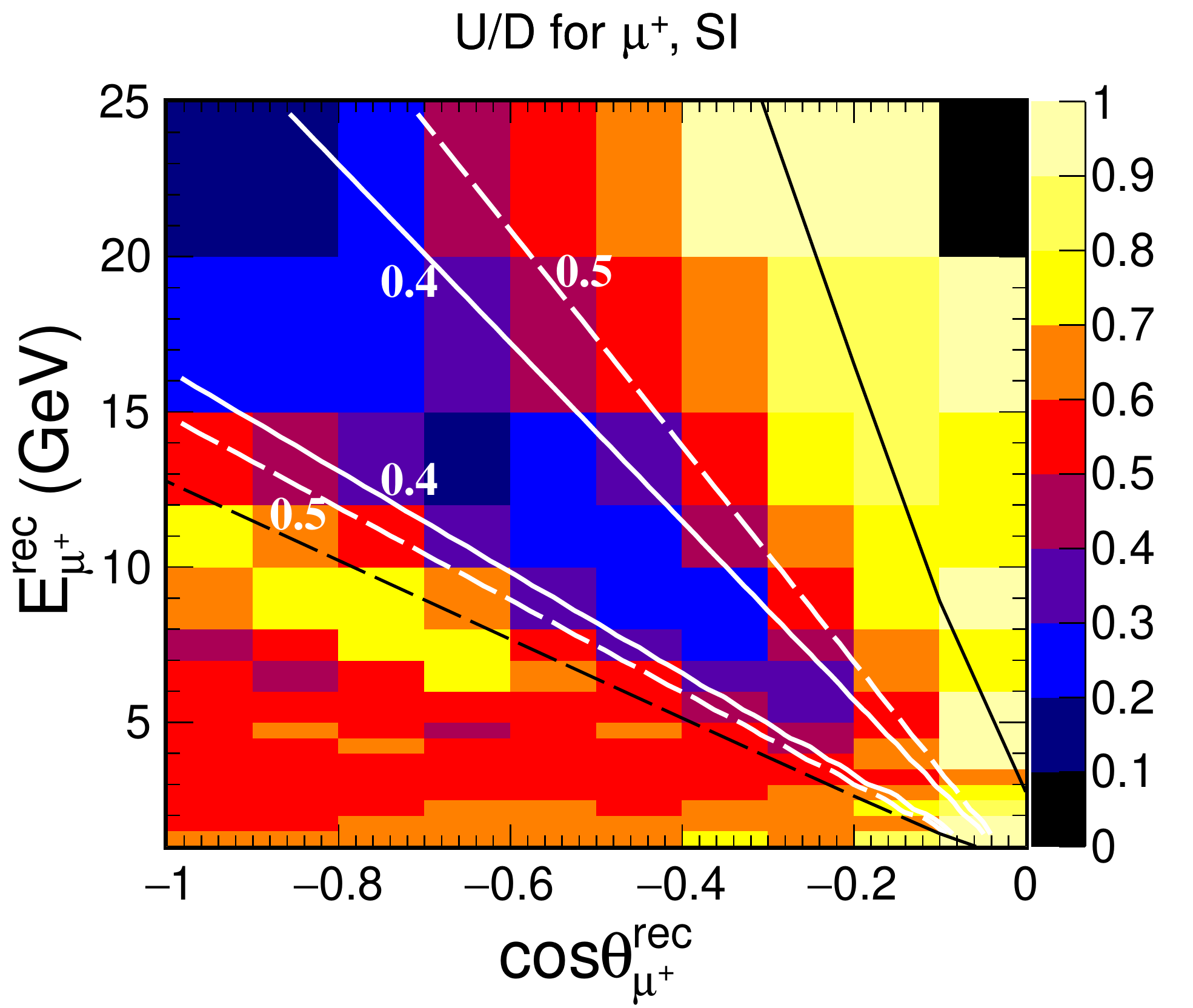}
  \mycaption{The distributions of U/D ratios
    in ($E_\mu^{\rm rec}$, $\cos\theta_\mu^{\rm rec}$) plane,
    for $\mu^-$ and $\mu^+$ events, in left and right panels, respectively.
    This is the average of 100 independent simulated data sets,
    each for 10-year exposure at ICAL, with $\varepsilon_{\mu\tau}=0$.
    The white solid and dashed lines correspond to the contours with
    fitted U/D ratio equal to 0.4 and 0.5, respectively, obtained after the fitting
    of oscillation valley with the function $F_0$ as given in
    Eq.~\ref{eq:fitting-func-sm}. The black solid and dashed lines
    correspond to $\log_{10}[L_\mu^\text{rec}/E_\mu^\text{rec}] =$ 2.2 and 3.0, respectively.
    We consider normal mass ordering, and the benchmark oscillation
    parameters given in Table~\ref{tab:osc-param-value}.
    }
    \label{fig:valley-sm-u-by-d}
\end{figure}
%----------------------------------

Fig.~\ref{fig:valley-sm-u-by-d} shows the distributions of the ratio of
upward-going and downward going events at ICAL in the
$(E_\mu^{\rm rec}$, $\cos\theta_\mu^{\rm rec})$ plane for 10 years of ICAL data.
The left and right panels show the U/D ratios
for $\mu^-$ and $\mu^+$ events, respectively.
For the sake of demonstration (in order to reduce the statistical
fluctuations in the figure and emphasize the physics),
we show the binwise average of 100 independent data sets,
each with 10-year exposure of ICAL.
The blue bands in Fig.~\ref{fig:valley-sm-u-by-d} correspond to the
reconstructed oscillation valley.
Note that a good reconstruction of the oscillation valley
would be an evidence for the fidelity of the ICAL detector.

To get the alignment of the oscillation valley, we fit the U/D distribution
for $\mu^-$ or $\mu^+$ independently with a functional form
\begin{equation}
  F_0(E_\mu^{\rm rec}, \cos\theta_\mu^{\rm rec}) = Z_0 + N_0 \cos^2
  \left(m_0 \frac{\cos\theta_\mu^{\rm rec}}{E_\mu^{\rm rec}} \right),
 \label{eq:fitting-func-sm}
\end{equation}
where $Z_0$, $N_0$, and $m_0$ are the independent parameters to be determined
from the fitting of U/D distributions. The parameter $Z_0$ quantifies the
minimum depth of the fitted U/D ratio, $N_0$ is the normalization constant,
whereas $m_0$ is the slope of the oscillation valley. Since more than
$95\%$ of the events at ICAL are contributed from the $\nu_\mu \to \nu_\mu$ and
$\bar\nu_\mu \to \bar\nu_\mu$ survival probabilities, we expect that the function
$F_0$, that resembles Eq.~\ref{eq:pmumu-nu-final-omsd}, would be suitable for
fitting the oscillation valley in the $(E_\mu^{\rm rec}$, $\cos\theta_\mu^{\rm rec})$
plane.
Here, the slope $m_0$ can be used directly to calibrate $\Delta m^2_{32}$.
The parameters $N_0$ and $Z_0$ also contain information about the atmospheric
mixing parameters, however, they cannot be connected easily
to the determinations of these parameters.

%%%%%%%%%%%%%%%%%%%%%%%%%%%%%%%%%%%%%%%%%%%%%%%%%%%%%%%%%%%%%%%%%%%%
\begin{figure}[t]
  \centering
  \includegraphics[width=0.48\linewidth]{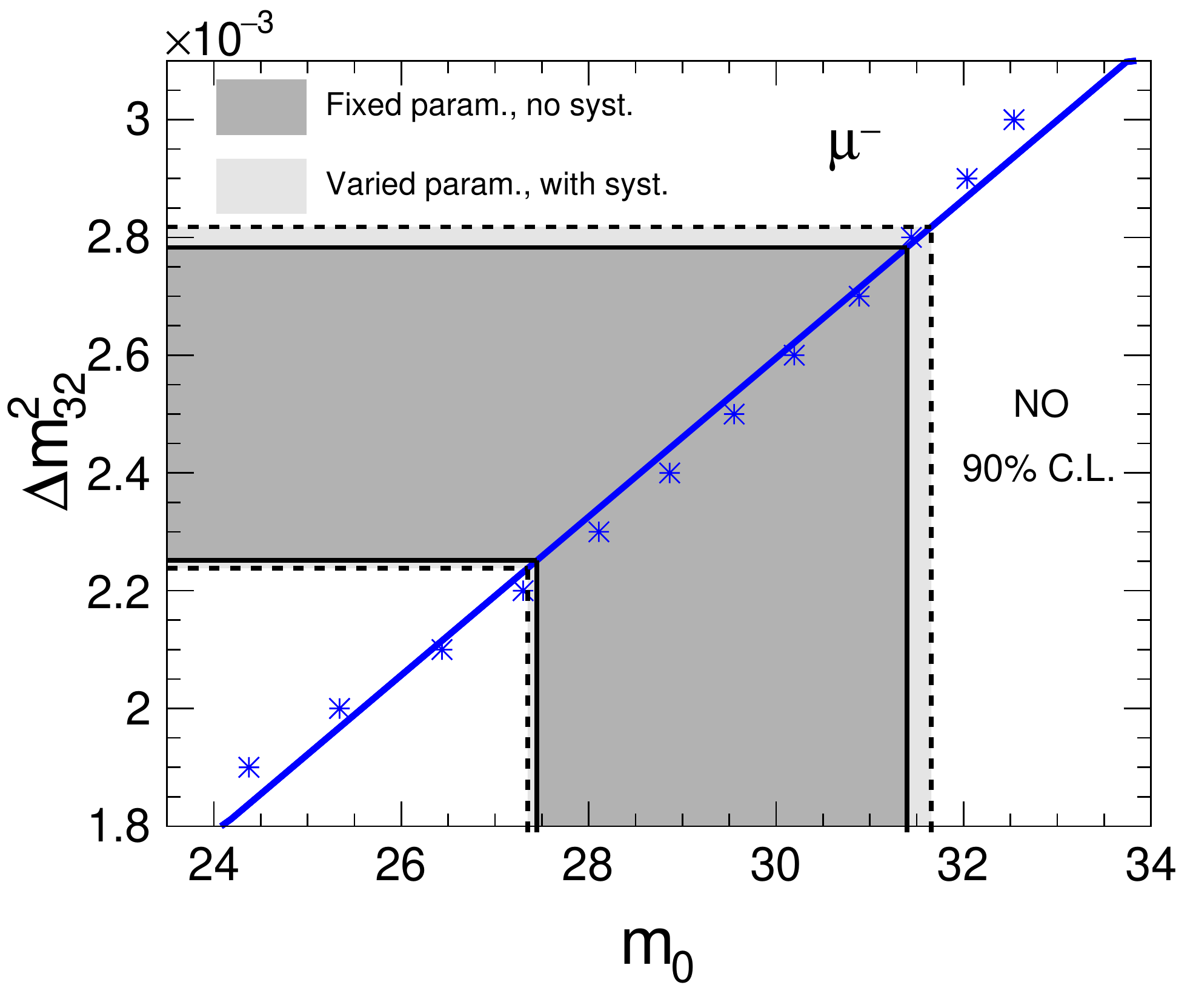}
  \includegraphics[width=0.48\linewidth]{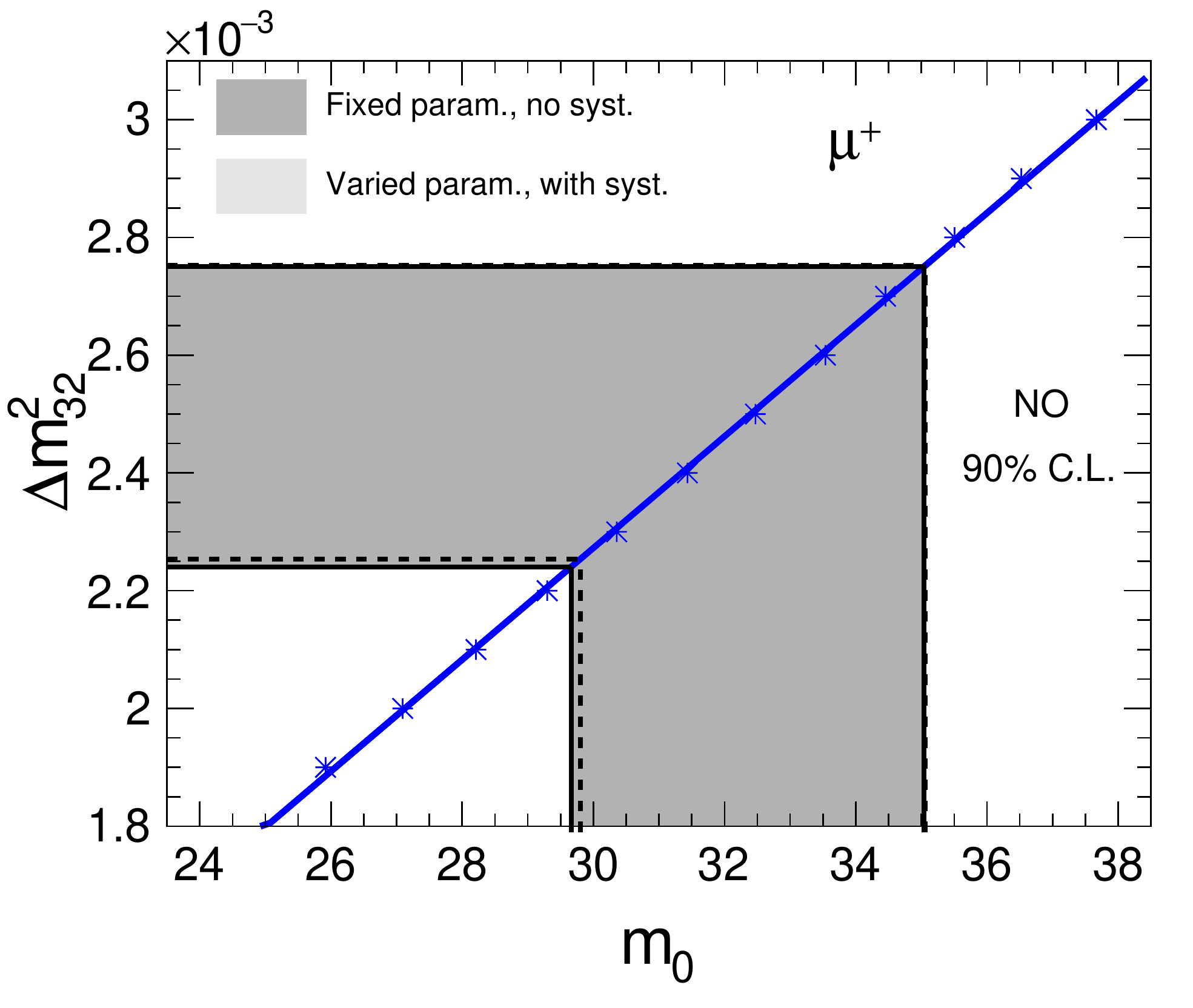}
  \mycaption{
    The blue stars and the blue lines
    indicate the calibration curves of $\Delta m^2_{32}$
    against $m_0$, obtained using 1000-year MC data with normal mass ordering. 
    The gray bands represent the 90\% C.L. allowed ranges of $m_0$
    (vertical bands), and hence of $\Delta m^2_{32}$ (horizontal bands), with a given input value of $\Delta m^2_{32}$ = 2.46$\times 10^{-3}$ eV$^2$,
    for an exposure of 10 years at ICAL.
    The results for $\mu^-$ and $\mu^+$ events are shown separately,
    in the left and right panels, respectively.
    The light (dark) gray  bands show the ranges when the errors on
    other oscillation parameters and the impact of systematic uncertainties 
    are included (excluded), as discussed in Sec.~\ref{subsec:eps-limit-delta-d}. For fixed-parameter case, we use the benchmark oscillation
      parameters given in Table~\ref{tab:osc-param-value}.
    }
    \label{fig:calibration-sm-dms}
\end{figure}
%%%%%%%%%%%%%%%%%%%%%%%%%%%%%%%%%%%%%%%%%%%%%%%%%%%%%%%%%%%%%%%%%%%%

Figure~\ref{fig:valley-sm-u-by-d} also shows the contour lines for the
fitted U/D ratio of 0.4 and 0.5. 
The contour lines for $\mu^+$ are seen to reproduce the data better than
those for $\mu^-$. The reason behind this is the matter effects, which appear
in $\mu^-$ data since the mass ordering is assumed to be NO here.
Note that the fitting function $F_0$ is designed for only the vacuum
oscillation, and we focus only on the region where vacuum oscillation
is expected to dominate. In order to be safe from matter effects,
we remove the events with $\log_{10}[L_\mu^\text{rec}/E_\mu^\text{rec}]>3.0$,
where the matter effects are significant. We also discard the events with
$\log_{10}[L_\mu^\text{rec}/E_\mu^\text{rec}]<2.2$, since these would have
barely oscillated. The latter cut also gets rid of most of the events
near horizon, where the distance traveled by the neutrinos has large errors.
In order to minimize the data from the bins with clearly large fluctuations,
we use another cut on the maximum value of the U/D ratio. This cut is
taken to be U/D $< 0.9$ for both $\mu^-$ and $\mu^+$ events.

We calibrate for $\Delta m^2_{32}$ by fitting the U/D ratio of 1000-year MC
data with input $\Delta m^2_{32}$ values in the range
$(1.9 - 3.0)\times 10^{-3}$ eV$^2$, and obtaining the corresponding  value of
$m_0$. 
The statistical fluctuations are estimated by simulating 100 independent
data sets of 10 years for a given input value of $\Delta m^2_{32}$ $= 2.46\times 10^{-3}$ eV$^2$, and
fitting for the U/D ratio independently with Eq.~\ref{eq:fitting-func-sm}.
The 100 values of $m_0$ thus obtained provide the expected uncertainties
on $\Delta m^2_{32}$. In Fig.\,\ref{fig:calibration-sm-dms}, the gray bands
show the $90\%$ C.L. ranges expected to be inferred by this method in
10 years.
The figure also shows the (small) deterioration in the $90\%$ C.L. range of
$\Delta m^2_{32}$ due to the incorporation of the present uncertainties
in the oscillation parameters, and all the five systematic errors,
as discussed in Sec.~\ref{subsec:eps-limit-delta-d}. With systematic uncertainties and error in other oscillation parameters, we get the expected $90\%$ C.L. allowed range for $\Delta m^2_{32}$ from $\mu^-$ events as (2.24 -- 2.82)$\times 10^{-3}$ eV$^2$ and from $\mu^+$ events as (2.25 -- 2.75)$\times 10^{-3}$ eV$^2$.

An advantage of the method proposed in this section over the original
proposal in \cite{Kumar:2020wgz} is that this is a one-step fitting procedure
as opposed to the two-step fitting procedure therein.
Note that the results obtained by this method are also slightly
better than the ones obtained in~\cite{Kumar:2020wgz}, as the shape
of the valley is explicitly fitted to. 
Moreover, we can extend this method to measure other characteristics of
the oscillation valley, such as identifying the signature of NSI,
which we shall discuss in the next section.

%%%%%%%%%%%%%%%%%%%%%%%%%%%%%%%%%%%%%%%%%%%%%%%%%%%%%%%%%%%%%%%%%
%%%%%%%%%%%%%%%%%%%%%%%%%%%%%%%%%%%%%%%%%%%%%%%%%%%%%%%%%%%%%%%%
\section{Identifying NSI through Oscillation Valley}
\label{sec:reco-valley}
%%%%%%%%%%%%%%%%%%%%%%%%%%%%%%%%%%%%%%%%%%%%%%%%%%%%%%%%%%%%%%%%
%%%%%%%%%%%%%%%%%%%%%%%%%%%%%%%%%%%%%%%%%%%%%%%%%%%%%%%%%%%%%%%

So far, we have discussed the characteristics of the oscillation valley
in the absence of NSI ($\varepsilon_{\mu\tau}=0$), and the measurement of
$\Delta m^2_{32}$ in this scenario. In this section, we explore the features
of the oscillation valley in the presence of NSI
(non-zero $\varepsilon_{\mu\tau}$).  We have already observed in
Sec.~\ref{subsec:osc-valley} that in the presence of
non-zero $\varepsilon_{\mu\tau}$, the oscillation valley in the oscillogram
has a curved shape that depends on the extent of $\epsmutau$, its sign,
and whether one is observing the $\mu^-$ or $\mu^+$ events.
Therefore, the curvature of oscillation valley is expected to be
a useful parameter for the determination of $\varepsilon_{\mu\tau}$.

%%%%%%%%%%%%%%%%%%%%%%%%%%%%%%%%%%%%%%%%%%%%%%%%%%%%%%%%%%%%%%%%%%
\subsection{Curvature of oscillation valley as a signature of NSI}
%\subsection{Extension of procedure in presence of NSI}
\label{subsec:curv-valley}
%%%%%%%%%%%%%%%%%%%%%%%%%%%%%%%%%%%%%%%%%%%%%%%%%%%%%%%%%%%%%%%%%%%

%%%%%%%%%%%%%%%%%%%%%%%%%%%%%%%%%%%%%%%%%%%%%%%%%%%%%%%%%%%%%%%%%
\begin{figure}[t]
  \centering
  \includegraphics[width=0.49\linewidth]{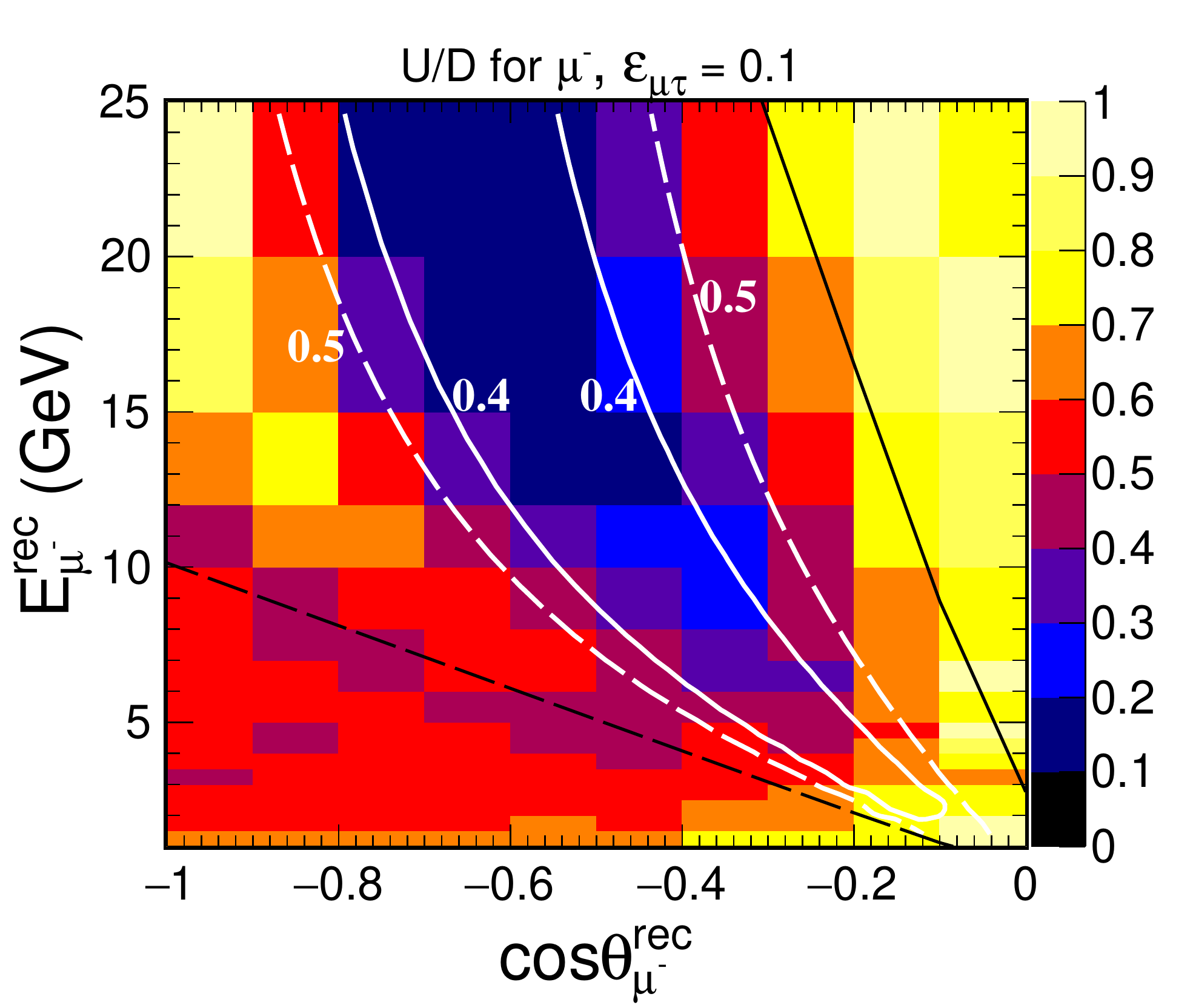}
  \includegraphics[width=0.49\linewidth]{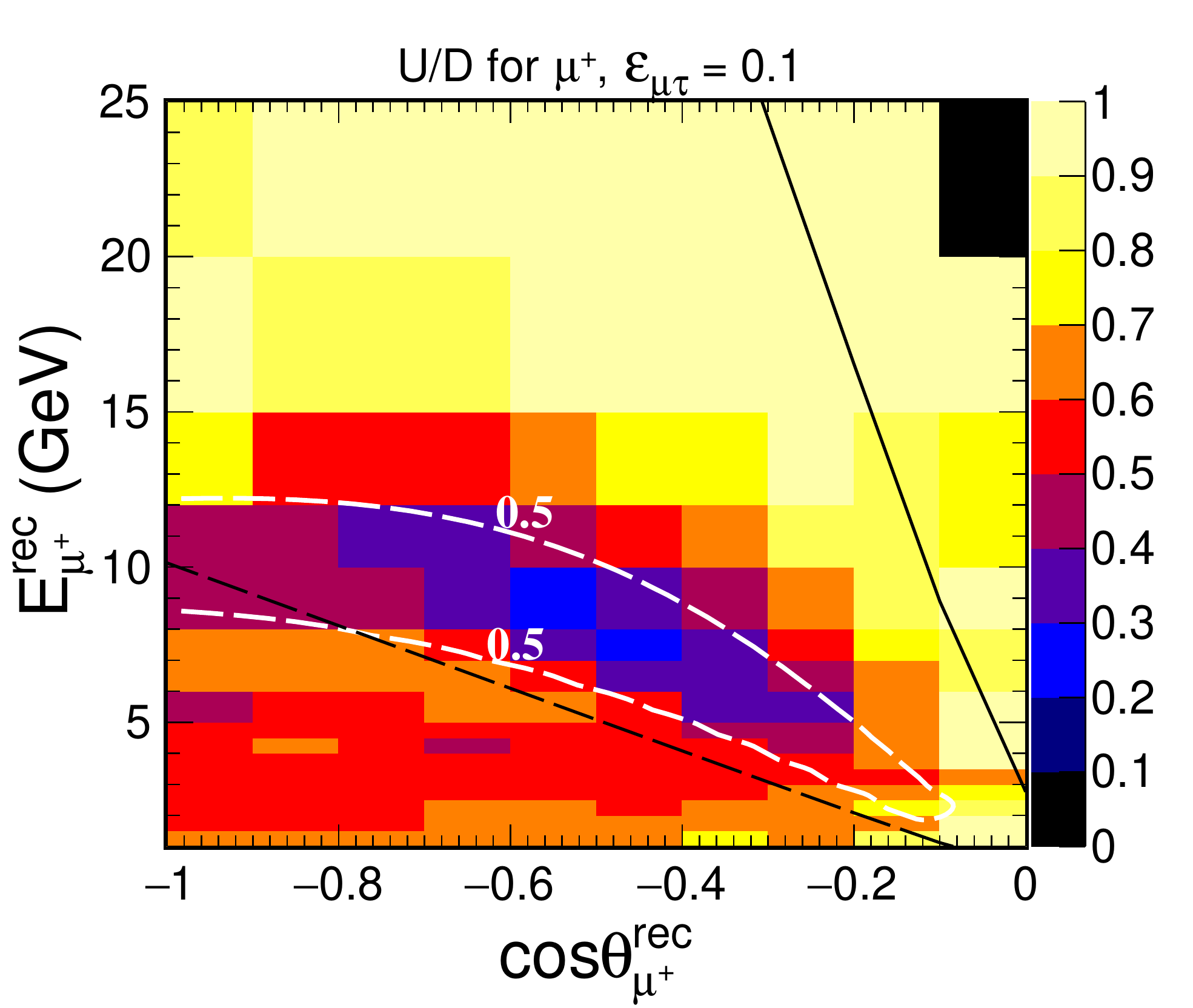}
  \includegraphics[width=0.49\linewidth]{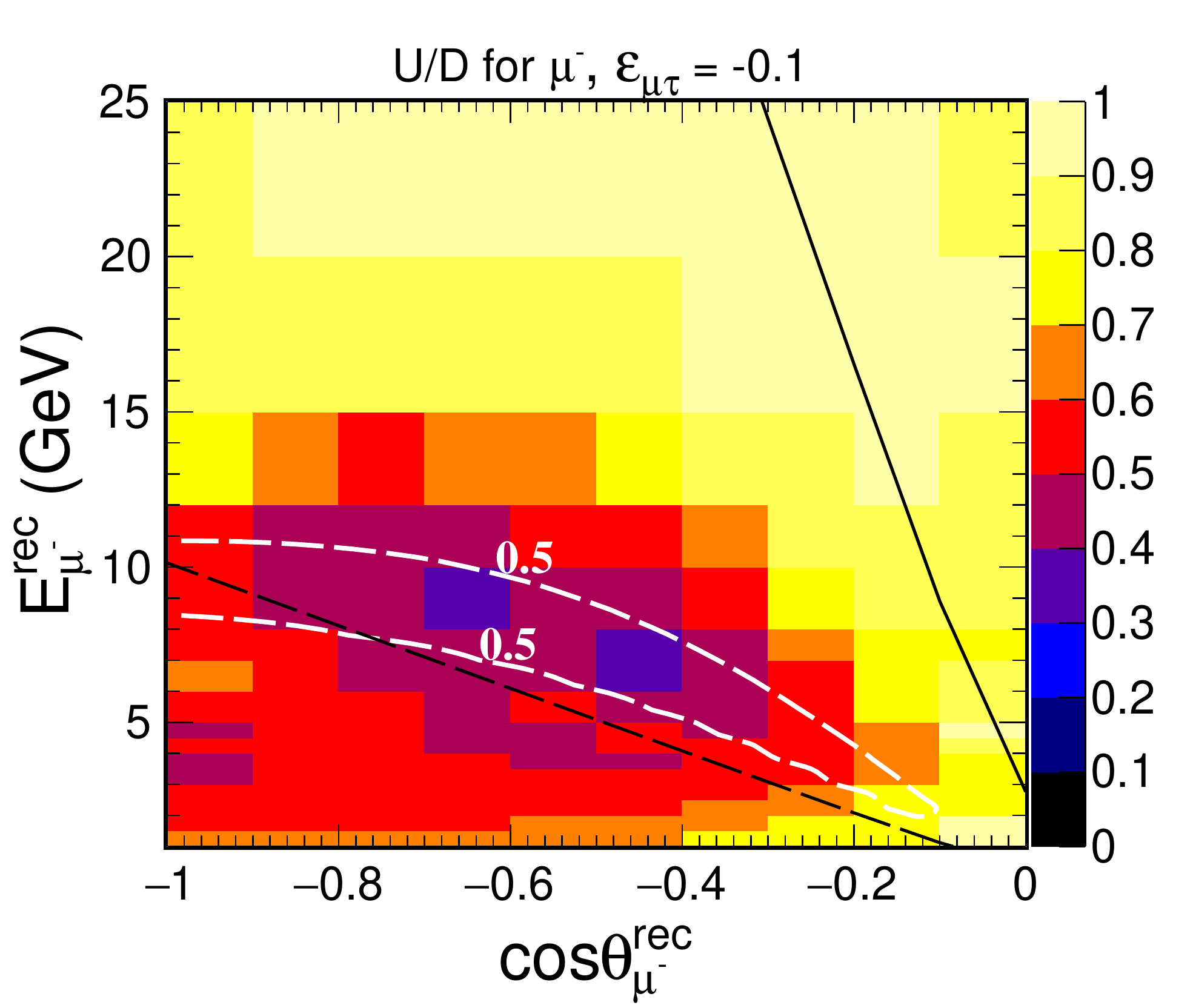}
  \includegraphics[width=0.49\linewidth]{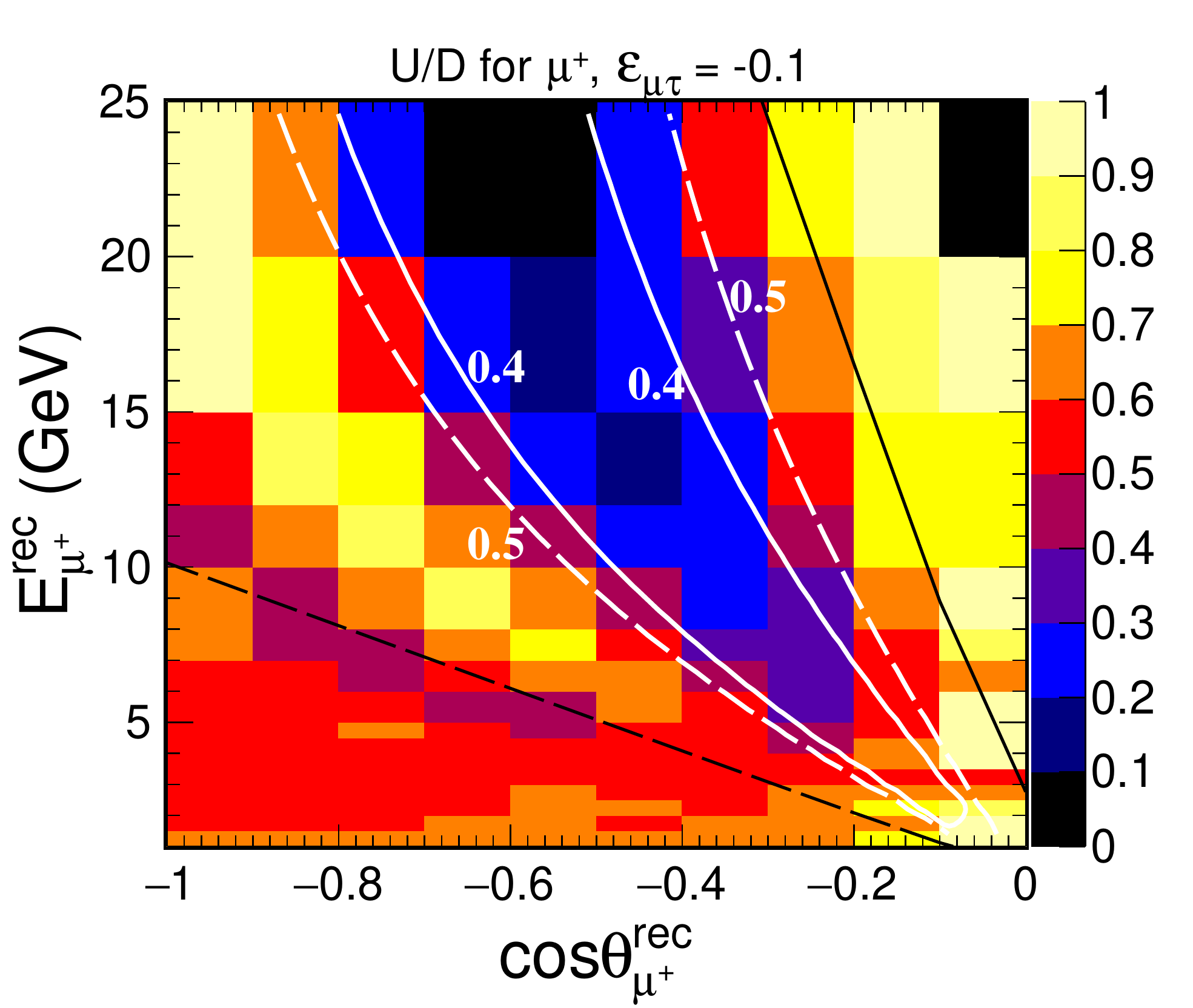}
  \mycaption{The distributions of U/D ratio in
    ($E_\mu^{\rm rec}$, $\cos\theta_\mu^{\rm rec}$) plane with non-zero
    $|\varepsilon_{\mu\tau}|$ (+0.1 in upper panels and -0.1 in lower panels)
    as expected at ICAL in 10 years. Left and right panels are for
    $\mu^-$ and $\mu^+$ events respectively.  The white solid and dashed
    lines correspond to contours with fitted U/D ratio equal to 0.4 and 0.5,
    respectively, obtained after fitting of the oscillation valley with
    function $F_\alpha$ (see Eq.~\ref{eq:fitting-func-nsi}).
    The black solid and dashed lines correspond to
    $\log_{10}[L_\mu^\text{rec}/E_\mu^\text{rec}] =$ 2.2 and 3.1, respectively.
    We consider normal mass ordering, and the benchmark oscillation
    parameters given in Table~\ref{tab:osc-param-value}.
    }    
    \label{fig:valley-nsi-u-by-d}
\end{figure}
%%%%%%%%%%%%%%%%%%%%%%%%%%%%%%%%%%%%%%%%%%%%%%%%%%%%%%%%%%%%%%%%%%%%%

Fig.\,\ref{fig:valley-nsi-u-by-d} presents the distributions of U/D ratios
of $\mu^-$ and $\mu^+$ events separately in the
$(E_\mu^{\rm rec}\,,\cos\theta_\mu^{\rm rec})$ plane, for two non-zero
$\varepsilon_{\mu\tau}$ values ({\it i.e.} $\pm 0.1$).
For the sake of demonstration (in order to reduce the statistical
fluctuations in the figure and emphasize the physics),
we show the binwise average of 100 independent data sets,
each with 10-year exposure of ICAL.
In contrast to the linear nature of oscillation valley with SI
($\varepsilon_{\mu\tau} =0$), the valley is curved in the presence of NSI.
The direction of bending of the oscillation valley is decided by sign of
$\varepsilon_{\mu\tau}$ as well as whether it is $\mu^-$ or $\mu^+$. 
An important point to note is that the nature of
curvature of the oscillation valley, that we observed in the
($E_\nu, \cos\theta_\nu$) plane with neutrino variables 
(see Fig.~\ref{fig:oscillogram}) with NSI, is preserved in the reconstructed
oscillation valley in the plane of reconstructed muon observables
($E_\mu^{\rm rec}, \cos\theta_\mu^{\rm rec}$).
This is due to the excellent energy and angular resolutions of the ICAL
detector for muons in the multi-GeV energy range.   

We retrieve the information on the bending of the oscillation valley
by fitting with an appropriate function, which is a generalization of        
Eq.\,\ref{eq:fitting-func-sm}. We introduce an additional free parameter
$\alpha$ that characterizes the curvature of the oscillation valley.
The function we propose for fitting the oscillation valley in the presence
of $\varepsilon_{\mu\tau}$ is  
\begin{equation}
  F_\alpha(E_\mu^{\rm rec}, \cos\theta_\mu^{\rm rec}) = Z_\alpha + N_\alpha \cos^2
  \left(m_\alpha \frac{\cos\theta_\mu^{\rm rec}}{E_\mu^{\rm rec}}  + \alpha\,
  \cos^2\theta_\mu^{\rm rec}  \right),
 \label{eq:fitting-func-nsi} 
\end{equation}
where $Z_\alpha$, $N_\alpha$, $m_\alpha$, and $\alpha$ are the free parameters
to be determined from the fitting of the U/D ratio in the
($E_\mu^{\rm rec},\,\cos\theta_\mu^{\rm rec}$) plane.
Equation\,\ref{eq:fitting-func-nsi} is inspired by 
the survival probability as given in Eq.\,\ref{eq:pmumu-nu-final-omsd}.
In Eq.~\ref{eq:fitting-func-nsi}, the parameter $\alpha$ enters multiplied
with  $\cos^2\theta_\mu^{\rm rec}$, where one factor of $\cos\theta_\mu^{\rm rec}$
includes the $L$ dependence of oscillation.
The other factor of $\cos\theta_\mu^{\rm rec}$ is to take into account the
baseline dependence of the matter potential. While this is a rather crude
approximation, it is sufficient to provide us a way of calibrating
$\epsmutau$, as will be seen later in the section. 
In Eq.\,\ref{eq:fitting-func-nsi}, the parameters $Z_\alpha$ and $N_\alpha$
fix the depth of valley, while $m_\alpha$ and $\alpha$ determine the
orientation and curvature of oscillation valley, respectively. Therefore,
$m_\alpha$ and $\alpha$ together account for the combined effect of
the mass-squared difference $\Delta m^2_{32}$ and the NSI parameter
$\varepsilon_{\mu\tau}$ in oscillations.    

As in the analysis in Sec.~\ref{subsec:dmsq-si}, we restrict the range of
$\log_{10}[L_\mu^\text{rec}/E_\mu^\text{rec}]$ to 2.2 -- 3.1, to remove the
unoscillated part ($<2.2$) as well as the region with significant
matter effects ($>3.1$). The cut on the U/D value is applied at 0.9.
In Fig.\,\ref{fig:valley-nsi-u-by-d}, the contours of the fitted U/D ratio
of 0.4 and 0.5, shown with white solid and dashed lines, respectively,
clearly indicate that the function $F_\alpha$ works well to reproduce the
curved nature of the oscillation valley with non-zero $\varepsilon_{\mu\tau}$.
We see that the oscillation valley for $\mu^-$ with positive (negative)
$\varepsilon_{\mu\tau}$ closely resembles that for $\mu^+$ with negative
(positive) $\varepsilon_{\mu\tau}$. This happens due to the degeneracy in
the sign of $\varepsilon_{\mu\tau}$ and the sign of matter potential $V_{\rm CC}$
-- these appear in the combination $\epsmutau V_{\rm CC}$ in the survival
probability expression in Eq.~\ref{eq:pmumu-nu-final-omsd}.
One can see that for $\mu^+$ ($\mu^-$) with $\epsmutau =0.1$ ($-0.1$),
the contour with the fitted U/D ratio of 0.4 is absent. This is because
the valley is shallower in these cases. This feature corresponds to
a similar one observed in Fig.~\ref{fig:updown-lbye-10}, where the
oscillation dip is shallower for $\mu^+$ ($\mu^-$) with
$\epsmutau =0.1$ ($-0.1$).

%%%%%%%%%%%%%%%%%%%%%%%%%%%%%%%%%%%%%%%%%%%%%%%%%%%%%%%%%%%%%%%%%%%%%%
\subsection{Constraints on $\varepsilon_{\mu\tau}$ from oscillation valley}
\label{subsec:eps-limit-osc-valley}
%%%%%%%%%%%%%%%%%%%%%%%%%%%%%%%%%%%%%%%%%%%%%%%%%%%%%%%%%%%%%%%%%%%%%

We saw in the previous section that the shape of the oscillation valley
in the muon observables is well-approximated by the function in
Eq.\,\ref{eq:fitting-func-nsi}, with the two parameters $\alpha$ and $m_\alpha$.
Therefore, we propose to use these two parameters for retrieving
$\varepsilon_{\mu\tau}$. At ICAL, since the data on $\mu^-$ and $\mu^+$ events
are available separately, we can have two independent fits for  the U/D
ratios in $\mu^-$ and $\mu^+$ events. This would give us independent
determinations of the parameter pairs $(\alpha^-, m_\alpha^-)$ and
$(\alpha^+, m_\alpha^+)$, for $\mu^-$ and $\mu^+$ events, respectively.
One can then get the calibration of $\varepsilon_{\mu\tau}$ in the
plane of $\alpha$ and $m_\alpha$, for $\mu^-$ and $\mu^+$ events, separately.
For other experiments that do not have charge identification capability,
there would be a single calibration curve of  $\varepsilon_{\mu\tau}$ with
$\alpha$ and $m_\alpha$, based on the U/D ratio of muon events without
charge information. 

%%%%%%%%%%%%%%%%%%%%%%%%%%%%%%%%%%%%%%%%%%%%%%%%%%%%%%%%%%%%%%%%%%%%%
\begin{figure}[t]
  \centering
  \includegraphics[width=0.75\linewidth]{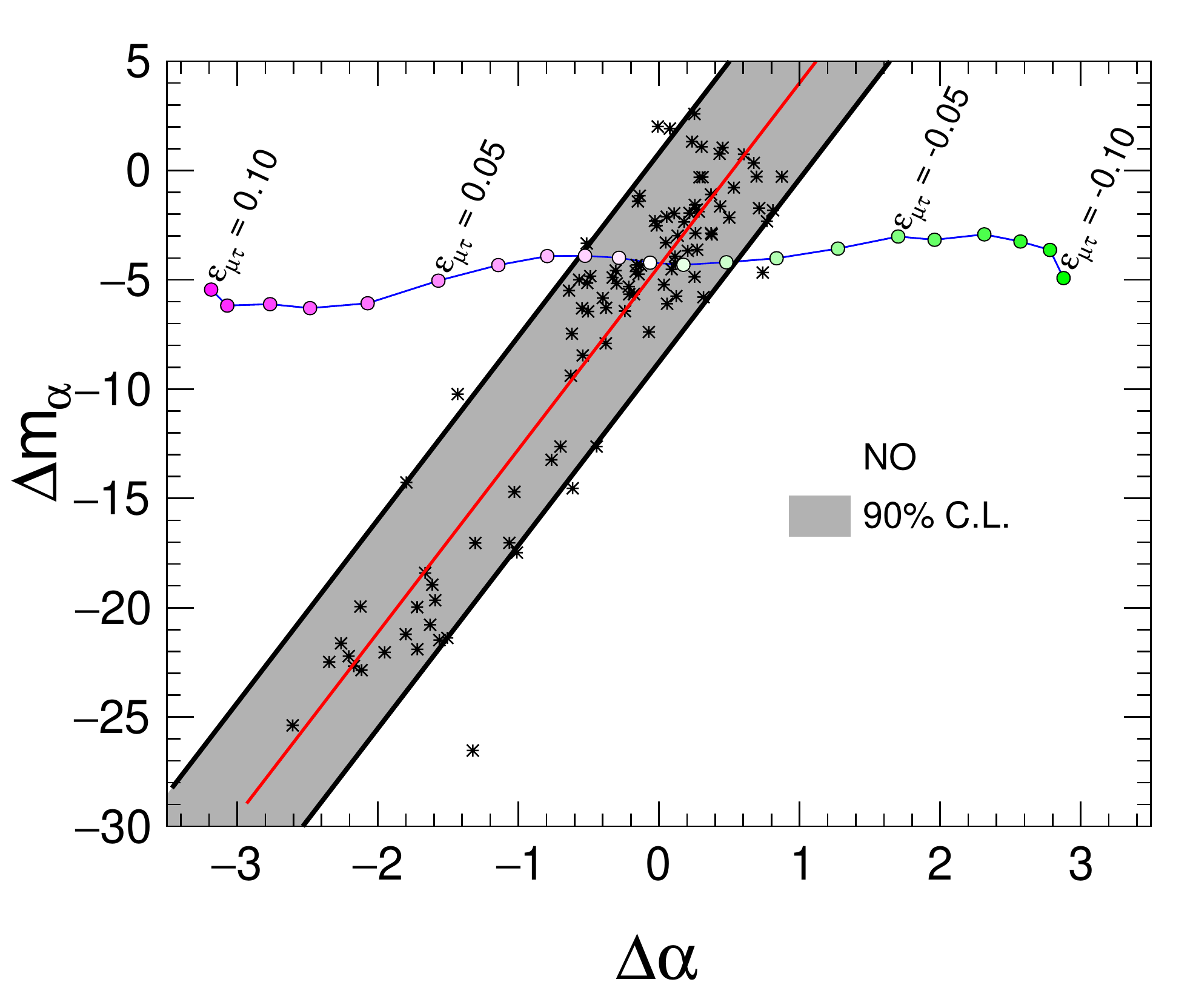}
  \mycaption{
      The blue curve with colored circles shows the calibration of $\epsmutau$
      in the $(\Delta \alpha, \Delta m_\alpha)$ plane, obtained from 1000-year MC.
      The black stars are the values of $(\Delta \alpha,\Delta m_\alpha$)
      obtained from 100 independent data sets, each for 10 years data of ICAL,
      when $\epsmutau=0$.
      The gray band represents the expected 90\% C.L. region in this plane  
      for a 10-year exposure at ICAL, when $\epsmutau=0$. 
      The part of the blue calibration line covered by the gray band
      gives the expected 90\% C.L. bounds on $\epsmutau$.
      We consider normal mass ordering, and the benchmark oscillation
      parameters given in Table~\ref{tab:osc-param-value}.
 }
 \label{fig:scatter-diff-m-alpha}
\end{figure}
%%%%%%%%%%%%%%%%%%%%%%%%%%%%%%%%%%%%%%%%%%%%%%%%%%%%%%%%%%%%%%%%%%%%%

Building up on our experience in the analysis of the oscillation dip
in the presence of NSI (see Sec.\,\ref{subsec:delta-d}), in this analysis
of oscillation valley, we use the following observables that quantify the
$\mu^-$ vs. $\mu^+$ contrast in the values of the parameters $\alpha$ and
$m_\alpha$: 
\begin{equation}
  \Delta \alpha = \alpha^- - \alpha^+\,, \hspace{1cm} 
  \Delta m_\alpha = m_{\alpha^-} - m_{\alpha^+} \,.
\end{equation}
The blue line with colored circles in Fig.\,\ref{fig:scatter-diff-m-alpha} shows the calibration
of $\varepsilon_{\mu\tau}$ in ($\Delta \alpha, \,\Delta m_\alpha$) plane,
obtained by using the 1000-year MC sample. It may be observed that the
calibration of $\varepsilon_{\mu\tau}$ is almost independent of $\Delta m_\alpha$,
and $\varepsilon_{\mu\tau}$ is almost linearly  proportional to $\Delta \alpha$.

In order to  estimate the extent to which $\epsmutau$ may be constrained
at ICAL with an exposure of 10 years, we simulate the statistical fluctuations
by generating 100 independent data sets, each corresponding to 10-year
exposure and $\epsmutau=0$. 
The black dots in Fig.\,\ref{fig:scatter-diff-m-alpha} are the values of
$\Delta \alpha$ and $\Delta m_\alpha$ determined from each of these data sets.
The figure shows the results, with the benchmark oscillation
parameters in Table~\ref{tab:osc-param-value}.

It is observed that with a 10-year data, the statistical fluctuations
lead to a strong correlation between the values of $\Delta \alpha$
and $\Delta m_\alpha$, and we have to employ a two-dimensional fitting
procedure. We fit the scattered points in $(\Delta \alpha$, $\Delta m_\alpha$)
plane with a straight line. The best fitted straight line is shown by
the red color in Fig.\,\ref{fig:scatter-diff-m-alpha}, which corresponds
to the average of these points, and indeed passes through the calibration
point obtained from 1000-year MC events for $\epsmutau=0$.
The 90$\%$ C.L. limit on $\varepsilon_{\mu\tau}$ is then obtained based on
the measure of the perpendicular distance of every point from the red line.
The region with the gray band in Fig.\,\ref{fig:scatter-diff-m-alpha}
contains 90\% of the points that are closest to the red line.
The expected 90\% C.L. bounds on $\epsmutau$ can then be determined from
the part of the blue calibration line covered by the gray band.
Any value of the pair $(\Delta \alpha, \Delta m_\alpha)$ that lies outside the
gray band would be a smoking gun signature for nonzero  $\epsmutau$
at 90$\%$ C.L.
Figure \,\ref{fig:scatter-diff-m-alpha} indicates that if
$\epsmutau$ is indeed zero, our method, as applied at the ICAL detector,
would be able to constrain it to the range $-0.022<\varepsilon_{\mu\tau}<0.021$
at 90\% C.L., with an exposure of 500 kt$\cdot$yr.

We explore the possible dilution of our results due to the
uncertainties in oscillation parameters and the five systematic
errors, following the same procedure described in
Sec.~\ref{subsec:eps-limit-delta-d}.  
This yields the expected 90\% C.L. bounds as
$-0.024<\varepsilon_{\mu\tau}<0.020$, thus keeping them almost unchanged.

%%%%%%%%%%%%%%%%%%%%%%%%%%%%%%%%%%%%%%%%%%%%%%%%%%%%%%%%%%%%%%%%%%%%
%%%%%%%%%%%%%%%%%%%%%%%%%%%%%%%%%%%%%%%%%%%%%%%%%%%%%%%%%%%%%%%%%%%%
\section{Summary and Concluding Remarks}
\label{sec:conclusion}
%%%%%%%%%%%%%%%%%%%%%%%%%%%%%%%%%%%%%%%%%%%%%%%%%%%%%%%%%%%%%%%%%%%%%
%%%%%%%%%%%%%%%%%%%%%%%%%%%%%%%%%%%%%%%%%%%%%%%%%%%%%%%%%%%%%%%%%%%%%

Atmospheric neutrino experiments are suitable for exploring the
non-standard interactions of neutrinos due to the accessibility to
high energies ($E_\nu$) and long baselines ($L_\nu$) through the Earth,
for which the effect of NSI-induced matter potential is large.
The NSI matter potential produced in the interactions of neutrino and
matter fermions may change the oscillation pattern.
The multi-GeV neutrinos, in particular, offer the advantage to access
the interplay of NSI and Earth matter effects.
In this paper, we have explored the impact of the NSI parameter
$\varepsilon_{\mu\tau}$ on the dip and valley patterns in neutrino
oscillation probabilities, and proposed a new method for extracting
information on $\epsmutau$ from the atmospheric neutrino data.

Detectors like ICAL, that have good reconstruction capability for
the energy, direction, and charge of muons, can reproduce the
neutrino oscillation dip and valley patterns in their reconstructed muon
data, from $\mu^-$ and $\mu^+$ events separately.
The observables chosen to reproduce these patterns are the ratios
of upward-going (U) and downward-going (D) events, which minimize
the effects of systematic uncertainties.
We analyze the U/D ratio in bins of reconstructed muon observables
$L_\mu^{\rm rec}/E_\mu^{\rm rec}$, and in the two-dimensional plane
$(E_\mu^{\rm rec}, \cos\theta_\mu^{\rm rec})$.
We show that nonzero $\epsmutau$ shifts the locations of oscillation dips
in the reconstructed $\mu^-$ and $\mu^+$ data in opposite directions
in the $L_\mu^{\rm rec}/E_\mu^{\rm rec}$ distributions.
At the same time, nonzero $\epsmutau$ manifests itself in the curvatures of
the oscillation valleys in the $(E_\mu^{\rm rec}, \cos\theta_\mu^{\rm rec})$ plane,
this bending being in opposite directions for $\mu^-$ and $\mu^+$ events.
The direction of shift in the dip location and the direction of
the bending of the valley also depend on the sign of
$\epsmutau$, as well as on the mass ordering.  ICAL, therefore, will be sensitive to sgn($\epsmutau$), once we know the neutrino mass ordering. We present our results in
the normal mass ordering scenario, the analysis for the inverted mass
ordering scenario will be exactly analogous. 

In the oscillation dip analysis, we introduce a new observable $\Delta d$,
corresponding to the difference in the locations of dips in
$\mu^-$ and $\mu^+$ event distribution.
We show the calibration of $\epsmutau$ against $\Delta d$ using
1000-year MC sample at ICAL, and demonstrate that it is almost independent of
the actual value of $\Delta m^2_{32}$. 
We incorporate the effects of statistical fluctuations corresponding to
10-year simulated data, uncertainties in the neutrino oscillation parameters,
and major systematic errors, by simulating multiple data sets.
Including all these features, it is possible to constrain the
NSI parameter in the range $-0.025 < \epsmutau < 0.024$ at 90\% C.L.
with 500 kt$\cdot$yr exposure.

In the oscillation valley analysis, we propose a function $F_\alpha$
that quantifies the curvature ($\alpha$) and orientation ($m_\alpha$) of
the oscillation valley in the ($E_\mu^{\rm rec}$, $\cos_\mu^{\rm rec}$) plane.
This analysis may be performed for the $\mu^-$ and $\mu^+$ events separately,
leading to independent measurements of $\epsmutau$.
However, we further find that the differences in these parameters for
the $\mu^-$ and $\mu^+$ events are quantities that are sensitive to
$\epsmutau$, and show the calibration for $\epsmutau$ in the
$(\Delta \alpha, \Delta m_\alpha)$ plane using 1000-year MC sample at ICAL.
Incorporating the effects of statistical fluctuations, uncertainties in the
neutrino oscillation parameters, and major systematic errors,
the NSI parameter can be constrained in the range 
$-0.022 < \epsmutau < 0.021$ at 90\% C.L.
with 500 kt$\cdot$yr exposure.

A special case of the function $F_\alpha$, with $\alpha=0$,
corresponds to a valley without any curvature. It is found that the
method of fitting for this function $F_0$ to the U/D ratio in the 
($E_\mu^{\rm rec}$, $\cos_\mu^{\rm rec}$) plane leads to a more robust and
precise measurement of the value of $\Delta m^2_{32}$
than the method proposed earlier~\cite{Kumar:2020wgz}.

Our analysis procedure is quite straightforward and transparent, and
is able to capture the physics of the NSI effects on muon observables 
quite efficiently. It builds upon the major features of the impact of
$\epsmutau$ on the oscillation dip, and identifies the
contrast in the shift in dip locations of $\mu^-$ and $\mu^+$ as the
key observable that would be sensitive to NSI.
It exploits the major pattern in the complexity of the two-dimensional
event distributions in the ($E_\mu^{\rm rec}$, $\cos_\mu^{\rm rec}$) plane by
fitting to a function approximately describing the curvature of the valley.
The two complementary methods of using the oscillation dips and valleys
provide a check of the robustness of the results. The shift of the dip location and the bending of the oscillation valley, and survival of these effects in the reconstructed muon data, are deep physical insights obtained through the analysis in this paper.

Note that, though the dip and the valley both arise from the first oscillation minimum,  
the information content in the valley analysis is more than that in the dip analysis. 
For example, (a) the features of the valley (like curvature) in Fig.~\ref{fig:valley-nsi-u-by-d} cannot be predicted from the shift in the dip as seen in Fig.~\ref{fig:updown-lbye-10},  and (b) the shift in the dip may have many possible sources, but the specific nature of bending of the valley would act as the confirming evidence for NSI. Therefore, physics unravelled from the valley analysis is much richer than that from the dip.

Note that muon charge identification plays a major role in our analysis.
In a detector like ICAL, the measurement of $\epsmutau$ is possible
independently in both the $\mu^-$ and $\mu^+$ channels.
The oscillation dips and valleys in $\mu^-$ and $\mu^+$ event samples
have different locations and shapes.
As a result, the information available in the dip and valley would be
severely diluted in the absence of muon charge identification.
But more importantly, the quantities sensitive to $\epsmutau$ in a robust
way turn out to be the ones that reflect the differences in the properties
of oscillation dips and valleys in $\mu^-$ and $\mu^+$ events.  
For example, $\Delta d$ is the observable identified by us, whose
calibration against $\epsmutau$ is almost independent of the actual
value of $\Delta m^2_{32}$.
For $\Delta d$ to be measured at a detector, the muon charge identification
capability is necessary.
The data from ICAL will thus provide the measurement of a crucial observable
that is not possible at other large atmospheric neutrino experiments
like Super-K or DeepCore/ IceCube.
At future long-baseline experiment like DUNE, that will have access
to 1 -- 10 GeV neutrino energy and separate data sets for neutrino and
antineutrino, our methodology to probe NSI using oscillation-dip location
can also be adopted, by replacing the U/D ratio with the ratio of
observed number of events and the predicted number of events without
oscillations.

We expect that further exploration of the features of the oscillation
dips and valleys, observable at ICAL-like experiments, would enable us
to probe neutrino properties in novel ways.

%==========================
\subsubsection*{Acknowledgments}
%==========================

This work is performed by the members of the INO-ICAL collaboration 
to suggest a new approach to probe Non-Standard Interactions (NSI) 
using atmospheric neutrino experiments which can distinguish between  
neutrino and antineutrino events. We thank S. Uma Sankar, Srubabati 
Goswami, and D. Indumathi for their useful and constructive comments  
on our work. We thank P. Denton, S. Petcov, G. Rajasekaran. and J. Salvado for useful communications. A. Kumar would like to thank the organizers of the virtual Mini-Workshop on Neutrino Theory at Fermilab, Chicago, USA during 21st to 23rd September, 2020 for giving an opportunity to present the main results of this work. A.K. would also like to thank the organizers of the XXIV DAE-BRNS High Energy Physics Online Symposium at NISER, Bhubaneswar, India during 14th to 18th December, 2020 for providing him an opportunity to give a talk based on this work. We acknowledge support of the Department of Atomic Energy (DAE), Government of India, under Project Identification No. RTI4002. S.K.A. is supported by the DST/INSPIRE Research Grant [IFA-PH-12] from the Department of Science and Technology (DST), India and the Young Scientist Project [INSA/SP/YSP/144/2017/1578] from the Indian National Science Academy (INSA). S.K.A. acknowledges the financial support from the Swarnajayanti Fellowship Research Grant (No. DST/SJF/PSA-05/2019-20) provided by the DST, Govt. of India and the Research Grant (File no. SB/SJF/2020-21/21) provided by the Science and Engineering Research Board (SERB) under the Swarnajayanti Fellowship by the DST, Govt. of India.

%===============================
\bibliographystyle{JHEP}
\bibliography{NSI-Atmospheric}
%================================

\end{document}